\documentclass[12pt]{article}
\pdfoutput=1
\usepackage[utf8]{inputenc}
\usepackage[DIV=13]{typearea}
\usepackage{ragged2e}
\usepackage{amsmath,amsfonts,bbm,mathrsfs,amssymb}
\usepackage{slashed,cancel}
\usepackage{cite}
\usepackage{bm} 
\usepackage{hyperref}
\hypersetup{colorlinks=true,urlcolor=magenta,anchorcolor=blue,citecolor=blue,filecolor=blue,
            linkcolor=magenta,menucolor=blue, linktocpage=true,pdfproducer=medialab}
\usepackage[english]{babel}
\usepackage{catchfile}
\usepackage{indentfirst}
\usepackage{graphicx}
\usepackage{float}
\usepackage{enumerate}
\usepackage{caption}
\usepackage{subfigure}
\usepackage{multirow}
\usepackage{tikz-feynman}
\usepackage[normalem]{ulem} 
\usepackage{booktabs}
\usepackage{physics}

\newcommand{\email}[1]{\href{mailto:#1}{\tt #1}}

\numberwithin{equation}{section}

\newcommand{\blue}[1]{\color{blue} #1 \color{black}}

\newcommand{\magenta}[1]{\color{magenta} #1 \color{black}}
\newcommand{\be}{\begin{equation}}
\newcommand{\ee}{\end{equation}}
\newcommand{\ba} {\begin{equation}\begin{aligned}}
\newcommand{\ea} {\end{aligned}\end{equation}}

\newcommand{\sL}{\mathscr{L}}

\newcommand{\cB}{\mathcal{B}}
\newcommand{\cG}{\mathcal{G}}

\newcommand{\cH}{\mathcal{H}}
\newcommand{\cO}{\mathcal{O}}
\newcommand{\cM}{\mathcal{M}}

\newcommand{\bc}{\mathbf{c}}
\newcommand{\bK}{\mathbf{K}}

\newcommand{\derp}{\partial}
\newcommand{\hc}{\text{h.c.}}

\newcommand{\ov}[1]{\overline{#1}}

\newcommand{\eV}{\ \text{eV}}

\newcommand{\TeV}{\ \text{TeV}}
\newcommand{\GeV}{\ \text{GeV}}
\newcommand{\MeV}{\ \text{MeV}}

\def\Tr{{\rm Tr}}

\def\BR{\mathcal{B}}

\def\vs{{\textit vs.} }
\newcommand*{\diff}[1]{\text{d}#1}
\newcommand\subsetsim{\mathrel{%
  \ooalign{\raise0.2ex\hbox{$\subset$}\cr\hidewidth\raise-0.8ex\hbox{\scalebox{0.9}{$\sim$}}\hidewidth\cr}}}

\def\Tr{{\rm Tr}}

\def\BR{\mathcal{B}}

\begin{document} 
\renewcommand*{\thefootnote}{\fnsymbol{footnote}}
\begin{titlepage}

\vspace*{-1cm}
\flushleft{\magenta{IFT-UAM/CSIC-22-109}}
\\[1cm]

\begin{center}
\bf\LARGE \blue{
The cost of an ALP solution to the}\\[2mm]
\boldmath
\bf\LARGE \blue{
neutral $B$-anomalies}\\[4mm]
\unboldmath
\centering
\vskip .3cm
\end{center}
\vskip 0.5  cm
\begin{center}
{\large\bf J.~Bonilla}~\footnote{\email{jesus.bonilla@uam.es}},
{\large\bf A.~de Giorgi}~\footnote{\email{arturo.degiorgi@uam.es}},
{\large\bf B.~Gavela}~\footnote{\email{belen.gavela@uam.es}},\\[2mm]
{\large\bf L.~Merlo}~\footnote{\email{luca.merlo@uam.es}}
and 
{\large\bf M.~Ramos}~\footnote{\email{maria.pestanadaluz@uam.es}}
\vskip .7cm
{\footnotesize
Departamento de F\'isica Te\'orica and Instituto de F\'isica Te\'orica UAM/CSIC,\\
Universidad Aut\'onoma de Madrid, Cantoblanco, 28049, Madrid, Spain
}
\end{center}
\vskip 2cm
\begin{abstract}
\justify

The neutral anomalies in $B$ decays are analysed in terms of the tree-level exchange of an axion-like-particle (ALP), within the  effective field theory framework. The complete two-dimensional  parameter space for ALP  couplings to electrons and muons is explored. The solutions to $R_K$ and to the two energy bins of $R_{K^\ast}$ are confronted with the impact of ALP exchange on other observables (meson oscillations,  leptonic  and semileptonic decays of $B$ mesons including searches for new resonances, astrophysical constraints), as well as with the theoretical domain of validity of the effective theory. Solutions based on ALPs heavier than $B$ mesons, or lighter than twice the muon mass, are shown to be excluded. In contrast, the exchange of on-shell ALPs provides solutions to   $R_K$ and/or $R_{K^\ast}$  within $2\sigma$ sensitivity which are technically compatible with those constraints. Furthermore, a ``golden ALP mass''  is identified at the frontier between the two energy bin windows of $R_{K^\ast}$, which could simultaneously explain these two $R_{K^\ast}$ anomalies together with  $R_K$; this calls for the convenience of different energy binning which would easily clear up this (unlikely) possibility. The impact of smearing on data analysis is also discussed. When loop effects are taken into account, the solutions found can  be in addition compatible with the data on the $g-2$ of the electron but not simultaneously with those on the $g-2$ of the muon.  Furthermore, loop effects may require fine-tunings of some coupling values.

\end{abstract}
\end{titlepage}
\setcounter{footnote}{0}

\pdfbookmark[1]{Table of Contents}{tableofcontents}
\tableofcontents

\renewcommand*{\thefootnote}{\arabic{footnote}}

\newpage
%
%
\section{Introduction}
\label{sec:intro}

Despite the huge experimental and theoretical effort  in direct searches at colliders and low-energy facilities,  no new particle has been observed since the  discovery of the Higgs boson~\cite{Englert:1964et,Higgs:1964ia,Higgs:1964pj}  at the  LHC~\cite{ATLAS:2012yve,CMS:2012qbp} a decade ago. Although this discovery constitutes a superb confirmation  of the Standard Model of particle physics (SM), an explanation of the origin of neutrino masses, the nature of Dark Matter, the baryon asymmetry of the Universe and a quantum-level description of gravity are lacking. 

Furthermore, in  recent years anomalies associated with the $B$ mesons have been observed as compared with SM expectations. Those include deviations in both neutral and charged current processes. Neutral current anomalous behaviour manifests in the angular distribution of  $B^0\to K^{0\ast}\mu^+\mu^-$ decay~\cite{LHCb:2013ghj,LHCb:2015svh,LHCb:2020lmf,LHCb:2020gog}, and in the  observed Lepton Flavour Universality (LFU)-violating quotient of the branching ratios for $B^\pm\to K^\pm\mu^+\mu^-$  \vs $B^\pm\to K^\pm e^+ e^-$ and for $B^0\to K^{0\ast}\mu^+\mu^-$  \vs $B^0\to K^{0\ast}e^+ e^-$
\cite{LHCb:2014vgu,LHCb:2017avl,LHCb:2019hip,LHCb:2021trn}. The LFU ratios are particularly clean observables theoretically and experimentally~\cite{Hiller:2003js,Bordone:2016gaq,Isidori:2020acz} and therefore represent an excellent window to new physics (NP).  Their generic expression in terms of the dilepton invariant mass $q^2$ reads
\be
R_X\equiv\dfrac{\displaystyle\int_{q^2_\text{min}}^{q^2_\text{max}}\dfrac{\dd\Gamma\left(B\to X_s\,\mu^+\mu^-\right)}{\dd q^2}\dd q^2}{\displaystyle\int_{q^2_\text{min}}^{q^2_\text{max}}\dfrac{\dd\Gamma\left(B\to X_s\,e^+e^-\right)}{\dd q^2}\dd q^2}\,,
\ee
where $X_s$ stands for either a $K$ or a $K^\ast$ meson, and where --here and in what follows-- the meson electric charges are implicit. Their most recent  and precise determination results in
\begin{align}
R_K=&0.846^{+0.042}_{-0.039}{}^{+0.013}_{-0.012}\qquad\quad\text{for}\quad1.1\GeV^2\leq q^2\leq6.0 
\GeV^2\qquad \text{central bin}\quad\text{\cite{LHCb:2021trn}}
\label{ExperimentalRK}\\[2mm]
R_{K^\ast}=&
\begin{cases}
0.69^{+0.11}_{-0.07}{}\pm0.05 & \quad\text{for}\quad1.1\GeV^2\leq q^2\leq6.0\GeV^2\qquad \text{central bin}
\\[2mm]
0.66^{+0.11}_{-0.07}{}\pm0.03 & \quad\text{for}\quad0.045\GeV^2\leq q^2\leq1.1\GeV^2\quad \text{low bin}
\end{cases}
~\text{\cite{LHCb:2017avl}}
\label{ExperimentalRKstar}
\end{align}
where  $R_K$ refers to data from $B^+$ meson decays and $R_{K^\ast}$ to data from $B^0$ decays, and where central/low bin refers to the higher/lower bin in $q^2$  for which experimental data are available. The SM prediction for $R_K$ and $R_{K^\ast}$ at the central bin region is $1.00\pm0.01$~\cite{Hiller:2003js,Bobeth:2007dw,Bordone:2016gaq}, while for $R_{K^\ast}$ at the low bin region is $0.92\pm 0.02$~\cite{Capdevila:2017bsm}. The measured deviations from these values represent the so-called neutral $B$-anomalies, with a significance of $3.1\sigma$, $2.5\sigma$ and $2.3\sigma$, respectively. Furthermore, anomalies in charged current processes have appeared in the form of LFU violation in the quotients of  $B$ semileptonic decay rates to $\tau$ leptons \vs those to electrons and muons. 

The not very high significance of each individual channel/measurement calls for caution: a purely experimental resolution --statistical fluctuation or systematic effect--  is not excluded. Nevertheless, the  different deviations are intriguingly consistent with each other once treated in an effective field theory description, as first formulated in Ref.~\cite{Descotes-Genon:2013wba} and recently updated in Refs.~\cite{Ciuchini:2020gvn,Alguero:2021anc,Altmannshofer:2021qrr,Geng:2021nhg,Hurth:2021nsi,Cornella:2021sby}. Altogether,  they could be interpreted as due to NP with a global statistical significance of $4.3\sigma$~\cite{Isidori:2021vtc}. Although no single flavour measurement exhibits a $5\sigma$ deviation from the SM, the emerging pattern could point to NP that violates lepton flavour universality, in particular in what concerns $R_{K^{(\ast)}}$. Other promising channels to test LFU are associated with  $\Lambda_b^0\to pK^-\ell^+\ell^-$, $B^+\to K^+\pi^+\pi^-\ell^+\ell^-$ and $B^0\to K^+\pi^-\ell^+\ell^-$ decays, which however are delicate observables as it is not known how the NP affects the hadronic structure of the final states involved (see Ref.~\cite{Isidori:2021tzd} for a possible strategy to overcome this problem).

The neutral LFU ratios are loop-level processes within the SM and the size of the observed deviations thus opens the possibility to explain those anomalies via  tree-level exchanges of NP particles. The first attempts in this direction in the last decade mainly focused on $Z'$ models~\cite{Descotes-Genon:2013wba,Altmannshofer:2013foa,Buras:2013qja,Crivellin:2015xaa,Celis:2015ara} or on lepto-quark scenarios~\cite{Alonso:2015sja,Calibbi:2015kma,Barbieri:2015yvd,Buttazzo:2017ixm}. In this paper, we will instead investigate the possibility that an axion-like-particle (ALP) reduces and eventually solves these neutral $B$-anomalies.

 Axions have been originally introduced as the pseudo-Goldstone-bosons (pGBs) which result from the dynamical solution to the SM strong CP problem~\cite{Peccei:1977hh,Peccei:1977ur,Weinberg:1977ma,Wilczek:1977pj} through a global chiral $U(1)$ symmetry --classically exact but anomalous at the quantum level. However, pGBs  also appear in a plethora of theories that extend the SM even if not linked to a solution to the strong CP problem. These include among others  the Majoron which stems from dynamical explanations of the lightness of active neutrino masses~\cite{Gelmini:1980re}, pGBs from supersymmetric frameworks~\cite{Bellazzini:2017neg}; the Higgs boson itself which can have a pGB nature as in Composite-Higgs models~\cite{Georgi:1984af}; and pGBs associated to extra-dimensional theories and string theories, which typically exhibit hidden $U(1)$'s~\cite{Cicoli:2013ana}. Frequently these pGBs have anomalous couplings to gauge currents and are described by the generic name of ALPs.
 
\begin{figure}[h!]
\centering
\includegraphics[width=0.7\textwidth]{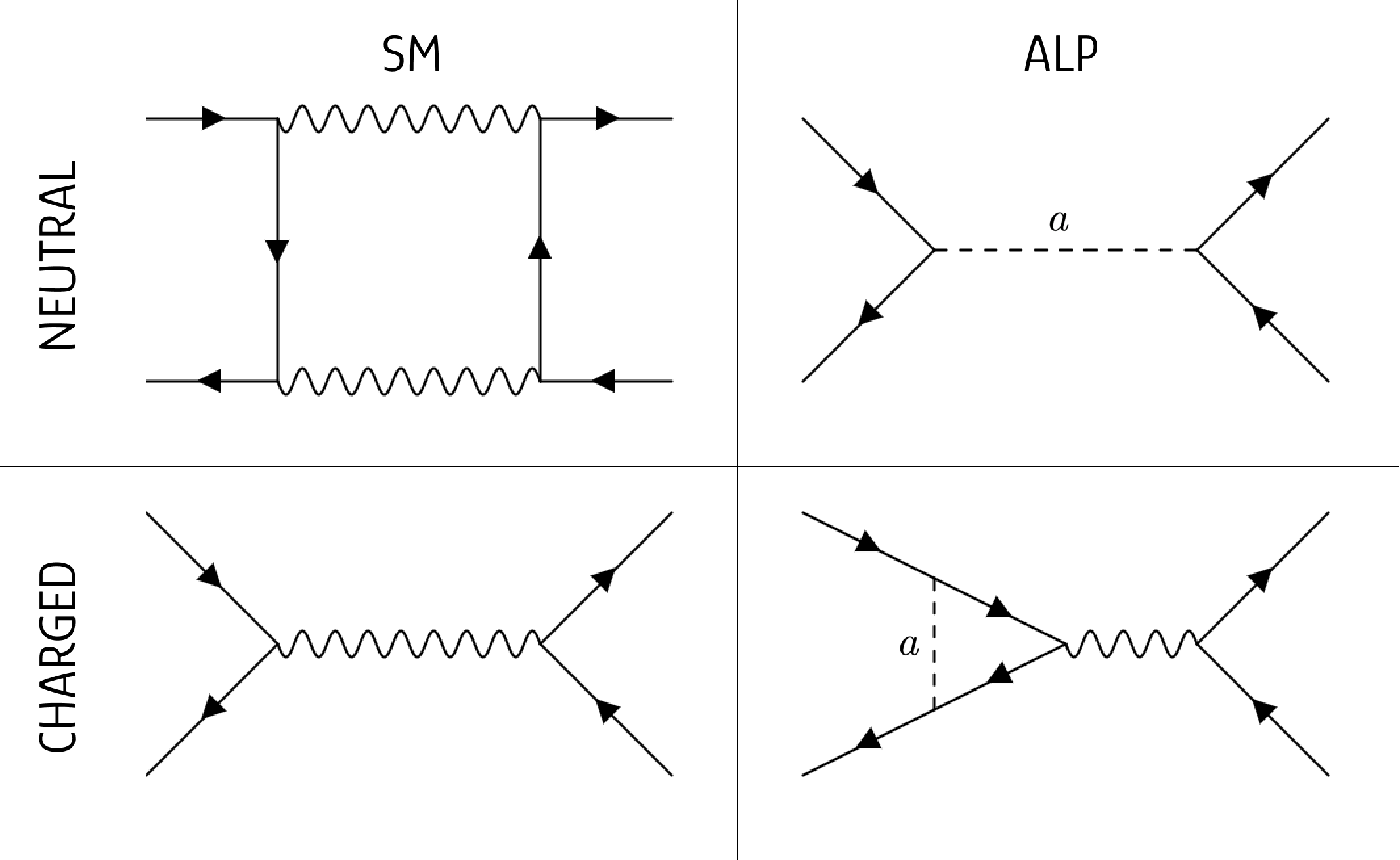}
\caption{\em Sketchy illustrations of flavour-changing neutral and charged currents in the SM \vs those induced by flavour-non-diagonal ALP  couplings. The wiggly lines denote SM electroweak gauge bosons, and all processes depicted are assumed to change flavour.}
\label{fig:sketchy-diagram}
\end{figure}

In the SM, flavour-changing charged currents  appear already as tree-level exchanges while neutral ones are one-loop suppressed processes.  The exchange of ALPs exhibits generically the opposite pattern: it induces flavour-changing 
neutral currents already at tree-level while charged ones require one-loop transitions, see Fig.~\ref{fig:sketchy-diagram}. We focus below on whether the tree-level exchange of ALPs can account for the neutral $B$-anomalies.  Although this question has been previously formulated in Ref.~\cite{Bauer:2021mvw}, we present for the first time the study of the complete parameter space, determining new solutions. 

Both the case of a heavy ALP and of a light ALP are considered (heavy/light as compared to $B$ meson masses). 
All the possible ranges for ALP masses will be carefully explored, and the  compatibility of each of the neutral  $B$-anomalies --$R_K$ and $R_{K^\ast}$-- with the SM prediction within the $1\sigma$ and $2\sigma$ levels will be determined, exposing the technical and theoretical cost of the solutions found. Moreover, we will explore whether there are specific values of the ALP mass for which all three neutral $B$-anomalies --i.e. $R_K$, $R_{K^\ast}$ central bin and $R_{K^\ast}$ low bin-- could be simultaneously explained.  The impact of the systematic errors  in the analysis will be discussed, including the survival prospects for the solutions with the improvement of the experimental sensitivities. Moreover, we will check the consistency of ALP solutions to the neutral $B$ anomalies  with the constraints from other flavour observables, in particular with the data in the branching ratios for $B_s\to\ell^+\ell^-$ and $B\to K^{(\ast)}\ell^+\ell^-$ decays, and astrophysical constraints. In addition, 
the impact and compatibility of the ALP solutions with the data on the anomalous magnetic moment of the muon and of the electron will be discussed.

The charged $B$-anomalies in terms of ALP exchange will not be contemplated in this work. They would be one-loop  processes --see Fig.~\ref{fig:sketchy-diagram}: consistent solutions would require assuming flavour-blind  ALP-fermion couplings, so that both the neutral and the charged $B$-anomalies would be induced only at loop-level. This is a very different setup, outside the scope of this work.
 
The structure of the paper can be inferred from the Table of Contents. 

%
%
\section{The ALP Lagrangian}
\label{sec:ALPLag}

The construction of the ALP effective Lagrangian goes back to the late $'80$s with the seminal works in Refs.~\cite{Georgi:1986df,Choi:1986zw}. It later underwent renewed interest~\cite{Brivio:2017ije,Chala:2020wvs,Bauer:2020jbp,Bonilla:2021ufe} associated with an intense effort to investigate in detail its parameter space~\cite{Freytsis:2009ct,Mimasu:2014nea,Jaeckel:2015jla,Izaguirre:2016dfi,Brivio:2017ije,Bauer:2017ris,Craig:2018kne,Frugiuele:2018coc,Bauer:2018uxu,Ebadi:2019gij,Merlo:2019anv,Gavela:2019cmq,Bauer:2019gfk,Bauer:2021wjo,Bauer:2021mvw,Bonilla:2022pxu}. The ALP $a$ is {\it defined here as a pseudoscalar field, singlet of the SM charges,  and described by  a Lagrangian invariant under the shift symmetry $a \to a + constant$, plus anomalous couplings which may break the shift invariance together with 
a small mass  term~\footnote{There is a certain arbitrariness in the definition of an ALP. The customary underlying idea is inspired by the case of true axions:  a global symmetry which is classically exact --and spontaneously realised-- but explicitly broken only at the quantum level. This explicit breaking is precisely that given  by the presence of gauge anomalous couplings  (in fact, in true axion models that mass is  a byproduct of  the anomalous couplings of ALP to the strong gauge sector of the theory~\cite{Weinberg:1977ma,Wilczek:1977pj,   Rubakov:1997vp,Berezhiani:2000gh,Gianfagna:2004je,Hsu:2004mf,Hook:2014cda,Fukuda:2015ana,Chiang:2016eav,Dimopoulos:2016lvn,Gherghetta:2016fhp,Kobakhidze:2016rwh,Agrawal:2017ksf,Agrawal:2017evu,Gaillard:2018xgk,Buen-Abad:2019uoc,Hook:2019qoh,Csaki:2019vte,Gherghetta:2020ofz, Hook:2018jle,DiLuzio:2021pxd,DiLuzio:2021gos}). In all generality, their presence is expected to source  a potential  for the ALP, and thus a mass.
  The ALP mass is usually represented by a --more general-- explicit mass term in the Lagrangian. Consistent with this underlying idea, all other shift-breaking operators are customarily expected to be even more suppressed than the mass term, and disregarded in phenomenological studies of ALP Lagrangians at leading order.}
 $m_a\ll f_a$}, where $f_a$ is the NP ALP scale. We focus in this paper on the CP-even ALP Lagrangian at next-to-leading order (NLO) of the linear expansion, that is up to $\cO(1/f_a)$ terms; this corresponds to operators with mass dimension up to five. The complete Lagrangian can be written as
\be
\sL=\sL_\text{SM}+\sL_a\,,
\ee
where $\sL_\text{SM}$ denotes the SM Lagrangian, 
\be
\begin{split}
\sL_\text{SM}=&-\dfrac14 X_{\mu\nu}X^{\mu\nu}+\sum_{\rm f} \ov{\rm f}\,i\,\slashed{D}\,{\rm f}+D_\mu\Phi^\dag D^\mu\Phi-V\left(\Phi^\dag \Phi\right)+\\
&-\left[\ov{Q'_L}\,Y_d\,\Phi\, d'_R+
\ov{Q'_L}\,Y_u\,\widetilde\Phi \,u'_R+
\ov{L'_L}\,Y_e\,\Phi\, e'_R+
\hc\right]\,,
\end{split}
\ee
$X_{\mu\nu}$ denotes the SM field strengths for the strong and electroweak (EW) gauge bosons, $\{X\equiv G, W, B\}$ respectively, and the sum over colour and weak gauge indices has been left implicit. The SM Higgs boson is denoted by $\Phi$ with $\widetilde{\Phi}\equiv i\sigma^2\Phi^\ast$, and $V\left(\Phi^\dag \Phi\right)$ is the Higgs potential.  The index  $\rm f$ runs over the SM chiral fermion fields $\text{f} \equiv \left\{ Q'_L , \, u'_R , \, d'_R , \, L'_L , \, e'_R \right\}$, where the primes refer to the flavour basis. In turn, a complete set of independent and non-redundant ALP-SM couplings is encoded in
\be
\sL_a=\dfrac12\derp_\mu\, a\,\derp^\mu\,a-\dfrac{m_a^2}{2}a^2+\sL_a^X+\sL_{\partial a}^\psi\,,
\label{non-redundant-L}
\ee 
where $\sL_a^X$ encompasses the couplings of the ALP to anomalous currents,\footnote{The coefficients of gauge anomalous terms are often defined with a suppression factor with respect to the notation used all throughout this paper, i.e. $c_i\to \alpha_i/(4\pi) c_i$, where $\alpha_i$ denotes the corresponding gauge field fine structure constant.  }
\begin{equation}
 \sL_a^X\,= \,-c_{\widetilde{W}}\dfrac{a}{f_a}W_{\mu\nu}^i\tilde W^{i\mu\nu}-c_{\widetilde{B}}\dfrac{a}{f_a}B_{\mu\nu}  \tilde B^{\mu\nu}-c_{\widetilde{G}}\dfrac{a}{f_a}G_{\mu\nu}^a\tilde G^{a\mu\nu}\,,
\label{general-NLOLag-lin}
\end{equation}
while the ALP-fermion couplings contained in $\sL_{\partial a}^\psi$  are derivative ones, i.e. invariant under constant shifts of the $a(x)$ field,  
\be
\sL_{\partial a}^\psi=
\dfrac{\partial_\mu a}{f_a} \Big[
\ov{Q}_L'\gamma^\mu \bc'_QQ_L'+
\ov{u}_R'\gamma^\mu \bc'_uu_R'+
\ov{d}_R'\gamma^\mu \bc'_dd_R'+
\ov{L}_L'\gamma^\mu \bc'_LL_L'+
\ov{e}_R'\gamma^\mu \bc'_ee_R' \Big] \,,
\label{ALPLag}
\ee
where $\bc'_{\rm f}$  are hermitian $3\times 3$ matrices in flavour space containing the Wilson coefficients of the corresponding operators: note that four of  the couplings contained in these matrices are not independent, as they can be removed applying the conservation of baryon number and of the three independent lepton numbers (disregarding neutrino masses)\footnote{The ALP-neutrino couplings will be argued to be irrelevant for the present tree-level analysis, and therefore neutrino masses and the PMNS mixing matrix are to be neglected throughout this work without loss of generality.}~\cite{Bonilla:2021ufe,Bauer:2021mvw}. Furthermore, a possible shift-invariant bosonic operator, $\mathcal{O}_{a\Phi} \equiv {\partial_\mu a}(\Phi^\dag i\overleftrightarrow{D_\mu}\Phi)/{f_a}$,  has {\it not} been included in Eq.~(\ref{non-redundant-L}) as this would also be redundant given the choice made to consider all possible fermionic couplings (minus four)  in $\sL_{\partial a}^\psi$.\footnote{Alternatively,  one of the operators in $\sL_{\partial a}^\psi$ could be substituted by $\mathcal{O}_{a\Phi}$ if wished, see for instance Ref.~\cite{Bonilla:2021ufe}.}
 Finally, the  condition of CP conservation implies that  $\bc'_{\rm f}=\bc^{\prime T}_{\rm f}$ and thus all fermionic couplings are real. 

The description above is explicitly invariant under the SM gauge group $SU(3)\times SU(2) \times U(1)$ gauge group. At low energies after electroweak symmetry breaking, the total Lagrangian Eq.~(\ref{non-redundant-L}) can be rewritten in the  mass basis, in which $\sL_a^X$ reads
\begin{equation}
\begin{split}
  \sL_a^X\,= & \,- c_{a\gamma \gamma}\,\dfrac{a}{f_a}F_{\mu\nu}\tilde F^{\mu\nu} - c_{a\gamma Z}\, \dfrac{a}{f_a} F_{\mu\nu}  \tilde Z^{\mu\nu}+
 \\
 & \, - c_{a ZZ}\,\dfrac{a}{f_a}Z_{\mu\nu}\tilde Z^{\mu\nu} - 2 c_{\widetilde{W}}\dfrac{a}{f_a}W_{\mu\nu}^+\tilde W^{-\mu\nu} -c_{\widetilde{G}}\dfrac{a}{f_a}G_{\mu\nu}^a\tilde G^{a\mu\nu}  \,.
\end{split}
\label{ferm-Lag}
\end{equation}
where  $\{F^{\mu\nu}, Z^{\mu\nu}, W_{\mu\nu}, G_{\mu\nu} \}$ denote respectively the electromagnetic, $Z$-boson, $W$-boson and gluonic field strengths, and
\begin{equation}
c_{a\gamma \gamma}\equiv c_w^2 c_{\widetilde{B}} + s_w^2 c_{\widetilde{W}} \,,\qquad
c_{a\gamma Z}\equiv2 c_s s_w \left( c_{\widetilde{W}} - c_{\widetilde{B}} \right)\,,\qquad
c_{a Z Z}\equiv s_w^2 c_{\widetilde{B}} + c_w^2 c_{\widetilde{W}} \,,
\end{equation}
where  the sine and cosine of the Weinberg angle are respectively denoted $s_w$ and $c_w$. The chirality-conserving fermionic Lagrangian $\sL_{\partial a}^\psi$ can also be written straightforwardly in the mass basis. Nevertheless, 
for practical purposes it is useful to use the equations of motion (EOM) supplemented with the anomaly contribution, to rewrite $\sL_{\partial a}^\psi$ in terms of chirality-flip fermion couplings plus anomalous terms, i.e.
\be
\begin{split}
\sL_{\partial a}^\psi = \sL^\psi_a  
&
- \Delta c_{a\gamma \gamma}  \dfrac{a}{f_a}F_{\mu\nu}  \tilde F^{\mu\nu} 
- \Delta c_{a\gamma Z} \dfrac{a}{f_a}F_{\mu\nu}  \tilde Z^{\mu\nu} +
\\
&
- \Delta c_{a ZZ} \dfrac{a}{f_a}Z_{\mu\nu} \tilde Z^{\mu\nu} 
- 2 \Delta c_{a\widetilde{W}} \dfrac{a}{f_a}W_{\mu\nu}^+\tilde W^{- \mu\nu} - \Delta c_{a\widetilde{G}}\dfrac{a}{f_a}G_{\mu\nu}^a\tilde G^{a\mu\nu} \,,
\label{ALPLag3}
\end{split}
\ee
where $ \sL^\psi_a  $ is a chirality-flip fermion Lagrangian that expressed in the mass basis reads 
\be
\sL^\psi_a= -\dfrac{i a}{f_a} \sum_{\psi=u,d,e} \sum_{i,j}  \left[ (m_{\psi_i} -m_{\psi_j})\left(\bK_\psi^{S}\right)_{ij} \ov\psi_i \, \psi_j +
(m_{\psi_i} + m_{\psi_j})
\left(\bK_\psi^{P}\right)_{ij} \ov\psi_i \gamma_5 \psi_j \right]\, +\ldots
\label{Lag-fermion-flip}
\ee
 where dots indicate ALP-fermion-Higgs interactions  left implicit as they will not be used in this paper. In this equation,   $m_{\psi_i}$ denotes the mass of the four-component fermion field $\psi_i$, and the ${\bf K}_\psi$ coefficient matrices are defined as combinations of coefficients ${\bf c}_\psi$, possibly weighted down by the CKM mixing matrix. For instance, choosing a basis in which the down sector masses are diagonal, it follows that 
\be
\bK_u^{S,P}
\equiv 
\dfrac{ \bc_u \pm V_{\text{CKM}} \bc_Q V_{\text{CKM}}^\dagger}{2}\,,
\qquad
\bK_d^{S,P}
\equiv \dfrac{ \bc_d\pm \bc_Q }{2}\,,
\qquad
\bK_e^{S,P} 
\equiv \frac{ \bc_e \pm \bc_L }{2} \,,
\label{Kdef}
\ee
with the sum  (difference) of the  operator coefficients ${\bf c}_\psi$ corresponding to the scalar (pseudoscalar) components of ${\bf K}_{\psi}$.
 The relation of the various ${\bf c}_\psi$  to the coefficient matrices in the flavour basis --see Eq.~(\ref{ALPLag})-- is given by 
\be
\begin{gathered}
V_u^\dagger \bc_u' V_u \equiv \bc_u \,, \qquad V_d^\dagger \bc_d' V_d \equiv \bc_d \,,  \qquad V_e^\dagger \bc_e' V_e \equiv \bc_e \,,
\\
U_d^\dagger \bc_Q' U_d \equiv \bc_Q \,,  \qquad U_e^\dagger \bc_L' U_e \equiv \bc_L \,,
\end{gathered}
\ee
where $U_\psi$ and $V_\psi$ are the unitary rotations associated to the left- and right-handed fermions, respectively, which allow to diagonalise  the fermion masses,
\be
M_\psi\equiv\dfrac{v}{\sqrt2}U^\dag_\psi\,Y_\psi\,V_\psi\,,
\ee
where $v=246\GeV$ is the EW vacuum expectation value (vev) and $M_\psi$ denote  real diagonal fermion mass matrices.  In turn, the CKM matrix is given by~\footnote{Unlike for the SM, some combinations of the matrices $V$ that rotate right-handed fields are now {\it a priori} physical. }
\be
V_{\text{CKM}} =U_u^\dagger U_d\,.
\label{CKM}
\ee
The contributions to anomalous couplings which appear in Eq.~(\ref {ALPLag3}) (and are a consequence of the chiral rotation performed) are given by
\begin{equation}
\begin{gathered}
\Delta c_{a\gamma \gamma}\equiv c_w^2 K_B + s_w^2 K_W \qquad\qquad
\Delta c_{a\gamma Z}\equiv 2 c_w s_w \left( K_W - K_B \right)\\
\Delta c_{a Z Z}\equiv  s_w^2 K_B + c_w^2 K_W\qquad\qquad
\Delta c_{a\widetilde{W}} \equiv  K_W\qquad \qquad
\Delta c_{a\widetilde{G}} \equiv K_G\,,
\end{gathered}
\label{Deltas}
\end{equation}
with the $K_X$ coefficients given by the following combinations of fermionic couplings:
\be
\begin{split}
& K_B \equiv \frac{\alpha_{em}}{8 \pi c_w^2} \Tr \left( \frac{1}{3} \bc_Q - \frac{8}{3} \bc_u - \frac{2}{3} \bc_d + \bc_L - 2 \bc_e \right) \,,
\\
& K_W \equiv \frac{\alpha_{em}}{8 \pi s_w^2} \Tr \Big( 3 \bc_Q + \bc_L \Big)\,, \\
& K_G \equiv \frac{\alpha_s}{8 \pi} \Tr \Big( 2 \bc_Q - \bc_u - \bc_d \Big) \,,
\label{Ks}
\end{split}
\ee
where $\alpha_{em}=e^2/4\pi$ and $\alpha_s=g_s^2/4\pi$, $e$ denotes the electric charge and $g_s$ the strong gauge coupling. These $\Delta c_i$ corrections can be reabsorbed  in the arbitrary coefficients in Eq.~(\ref{ferm-Lag}), $ c_i\rightarrow c_i+\Delta c_i$  (e.g.  $ c_{a\gamma \gamma}\rightarrow c_{a\gamma \gamma}+\Delta c_{a\gamma \gamma}$ etc.) so that in all generality the complete ALP Lagrangian in Eq.~(\ref{non-redundant-L}) can be rewritten as 
\be
\sL_a=\dfrac12\derp_\mu\, a\,\derp^\mu\,a-\dfrac{m_a^2}{2}a^2+\sL_a^X +\sL_{ a}^\psi \,,
\label{non-redundant-L-flip}
\ee 
with arbitrary operator coefficients. Nevertheless, this analysis illustrates that contributions from anomalous couplings are  automatic and  unavoidable when relating the initial chirality-conserving and explicitly shift-invariant ALP fermionic basis to chirality-flip fermion operators. In practice, for fermionic processes involving only tree-level ALP exchanges, it is completely equivalent to use either the chirality-conserving fermion Lagrangian $\sL_{\partial a}^\psi$ in Eq.~(\ref{ALPLag}) (or its mass-basis version) or the chirality-flip one $\sL^\psi_a$  in Eq.~(\ref{Lag-fermion-flip}). On the contrary, consistency requires to take into account the complete  combination of couplings in Eq.~(\ref{ALPLag3}) for some loop-level analyses of ALP exchanges involving fermions. For most of this work we focus only on tree-level exchange of ALPs,  and  $\sL^\psi_a$ alone will thus suffice unless otherwise specified. 

Eq.~(\ref{Lag-fermion-flip}) shows then that only pseudoscalar couplings contribute at tree-level of the EFT to flavour-diagonal interactions, while both scalar and pseudoscalar contributions are present for the off-diagonal ones. Moreover, all  tree-level ALP-fermion interactions are proportional to the masses of the fermions involved: the naive expectation is that couplings with light fermions are subdominant with respect to couplings with heavier fermions. In particular, the flavour-conserving ALP interactions pertinent to our analysis with electrons are much smaller than those with muons.  It also follows from  Eq.~(\ref{Lag-fermion-flip}) that ALP-mediated $B\to K$ transition amplitudes are proportional to (the $23$ element of) the scalar coupling $\bK_d^{S}$, while  $B\to K^*$ ones are proportional to the pseudoscalar coupling $\bK_d^{P}$. 

It is pertinent to stress that  the only NP couplings to be considered below --and as customary in the literature-- are the ALP couplings to the quark bilinear $\bar{b}s$, and the lepton $\mu^+\mu^-$ and $e^+e^-$ channels, i.e,  in the notation of Eq.~(\ref{Lag-fermion-flip}):
\be 
\begin{split}
\sL^\psi_a\supset 
&
-\dfrac{i a}{f_a} \Big[ (m_{s} -m_b)\left(\bK_d^{S}\right)_{sb}  (\ov s \, b - \ov b \, s) +
(m_s + m_b)\left(\bK_d^{P}\right)_{sb} (\ov s \gamma_5 b + \ov b \gamma_5 s)+ \\
&\hspace{2cm}+ 2m_e  \left(\bK_e^{P}\right)_{ee} \ov e \gamma_5 e + 2m_\mu  \left(\bK_e^{P}\right)_{\mu\mu} \ov\mu \gamma_5 \mu\Big]
\,.
\end{split}
\label{Lag-fermion-flip-reduced}
\ee 
Such a specific choice of parameters is part of the theoretical cost required to explain the neutral $B$ anomalies through tree-level exchange of ALP couplings.  Nevertheless, it may be natural to disregard ALP interactions with up-type quarks, in spite of  fermionic ALP couplings being proportional to fermion masses,  as their contribution to the neutral $B$ anomalies is loop-suppressed.
 But  the opposite could be argued for, for instance, ALP coupling to taus, etc. 
Overall, to suppress all fermionic ALP couplings  in Eq.~(\ref{Lag-fermion-flip}) except those in Eq.~(\ref{Lag-fermion-flip-reduced})  is technically possible, as there are enough free parameters in the initial Lagrangian --Eq.~(\ref{non-redundant-L})-- as to allow for it. 

\subsection*{EFT validity} Finally, the question of the validity of the ALP EFT must be addressed.  In order for the ALP Lagrangian to be approximately shift-invariant,  the ALP mass $m_a$ must be small compared with the EFT scale $f_a$, $m_a\ll f_a$.  Furthermore, as the coupling dependence is of the form 
$c_i/f_a$,  $c_i<1 $ must hold for all Lagrangian coefficients as indicated by naive dimensional analysis. 

The absolute value of the ALP scale is also relevant. The consistency of formulating the effective field theory in terms of operators which are invariant under the EW symmetry, see Eqs.~\eqref{non-redundant-L}-\eqref{ALPLag}, implies to consider in all cases $f_a$ values larger than the EW scale, $f_a\gtrsim v$. We will adhere throughout this work to this condition, as an ALP scale below the EW scale is difficult to sustain in view of the non-observation of NP fields expected to accompany any renormalisable completion of the ALP scenario. Within this setup, we will explore two regimes: a ``heavy ALP''  and a ``light ALP'', where the  denomination heavy/light refers to the ALP mass size compared to $B$ meson masses.\\
\\

The next sections are dedicated to the phenomenological analysis of the different possible ranges for the ALP mass:  i) an ALP heavier than the $B$ mesons, ii) an ALP with a mass within the energy bin windows considered for the neutral-$B$ anomalies, and iii) a light ALP with mass  $1\MeV<m_a< 2m_\mu$ where $m_\mu$ denotes the muon mass. In the numerical computations, the exact values of the input parameters used can be read in Tab.~\ref{tab:parameters} of App.~\ref{sec:Input}.

%
%
\section{Heavy ALP}
\label{sec:HEAVYALP}

\subsection{The low-energy Lagrangian}
\label{sec:LEALPLag}

For an ALP heavier than the $B$ mesons,  the ALP can be safely integrated out to analyse its impact on $B$ transitions. The result is an effective Lagrangian valid at energies lower than $m_a$, which in the mass basis  can be decomposed as 
\be
\sL_a^\text{eff}= \sL_a^\text{eff-4f} +  \sL_a^\text{mixed} + \ldots
\label{Leff-total}
\ee
 where the dots encode pure gauge interactions --left implicit as they will have no impact on the results  in this section, $\sL_a^\text{mixed}$ encodes interactions involving two fermions and anomalous gauge currents, and $\sL_a^\text{eff-4f}$ encodes the effective  four-fermion couplings, which are specially significative for the analysis of $B$-anomalies and read 
\be
\sL_a^\text{eff-4f}=-\dfrac{1}{2(f_am_a)^2}\Bigg[\sum_{\psi}\sum_{i,j}  \left( (m_{\psi_i} -m_{\psi_j})\left(\bK_\psi^{S}\right)_{ij} \ov\psi_i \, \psi_j +
(m_{\psi_i} + m_{\psi_j})
\left(\bK_\psi^{P}\right)_{ij} \ov\psi_i \gamma_5 \psi_j \right)
\Bigg]^2\,.
\ee
Among  the effective operators in this last equation, only those composed of a $s$ and a $b$ quark fields together with flavour-diagonal leptonic currents are relevant to the tree-level phenomenological analysis of the neutral B-anomalies, i.e.
\be
\sL_a^\text{eff-4f}\supset-\dfrac{4 m_\ell}{2(f_a m_a)^2}\Big[(m_s - m_b )\left({\bf K}_d^S\right)_{sb} \, \ov{s} \, b +
(m_s + m_b) \left({\bf K}_d^P\right)_{sb} \, 
\ov{s}\, \gamma_5\, b \Big]
\Big[\left({\bf K}_e^P\right)_{\ell\ell} \, 
\ov{\ell}\, \gamma_5\, \ell \Big]\,.
\label{EffHeavyALPLag}
\ee
It follows that  only pseudoscalar leptonic  interactions remain, while both scalar and pseudoscalar contributions will contribute to quark currents. It is useful to re-express this Lagrangian in a more compact way as
\be
\sL_a^\text{eff-4f}
\supset -\dfrac{4G_F}{\sqrt2}\sum_{\ell=e,\mu,\tau}V_{tb}V^\ast_{ts}\left(C^\ell_{P_+}\,\cO^\ell_{P_+}+C^\ell_{P_-}\,\cO^\ell_{P_-}\right)\,,
\label{EffHeavyALPLagPheno}
\ee
where $G_F\equiv 1/(\sqrt2v^2)$ is the Fermi constant as extracted from  muon decay, and the operators $\cO^\ell_{P_\pm}$ are defined as 
\be
\cO^\ell_{P_+}\equiv \dfrac{\alpha_\text{em}}{4\pi}\left(\ov{s}\,b\right) \left(\ov{\ell}\,\gamma_5\, \ell\right)\,,
\qquad\qquad
\cO^\ell_{P_-}\equiv \dfrac{\alpha_\text{em}}{4\pi}\left(\ov{s}\,\gamma_5\,b\right) \left(\ov{\ell}\,\gamma_5\, \ell\right)\,,
\label{OPpmOperators}
\ee
where the $\pm$ subscripts remind the parity of the quark current component of the operators. The Wilson coefficients $C^\ell_{P_\pm}$ are then given by
\be
C^\ell_{P_\pm} \equiv \dfrac{2\sqrt{2}\pi}{\alpha_\text{em} G_F  V_{tb}V_{ts}^* }\dfrac{m_\ell}{(f_a m_a)^2}(m_s\mp m_b) \left({\bf K}^{S,P}_d\right)_{sb}\left({\bf K}^P_e\right)_{\ell\ell}\,.
\label{DefinitionCellPpm}
\ee
It follows from the parity structure that $\cO^\ell_{P_+}$ will contribute to $B\to K\ell^+\ell^-$  processes, while $\cO^\ell_{P_-}$ can instead mediate both $B\to K^\ast\ell^+\ell^-$  and $B_s\to\ell^+\ell^-$ decays.

Yet another notation for four-fermion couplings is that customarily used in EFT analyses of $B$-anomalies, in which the combinations of couplings that can result from the tree-level integration of a heavy ALP read~\cite{Alonso:2015sja,Buchalla:1995vs,Chetyrkin:1996vx}\footnote{In other words, other dimension-six effective operators mediating LFU violation and encoded  in $\sL^{\text{eff}}_{\Delta B=1}$, such as for example $C_9^{(\prime)}$ and $C_{10}^{(\prime)}$, 
are disregarded here because they cannot be generated by the tree-level exchange  of a heavy ALP.
}  
\begin{equation}
\sL^{\text{eff}}_{\Delta B=1} \supset -\dfrac{4G_F}{\sqrt{2}} \sum_{\ell=e,\mu,\tau}V_{tb}V_{ts}^*\left(C^\ell_P\cO^\ell_P+C^{\ell\prime}_P\cO^{\ell\prime}_P\right)\,,
\end{equation}
with 
\be
\cO^{\ell(\prime)}_P\equiv \dfrac{\alpha_\text{em}}{4\pi}\left(\ov{s} P_{R(L)} b\right) \left(\ov{\ell}\gamma_5 \ell\right)\,.
\ee
 The relation between this formulation and  the operators and Wilson coefficients in  Eq.~\eqref{EffHeavyALPLagPheno} is simply 
 \be
\cO^\ell_{P_\pm}\equiv \cO^\ell_{P}\pm \cO^{\ell\prime}_P\,, 
\qquad\qquad
C^\ell_{P_\pm} = \dfrac{C^\ell_P\pm C^{\ell\prime}_P}{2}\,.
\ee

\subsection{Phenomenology of a heavy ALP}
\label{sec:PhemHEAVYALP}

As shown above,  for a given lepton $\ell$, the  leading  four-fermion effective operators  induced by tree-level exchange of an ALP spans a two-parameter space $\{C^\ell_{P_+}, C^\ell_{P_-}\}$. Conversely, the parity analysis implies that typically these coefficients contribute to different observables, which can then be easily compared for electrons \vs muons (tau leptons are not considered here), e.g.
\begin{itemize}
\item $\cO^\ell_{P_+}$:  $\cO^e_{P_+}$ and $\cO^\mu_{P_+}$ will contribute to $R_K$. 
\item $\cO^\ell_{P_-}$:  $\cO^e_{P_-}$ and $\cO^\mu_{P_-}$ will contribute to $R_{K^\ast}$, as well as to $B_s\rightarrow \ell^+\ell^-$ decays.
\end{itemize}
Note that  both $C^\ell_{P_+}$ and $C^\ell_{P_-}$ are proportional to the coupling combination ${\bf K}^P_e\sim( \bc_e - \bc_L )$ in Eq.~(\ref{Kdef}), while they differ on the dependence on quark couplings, that is ${\bf K}^{S}_d\sim  (\bc_d+ \bc_Q) $ and ${\bf K}^{P}_d\sim  (\bc_d - \bc_Q) $, respectively. The proportionality to  ${\bf K}^P_e$  will also appear in other observables to be discussed below, namely the anomalous magnetic moments of the muon and of the electron, while conversely $B_s-\ov{B}_s$ oscillations only depend on ALP-quark couplings.

The different observables of interest to our analysis are discussed next in more detail in terms of the  contributions of the Wilson coefficients.

\boldmath
\subsubsection{$B\to K\ell^+\ell^-$, $R_K$, $\Delta M_s$ and magnetic moments}
\label{subsec:HeavyALPRK}
\unboldmath

The differential decay width for the $B\to K^{(\ast)} \ell^+ \ell^-$ processes can be written as 
\be
\dd\Gamma(B\to K^{(\ast)} \ell^+ \ell^-)=\dfrac{1}{(2\pi)^3}\dfrac{1}{32M_B^3}\left|\ov{\cM}\right|^2\dd q^2\dd Q^2\,,
\ee
where $\ov{\cM}$ is the matrix element of the process summed over the polarisations of the meson and leptons in the final state and $M_B$ is the $B$ meson mass, while the four-momenta are defined as $q^2\equiv(p_{\ell^+}+p_{\ell^-})^2\equiv(p-k)^2$ and $Q^2\equiv (p_{\ell^+}+k)^2=(p-p_{\ell^-})^2$, where $p$  denotes the four-momentum of the initial state $B$-meson, $k$ that of the $K^{(\ast)}$-meson, and $p_{\ell^\pm}$ those of the leptons $\ell^\pm$. For $m_a^2\gg q^2$, the $q^2$-dependence can be neglected on the ALP propagator and a simple integration over $q^2$ remains.

\boldmath
\subsubsection*{$B\to K\ell^+\ell^-$}
\unboldmath
As a step previous to the analysis of $R_K$, we discuss next the semileptonic $B$ decay widths into dilepton pairs, for which the experimental data available are shown In Tab.~\ref{tab:Proposal}. We compared the results obtained 
 for the $q^2$ integration over the dilepton mass regions of interest for the anomaly --$1\GeV^2 <q^2 < 7 \GeV^2$-- using two popular softwares, Flavio~\cite{Straub:2018kue} and EOS~\cite{vanDyk:2021sup},  with  the corresponding expression in Ref.~\cite{Bobeth:2007dw} --which is valid up to $\cO(m^3_\ell)$ corrections, 
\be
\BR(B\to K \ell^+ \ell^-)_{1.0\GeV^2}^{7.0\GeV^2}=
\left(\dfrac{\tau_{B^\pm}}{1.64\text{ps}}\right)
\left(1.91 +0.08\,C^{\ell2}_{P_+}
-\dfrac{m_\ell}{\GeV}
\dfrac{C^\ell_{P_+}}{1.46}
-\dfrac{m_\ell^2}{\GeV^2} \dfrac{C^{\ell2}_{P_+}}{5.18^2}\right)\,,
\ee
where $\tau_{B^\pm}$ is the lifetime of the meson $B^\pm$ and the $q^2$ interval of integration ${}_{1.1\GeV^2}^{7.0\GeV^2}$ is indicated. We found a good agreement, as the numerical differences can be understood as a consequence of the more recent input data used in the softwares. The analytic expression above makes it easy to understand why the first term in this expression quadratic  in the Wilson coefficients is the same for $e$ and $\mu$ in the approximation considered, and it can even dominate over the linear term. Indeed, using Flavio and EOS (which agree with each other to very high accuracy), we find for the integration over the $q^2$ range of the  central bin window 
\be
\begin{aligned}
\BR(B\to K e^+ e^-)_{1.1\GeV^2}^{6.0\GeV^2} & =10^{-7}\times\left(1.5-3.4\times 10^{-4}\, C^e_{P_+} +7.1\times 10^{-2}\,C^{e2}_{P_+}\right)\,,\\
\BR(B\to K \mu^+ \mu^-)_{1.1\GeV^2}^{6.0\GeV^2} & =10^{-7}\times\left(1.5-7.0\times 10^{-2}\,C^\mu_{P_+} +7.1\times 10^{-2}\,C^{\mu2}_{P_+}\right)\,,
\end{aligned}
\label{BKllHeavyALP}
\ee
with a theoretical error of $\mathcal{O}(15)\%$ at $1\sigma$. The corresponding $2\sigma$ bounds on the Wilson coefficients read:
\begin{equation}
C_{P_+}^e \in [-2.6,2.6]\quad\quad \text{and} \quad\quad C_{P_+}^\mu \in [-1.3,2.3]\,,
\label{HEAVYALPSemileptonicBoundK}
\end{equation}
taking into account both experimental and theoretical errors. These constraints are depicted in grey in Fig.~\ref{fig:RKHALP}.

\begin{table}[h!]
\centering{}
\renewcommand{\arraystretch}{1.4}
\resizebox{\textwidth}{!}{
\begin{tabular}{c|c|c|c|c|c}
Observable&
$q^2$ [GeV$^2$] & 
Values& 
Heavy & On Bin & Light  \\[1mm]
\hline
&&&&&\\[-4mm]
$\dd\BR/\dd q^2(B^+ \to K^+ e^+ e^-)$ &\multirow{2}{*}{$(1.1,\,6)$}& $(28.6^{+2.0}_{-1.7} \pm 1.4) \times 10^{-9}$~\cite{LHCb:2019hip} & \multirow{2}{*}{\checkmark}& \multirow{2}{*}{\checkmark}& \multirow{2}{*}{\checkmark}\\[1mm]
$\BR(B^+ \to K^{+} e^+ e^-)$ && $\left(14.01^{+0.98}_{-0.83}\pm 0.69\right) \times 10^{-8}$ &&&\\[1mm]
\hline
&&&&&\\[-4mm]
$\dd\BR/\dd q^2(B^+ \to K^{+} \mu^+ \mu^-)$ & \multirow{2}{*}{$(1.1,\,6)$} & $\left(24.2\pm0.7\pm 1.2\right) \times 10^{-9}$\cite{LHCb:2014cxe} & \multirow{2}{*}{\checkmark} & &\multirow{2}{*}{\checkmark}\\[1mm]
$\BR(B^+ \to K^{+} \mu^+ \mu^-)$ & & $\left(11.86\pm0.34\pm 0.59\right) \times 10^{-8}$ &&&\\[1mm]
\hline
&&&&&\\[-4mm]
$\BR(B^+ \to K^{+} a (\mu^+ \mu^-))$ &$(0.06,\,22.1)$& $< 1 \times 10^{-9}$~\cite{LHCb:2016awg} &&\checkmark&\\[1mm]
\hline
\hline
&&&&&\\[-4mm]
\multirow{2}{*}{$\cB(B^0\to K^{0\ast} e^+ e^-)$} &$(1.1,6)$&$(1.8\pm 0.6)\times 10^{-7}$\cite{Belle:2019oag}& 
\checkmark & \checkmark& \checkmark\\
&$(0.1,8)$&$(3.7\pm1.0)\times 10^{-7}$\cite{Belle:2019oag}&&\checkmark &\\[1mm]
\hline
&&&&&\\[-4mm]
\multirow{6}{*}{$\cB(B^0 \to K^{0\ast} a(e^+ e^-))$}&$(0.0004,0.05)$
&$<1.344\times 10^{-7}$& &\checkmark&\checkmark\\[1mm]
&$(0.05,0.15)$&$<1.22\times 10^{-8}$&&\checkmark& \\
&$(0.25,0.4)$&$<1.97\times 10^{-8}$&&\checkmark&\\
&$(0.4,0.7)$&$<1.74\times 10^{-8}$&&\checkmark&\\
&$(0.7,1)$&$<6.5\times 10^{-9}$&&\checkmark&\\[1mm]
\hline
&&&&&\\[-4mm] 
$\mathcal{B}(B^0 \to K^{0\ast} \mu^+ \mu^-)$ &$(1.1,6)$&$1.9^{+0.7}_{-0.6} \times 10^{-7}$\cite{Belle:2019oag}&\checkmark &  &\checkmark\\[1mm]
\hline
&&&&&\\[-4mm] 
$\cB(B^0 \to K^{0\ast} a (\mu^+ \mu^-))$ & $(0.05,18.9)$ & $<3\times 10^{-9}$\cite{LHCb:2015nkv} & &\checkmark & \\[1mm]
\hline
\end{tabular}
}
\caption{\em The checkmarks correspond to the strongest bounds used in each mass regime analysed in this work. Notice that the second entries in the first two lines have been obtained from the corresponding first entry simply integrating over the bin window spread. For $\cB(B^0 \to K^{0\ast} a(e^+ e^-))$, no bound can be extracted for $q^2\in[0.15,\,0.25]$ GeV$^2$ as the data are incompatible with the SM prediction at more than $2\sigma$.}
\label{tab:Proposal}
\end{table}

\boldmath
\subsubsection*{$R_K$}
\unboldmath 
It follows from the previous expressions that the LFU ratio $R_K$ can be written in terms of the two coefficients $C_{P_+}^\ell$:
\be
R_K=1+ \dfrac{0.21\,C_{P_+}^e-4.67\,C_{P_+}^\mu+4.73(\,C_{P_+}^{\mu2}-C_{P_+}^{e2})}{100-0.21\,C_{P_+}^e+4.73\,C_{P_+}^{e2}}\,.
\label{RKHeavyALP}
\ee
The large theoretical errors reported for the semileptonic decays may be expected to cancel in this observable in general,  and the largest source of uncertainty to determine the Wilson coefficients are the experimental errors. Given that experimentally $R_K<1$,  the second term in this equation should be negative.

Some naive conclusions can be obtained when leptonic NP contributions are assumed only for either the electron or the muon sector. For instance, let us consider the $2\sigma$ error range for $R_K$, $R_K\in[0.768,\,0.935]$~\cite{LHCb:2021trn}. In the absence of ALP-muon couplings, this would require  the effective Wilson coefficients to lie in the range 
\be
C^e_{P_+}\in[1.2,\,2.6]\vee[-2.6,\,-1.2]\, \qquad \text{for $C^\mu_{P_+}=0$\,.}
\ee
On the contrary, if NP in the lepton sector would contribute only to muon couplings, i.e. $C^e_{P_+}=0$, there is no value of $C^\mu_{P_+}$   that would solve the $R_K$ anomaly at the $2\sigma$ level. This is easily understood noting that such a solution would require the $\BR(B\to K \mu^+ \mu^-)$ in Eq.~\eqref{BKllHeavyALP} to be suppressed with respect to the SM value, which in turns requires the  term linear in $C^\mu_{P_+}$ to dominate over the quadratic one, i.e. $|C^\mu_{P_+}|<1$. However, due to the suppression provided by the numerical prefactor of that linear term, the NP contribution would not be then large enough  to generate a significant shift from the SM prediction. 
\begin{figure}[t] 
\centering
\includegraphics[width=0.48\textwidth]{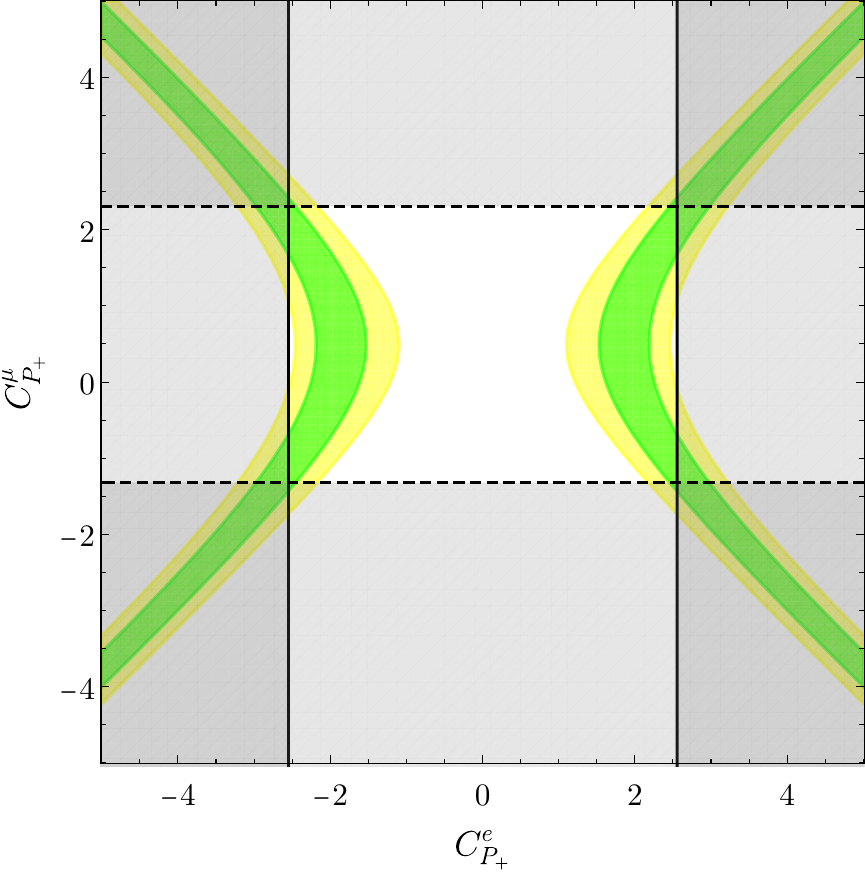}
\caption{\em Parameter space for $R_K$, for an ALP heavier than $B$ mesons.  In yellow and  green are respectively depicted the $1\sigma$ and $2\sigma$ solutions to the central bin of $R_K$. The grey regions around the frame of the figures are excluded  at $2\sigma$  by data on semileptonic $B\to K e^+e^-$ (solid black contours) and  $B\to K \mu^+\mu^-$ (dashed black contours) decays.}
\label{fig:RKHALP}
\end{figure}

In summary, it follows that $R_K$ could be explained through ALP-electron couplings alone (in addition to the $\text{ALP}$-$bs$ couplings), but not through ALP-muon couplings alone. 
 
The two-dimensional enlargement of the parameter space as spanned by the variables $\{C^e_{P_+}, C^\mu_{P_+}\}$ is depicted in  Fig.~\ref{fig:RKHALP}, which illustrates the $2\sigma$ region where both parameters could be simultaneously at play and account for $R_K$. Taking into account the bounds from semileptonic decays in Eq.~\eqref{HEAVYALPSemileptonicBoundK}, the allowed  area is given by 
\be
\begin{split} 
& C^e_{P_+}\in[1.2,\,2.6]\vee[-2.6,\,-1.2]
\\
& C^\mu_{P_+}\in[-1.3,\,2.3]\,.
\label{LimitsCl-RK}
\end{split}
\ee
Note that these two independent parameters can be traded by two specific combinations of the ALP-fermion couplings  defined in the mass basis  in the Lagrangian Eq.~\eqref{DefinitionCellPpm},
\be
\begin{aligned}
C^e_{P_+} \approx&\, -1.3\times 10^4\GeV^2\left(\dfrac{10\GeV}{m_a}\right)^2 \dfrac{\left(\bc_d+\bc_Q\right)_{sb}}{f_a}\dfrac{\left(\bc_e-\bc_L\right)_{ee}}{f_a}\,,\\
C^\mu_{P_+} \approx&\, -2.7\times 10^6\GeV^2\left(\dfrac{10\GeV}{m_a}\right)^2 \dfrac{\left(\bc_d+\bc_Q\right)_{sb}}{f_a}\dfrac{\left(\bc_e-\bc_L\right)_{\mu\mu}}{f_a}\,.
\end{aligned}
\label{Cs-simplified}
\ee
The limits obtained in Eq.~(\ref{LimitsCl-RK}) translate then into the following constraints, for instance for $m_a=10$ GeV:
\be
 \begin{split}
& \frac{\left(\bc_d+\bc_Q\right)_{sb} \left(\bc_e-\bc_L\right)_{ee}}{f_a^2} 
\in\left([0.93,\,2.02]\vee[-2.02,\,-0.93]\right) \times10^{-4}\GeV^{-2}\,,
 \\
 & \frac{\left(\bc_d+\bc_Q\right)_{sb} \left(\bc_e-\bc_L\right)_{\mu\mu}}{f_a^2}
 \in[-0.87,\,0.49]\times 10^{-6} \GeV^{-2}\,.
\end{split}
\label{first-results}
\ee
This result already implies that $\left(\bc_e-\bc_L\right)_{\mu\mu}$ needs to be about two orders of magnitude smaller than $\left(\bc_e-\bc_L\right)_{ee}$ to explain $R_K$ via heavy ALP exchange.
We consider next other relevant observables which are not describable in terms of $C^\ell_{P_\pm}$, but they are sensitive only to either the quark factor $\left(\bc_d+\bc_Q\right)_{sb}$ or the leptonic factors $\left(\bc_e-\bc_L\right)_{\ell\ell}$. Nevertheless, they will be shown to provide  further restrictions on the ALP explanation of $B$-anomalies.

\boldmath
\subsubsection*{$\Delta M_s$}
\unboldmath

\begin{figure}[h!] 
\centering%
\subfigure[{}\label{fig:DeltaMs}]%
{\includegraphics[width=0.47\textwidth]{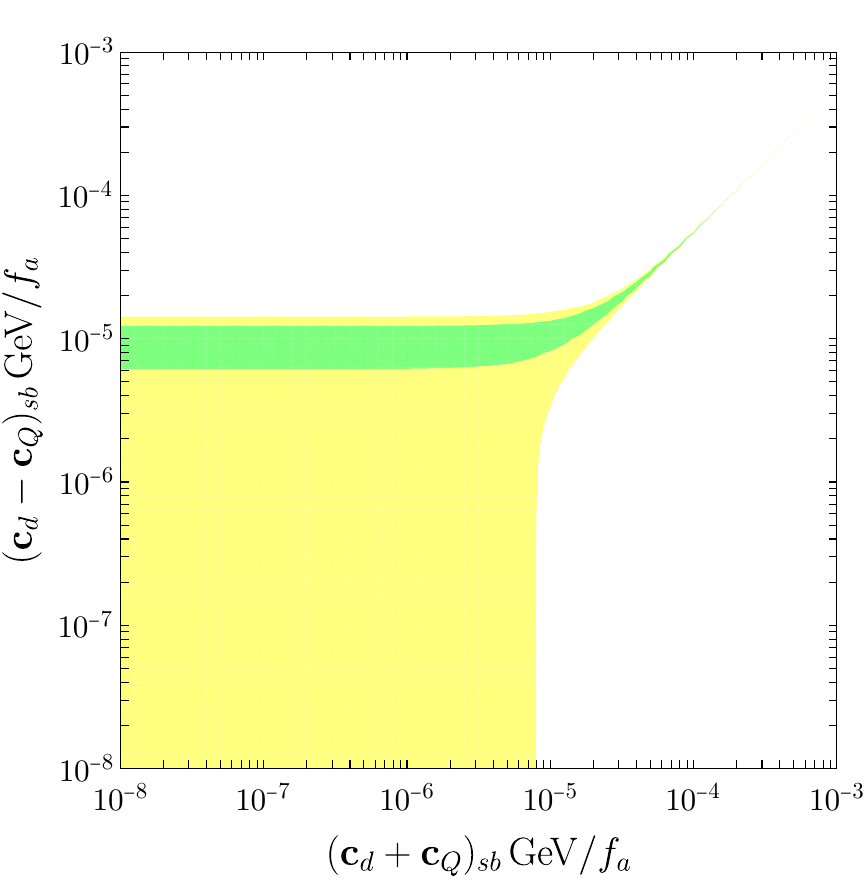}}\qquad
\subfigure[{}\label{fig:KellRKDeltaMs}]%
{\includegraphics[width=0.46\textwidth]{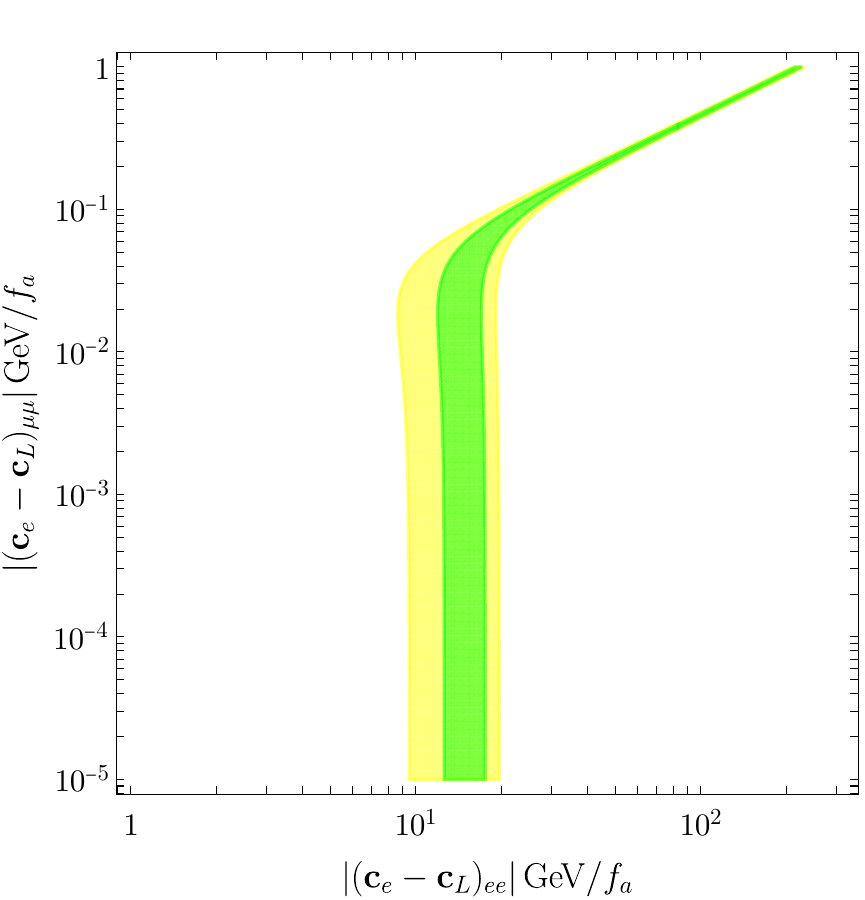}}
\caption{\em  ALP heavier than $B$ mesons. In Fig.~\ref{fig:DeltaMs},  $(\bc_d\pm \bc_Q)_{sb}/f_a$ parameter space allowed by  $B_s$ meson oscillation constraints, at $1\sigma$ ($2\sigma$) in green (yellow). This plot is symmetric with respect to any of the axes for negative values of the coordinates. In Fig.~\ref{fig:KellRKDeltaMs}, the parameter space for the combinations $(\bc_e-\bc_L)_{ee}/f_a$ and $(\bc_e-\bc_L)_{\mu\mu}/f_a$ that solve the $R_K$ anomaly in Fig.~\ref{fig:RKHALP} for the maximal allowed value ${(\bc_Q+ \bc_d)_{sb}/f_a=10^{-5}\GeV^{-1}}$. The ALP mass is fixed to $m_a=10\GeV$ in both plots.}
\end{figure}

The  $B_s$ meson mass difference $\Delta M_s$ measured in  $B_s-\ov{B}_s$ oscillations can be defined as
\be 
\Delta M_s=\dfrac{1}{M_{B_s}}\left|\left\langle\ov{B}_s\left|\cH_{\Delta B=2}^\text{eff.}\right|B_s\right\rangle\right|\,,
\ee
where $M_{B_s}$ is the mass of the $B_s$ meson and $\cH_{\Delta B=2}^\text{eff.}$ is the effective Hamiltonian describing $\Delta B=2$ transitions.  The data imply that~\cite{LHCb:2021moh}
\begin{equation}
\Delta M_s = (17.7656 \pm 0.0057)\,\text{ps}^{-1}\,,
\end{equation}
to be compared with the SM prediction that we take from Ref.~\cite{FermilabLattice:2016ipl}, $\Delta M_s^{\rm SM} = (20.1^{+1.2}_{-1.6})\,{\rm ps}^{-1}$. (Assuming the most recent results for the SM prediction~\cite{DiLuzio:2019jyq} that are compatible with the SM within $1\sigma$, the following conclusions do not change as the maximum NP couplings allowed by data at $2\sigma$ are in either case very alike. Nevertheless, the $1\sigma$ region in Fig.~\ref{fig:DeltaMs} would be in this case enlarged, constraining couplings $(\mathbf{c}_d+\mathbf{c}_Q)_{sb}/f_a\lesssim 6\times 10^{-5}\,\text{GeV}^{-1}$.)

The generic expression for the contribution of the tree-level exchange of a heavy ALP to $\Delta M_s$    has been recently presented in Ref.~\cite{Bauer:2021mvw}, and we thus refrain from repeating that analysis here. It suffices to mention that the corresponding bound applies to the two ratios
\be
\dfrac{\left(\bc_d\pm\bc_Q\right)_{sb}}{m_a\,f_a}\,.
\label{BoundMesonOscillationsHeavyALP}
\ee 
The values allowed by data for these combinations are  illustrated in Fig.~\ref{fig:DeltaMs}  for the benchmark ALP mass value  $m_a= 10$ GeV. They agree with those in Ref.~\cite{Bauer:2021mvw}.\footnote{Which expressed the result in terms of separate bounds for $c_d$ and $c_Q$. } The figure illustrates that large values of $(\bc_d\pm\bc_Q)_{sb}/f_a$ up to $\sim10^{-4}\GeV^{-1}$ are allowed on a fine-tuned region of the parameter space (the spikes in the figure).   Otherwise, in the $1\sigma$ region (in green)  the solutions constrain {$\left|(\bc_d-\bc_Q)_{sb}\right|/f_a$}, while only an upper bound results for the orthogonal combination, $\left|(\bc_d+\bc_Q)_{sb}\right|/f_a$, which is the one relevant for $R_K$,
\be
\left|(\bc_d-\bc_Q)_{sb}\right|/f_a\sim 10^{-5}\GeV^{-1}\,, \qquad \left|(\bc_d+\bc_Q)_{sb}\right|/f_a\lesssim 10^{-5}\GeV^{-1}\,.
\label{RK-quark-coup-limits}
\ee
At the $2\sigma$ level (in yellow) only an upper bound of $\mathcal{O}(10^{-5})\GeV$ can be extracted for both combinations.

A naive estimation of the impact of  $\Delta M_s$ on $R_K$ can now be achieved by comparing these constraints on ALP-quark couplings with the products of ALP-quark and ALP-lepton couplings relevant for $R_K$, see Eq.~(\ref{Cs-simplified}). For the illustrative case $(C^e_{P_+},\,C^\mu_{P_+})=(2,\,0)$ in Fig.~\ref{fig:RKHALP}, $m_a=10\GeV$  and the value $\left|(\bc_d+\bc_Q)_{sb}\right|/f_a=10^{-5}\GeV^{-1}$ which saturates the bound in Eq.~\eqref{RK-quark-coup-limits}, it would follow
\be
\dfrac{\left|\left(\bc_e-\bc_L\right)_{ee}\right|}{f_a}\approx 16\GeV^{-1}\,.
\label{ceeboundHeavyALP}
\ee 
This result leads right away to a clash with the validity of the EFT,  though, as $f_a$ is expected to be at least of the order of the electroweak scale,  see Eqs.~(\ref{non-redundant-L}) and (\ref{general-NLOLag-lin}), and  $f_a>m_a>M_{B_s}$.  This happens even for the smaller possible scale values, e.g. 
\be
f_a\approx 100\GeV\Longrightarrow\left|\left(\bc_e-\bc_L\right)_{ee}\right|\approx10^3\,,
\ee 
which are unacceptably large ALP-lepton couplings, well outside the perturbative regime of the EFT.\footnote{Such large values of the electron coupling can be attained e.g. selectively in electrophilic ALP models~\cite{Darme:2020gyx} where the electron coupling can be exponentially enhanced without increasing $m_a$ or $f_a$.} 

Note that  smaller  ALP-quark couplings would not soften the issue as they would require  even larger  lepton-ALP couplings. The situation improves but is still problematic for $\left|(\bc_Q+ \bc_d)_{sb}\right|/f_a$ values in the fine-tuned region, e.g. $10^{-4}\GeV^{-1}$, which would translate into $\left|\left(\bc_e-\bc_L\right)_{ee}\right|\approx10^2$. To vary  the ALP mass does not resolve the issue either, as the relevant combination of ALP-quark couplings  scales with the ALP mass, see Eq.~\eqref{BoundMesonOscillationsHeavyALP}: a larger $m_a$  would lead to larger values of the combination of leptonic couplings involved, worsening the EFT validity prospects.

The exercise above assumed no NP contribution from the muon sector. 
Fig.~\ref{fig:KellRKDeltaMs} considers the whole parameter space $\left|\left(\bc_e-\bc_L\right)_{ee}\right|$ \vs $\left|\left(\bc_e-\bc_L\right)_{\mu\mu}\right|$ that solves the $R_K$ anomaly, again for $m_a=10\GeV$ and  $\left|(\bc_d+ \bc_Q)_{sb}\right|/f_a=10^{-5}\GeV^{-1}$. The $\Delta M_s$ constraint in Eq.~\eqref{ceeboundHeavyALP} falls then within the green band of this plot. The figure shows that, necessarily,
\be
\dfrac{|\left(\bc_e-\bc_L\right)_{ee}|}{f_a} \ge  
10\GeV^{-1}
\label{LowerLimitCeRK}   
\ee
and thus the conclusion described above holds even for a non-vanishing $C^\mu_{P_+}$: it is possible to explain the $R_K$ anomaly consistently with data from semileptonic decays and $B_s$-meson oscillations, but the corresponding couplings are outside the range of validity of the ALP EFT.

\subsubsection*{Anomalous magnetic moment of the electron and the muon}
The measurement of the electric dipole moment of the electron with Caesium atoms~\cite{Parker:2018vye,Hanneke:2008tm,Hanneke:2010au} and the measurement with Rubidium atoms~\cite{Morel:2020dww} show deviations from the SM prediction in opposite directions. We will focus on the Caesium experimental determination, which is the one  that shows the largest tension with the SM of about $\sim2.4\sigma$,
\be
\Delta a_e\equiv a_e^\text{exp}-a_e^\text{SM}=-(88\pm36)\times10^{-14}\,,
\label{DeltaaeExp2sigma}
\ee
and consider the $2\sigma$ interval as a bound on the range allowed. 
In turn, for the data on $g-2$ for the muon, a longstanding $4.2\sigma$ anomaly~\cite{Aoyama:2020ynm} indicates
\be
\Delta a_\mu=\left(25.1\pm5.9\right)\times 10^{-10}\,,
\label{DeltaamuExp4sigma}
\ee
with consistent results across different experiments~\cite{Muong-2:2006rrc,Muong-2:2021ojo}.\footnote{The BMW lattice QCD collaboration computed recently the leading hadronic vacuum polarisation contribution to the muon $g-2$ with sub percent precision~\cite{Borsanyi:2020mff}, and using this result the tension would reduce to only $1.6\sigma$. Recent results from other lattice groups and lattice methodologies~\cite{Ce:2022kxy,Alexandrou:2022amy} are also converging towards a smaller tension with respect to the SM prediction, at least in the so-called ``intermediate'' range, while finding instead tensions in $e^+ e^-$ data. Waiting for further clarification, we will consider in this work the aforementioned value of $\Delta a_\mu$ in Eq.~\eqref{DeltaamuExp4sigma}.}

\begin{figure}[tbh] 
\centering
\subfigure[\em Double ALP Fermion Coupling\label{fig:DiagramsMagneticMomentLH}]%
{\includegraphics[width=0.4\textwidth]{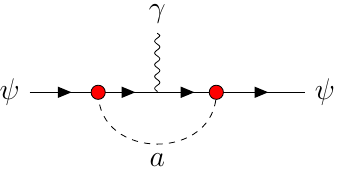}}\qquad
\subfigure[\em Single ALP Fermion Coupling\label{fig:DiagramsMagneticMomentRH}]%
{\includegraphics[width=0.4\textwidth]{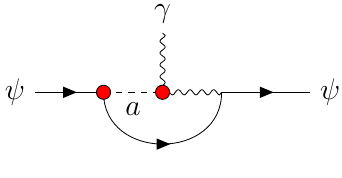}}
\caption{\em One-loop ALP-mediated diagrams contributing to $(g-2)$ of the fermion $\psi$.}
\label{fig:DiagramsMagneticMoment}
\end{figure}
ALP  exchange can contribute  to both $\Delta a_e$~\cite{Bauer:2021mvw}  and $\Delta a_\mu$. The effects appear  at one-loop, as depicted in  Fig.~\ref{fig:DiagramsMagneticMoment}. In the chirality-flip basis Eq.~(\ref{Lag-fermion-flip}),    the ALP-fermion couplings are mass dependent and their insertion in both internal vertices  --Fig.~\ref{fig:DiagramsMagneticMomentLH}-- is expected to be subdominant  with respect to the amplitudes containing one insertion of ALP-anomalous gauge couplings --Fig.~\ref{fig:DiagramsMagneticMomentRH}, and in particular of the ALP-photon anomalous coupling. In the limit $m_a \gg m_\ell$ it results
\be 
\Delta a_\ell^{\text{ALP}} \simeq \, A_\ell\,
\dfrac{c_{a\gamma\gamma} + \Delta c_{a\gamma\gamma}}{f_a}\, 
\dfrac{\left(\bc_e-\bc_L\right)_{\ell\ell}}{f_a}  \,,
\label{eq:heavyALP-g-2}
\ee
where the constant in front of this expression reads
\be 
 A_{\ell} \equiv  \dfrac{m_\ell^2}{2 \pi^2}\left( \log \dfrac{\Lambda^2}{m_a^2} - \dfrac{3}{2} \right) = 
\begin{cases}
1.02\times 10^{-7} \GeV^2\qquad  &\text{for the electron}
\\[2mm]
4.36\times 10^{-3} \GeV^2  &\text{for the muon}\,,
\end{cases}
 \ee
with  $\Lambda$ assumed to be of $\mathcal{O}(1\TeV)$ and  $\Lambda = 4 \pi f_a$ by naive dimensional analysis. In the formula above, $c_{a\gamma\gamma}$ denotes the tree-level arbitrary anomalous gauge coupling in the initial Lagrangian, Eq.~(\ref{non-redundant-L}).  In contrast,  $\Delta c_{a\gamma\gamma}$ is the anomalous contribution induced by the fermion rotation performed to pass to the chirality-flip basis, and it is given by a precise combination of fermion-ALP couplings,
 see Eqs.~(\ref{Deltas})-(\ref{non-redundant-L-flip}), i.e.
 \be
\Delta c_{a\gamma\gamma}\simeq \,  - \dfrac{\alpha_{em}}{4 \pi} \left[ \left(\bc_e-\bc_L\right)_{ee} + \left(\bc_e-\bc_L\right)_{\mu\mu} \right]\,,
\label{ElectronG2ee}
\ee
which using Eqs.~(\ref{DeltaaeExp2sigma}) and (\ref{DeltaamuExp4sigma}) leads respectively to the following $2\sigma$ constraints: 
\be
\dfrac{1}{f_a} \left[     
\left(\bc_e-\bc_L\right)_{ee} \left( \left(\bc_e-\bc_L\right)_{ee} + \left(\bc_e-\bc_L\right)_{\mu\mu} - \dfrac{4\pi}{\alpha_{em}} \, c_{a\gamma\gamma} \right)\right]^{1/2}\in [0.05,\,0.16] \GeV^{-1}\,,
\label{ElectronG2ee-bounds}
\ee
and
\be
\dfrac{1}{f_a} \left[     
\left(\bc_e-\bc_L\right)_{\mu\mu} \left( \dfrac{4\pi}{\alpha_{em}} \, c_{a\gamma\gamma} - \left(\bc_e-\bc_L\right)_{ee} - \left(\bc_e-\bc_L\right)_{\mu\mu} \right)\right]^{1/2}\in [0.023,\,0.038] \GeV^{-1}\,,
\label{ElectronG2mumu-bounds}
\ee
Were the bare anomalous coupling $c_{a\gamma\gamma}$ to vanish, an explanation of  $R_K$ in terms of heavy ALP exchange would be excluded, because the data on  $R_K$ and $\Delta a_{e}$ cannot be simultaneously accommodated, given the bound in Eq.~(\ref{LowerLimitCeRK}) and the fact that \mbox{$\left(\bc_e-\bc_L\right)_{\mu\mu} \ll \left(\bc_e-\bc_L\right)_{ee}$} --see discussion after Eq.~(\ref{first-results}). 

Strictly speaking, though, the possibility to explain through heavy ALP exchange both $R_K$ and $\Delta a_\ell$ cannot be completely excluded because $c_{a\gamma\gamma}$ is arbitrary. Its value can  be fine-tuned to fit for instance $R_K$ data and the $\Delta a_e$ bound. Note that the coupling values then required  would  not allow to account in addition for the $\Delta a_\mu$ anomaly. Indeed,  the expressions for $\Delta a_\mu^{\text{ALP}}$ and $\Delta a_e^{\text{ALP}}$ would then imply the constraint --at the $2\sigma$ level-- 
\be
\dfrac{\left(\bc_e-\bc_L\right)_{\mu\mu}}{\left(\bc_e-\bc_L\right)_{ee}}\simeq\dfrac{\Delta a_\mu^{\text{ALP}}}{\Delta a_e^{\text{ALP}}}\dfrac{A_e}{A_\mu}\in-[0.02,\,0.54]\,,
\label{CorrelationDeltaamue}
\ee
which is inconsistent with the hierarchy between $\left(\bc_e-\bc_L\right)_{ee}/f_a$ and $\left(\bc_e-\bc_L\right)_{\mu\mu}/f_a$ shown in Fig.~\ref{fig:KellRKDeltaMs}. 
 
Building on the same freedom on the value of the initial $c_{a\gamma\gamma}$, one may still wonder whether it is technically possible a solution in which 
the amplitude of  the second diagram in Fig.~\ref{fig:DiagramsMagneticMoment}  cancels for either $\Delta a_e$ or $\Delta a_\mu$,  forcing $c_{a\gamma\gamma} + \Delta c_{a\gamma\gamma}=0$, so as to  explain then  the experimental value of that observable  in terms of  just the  first diagram in that figure (which has been neglected up to now).  This option leads to a dead end as well: i) for   $\Delta a_e$, because its  $(m_e)^4 / ( f_a m_a)^2$ suppression  makes it totally negligible;\footnote{The same applies if the  $\Delta a_e$ value inferred from Rubidium was considered instead, in spite of its weaker strength.} ii)  for $\Delta a_\mu$, because the $(m_\mu)^4 / ( f_a m_a)^2$ contribution is always negative, contrary to the experimental $\Delta a_\mu>0$ value, and also the prediction for $\Delta a_e$ would be incompatible with observation.
 
In summary, no simultaneous explanation  in terms of tree-level heavy ALP exchange is possible for the three observables in the set $\{R_K, \Delta a_{e}, \Delta a_{\mu}\}$. Furthermore, although such an explanation is possible for the $\{R_K,  \Delta a_{e}\}$ set, the data on $R_K$ would always require a strong  disregard of the EFT validity condition.

\boldmath
\subsubsection{$R_{K^\ast}$, $B\to K^\ast\ell^+\ell^-$,  and  $B_s\to \ell^+ \ell^-$}
\unboldmath
The analysis of $R_{K^\ast}$ can be done in analogy with that for  $R_{K}$ and $B\to K \ell^+\ell^-$ above, although the data on $B_s\to \ell^+ \ell^-$ will add an extra essential ingredient because the purely leptonic decays share with $R_{K^\ast}$ the same dependence on the effective couplings $C^\ell_{P_-}$ only. 

\boldmath
\subsubsection*{$B\to K^\ast\ell^+\ell^-$}
\unboldmath  

\begin{figure}[t] 
\centering
\includegraphics[width=0.49\textwidth]{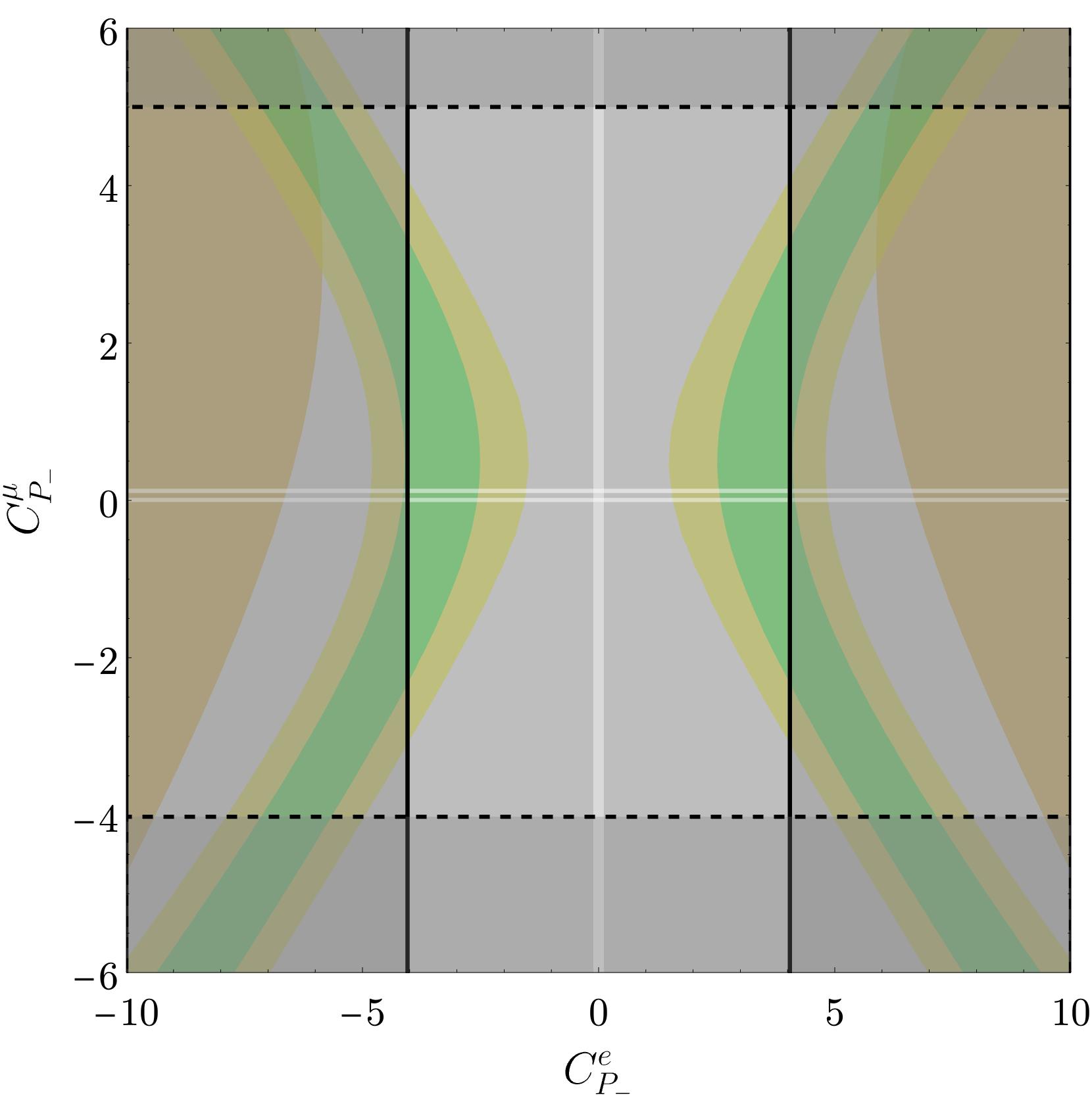}
\caption{\em Parameter space for $R_{K^{\ast}}$, for an ALP heavier than $B$ mesons.  In yellow and  green are respectively depicted the $1\sigma$ and $2\sigma$ solutions to the central bin, while in orange are indicated the $1\sigma$ solutions to the low bin. The grey regions around the frame of the figures are excluded  at $2\sigma$  by data on semileptonic $B\to K^{\ast}e^+e^-$ (solid black contours) and  $B\to K^{\ast}\mu^+\mu^-$ (dashed black contours) decays. The regions excluded by purely leptonic $B_s$ decays reach the central area and are also depicted in grey (horizontally for the electron channel and vertically for the muon one) leaving available the narrow white strips.}
\label{fig:RKstarHALP}
\end{figure}

Using the EOS software, we obtain
\be
\begin{aligned}
\BR(B\to K^\ast \mu^+ \mu^-)_{1.1\GeV^2}^{6.0\GeV^2} & =10^{-7}\times\left(1.9-7.4\times 10^{-2}\,C^\mu_{P_-} +7.5\times 10^{-2}\,C^{\mu2}_{P_-}\right)\,,\\
\BR(B\to K^\ast e^+ e^-)_{1.1\GeV^2}^{6.0\GeV^2} & =10^{-7}\times\left(1.9-3.6\times 10^{-4}\, C^e_{P_-} +7.5\times 10^{-2}\,C^{e2}_{P_-}\right)\,,
\end{aligned}
\label{BKstarllHeavyALPHigh}
\ee
and
\be
\begin{aligned}
\BR(B\to K^\ast \mu^+ \mu^-)_{0.045\GeV^2}^{1.1\GeV^2} & =10^{-7}\times\left(1.2-9.3\times 10^{-3}\,C^\mu_{P_-} +1.5\times 10^{-3}\,C^{\mu2}_{P_-}\right)\,,\\
\BR(B\to K^\ast e^+ e^-)_{0.045\GeV^2}^{1.1\GeV^2} & =10^{-7}\times\left(1.3-4.8\times 10^{-5}\, C^e_{P_-} +1.6\times 10^{-3}\,C^{e2}_{P_-}\right)\,,
\end{aligned}
\label{BKstarllHeavyALPLow}
\ee
respectively for the central and low energy bin regions, see Eq.~(\ref{ExperimentalRKstar}). The theoretical errors are estimated to be at the $15\%$ level at $1\sigma$ (similarly to those for the semileptonic $B\to K$ decays earlier on).  The comparison of these equations with those for $B\to K$ semileptonic transitions in Eqs.~\eqref{BKllHeavyALP} shows 
a very similar structure, with in particular the terms quadratic on the Wilson coefficients being positive, and a very similar pattern for the prefactors of linear \vs quadratic terms. A comparison with the experimental data in Tab.~\ref{tab:Proposal} results in the following $2\sigma$ bounds on the Wilson coefficients, 
\begin{equation}
C_{P_-}^e \in [-4.0,4.0]\quad\quad \text{and} \quad \quad  C_{P_-}^\mu \in [-4.0,5.0]\,,
\end{equation} 
which are illustrated in Fig.~\ref{fig:RKstarHALP} as grey shaded regions delimitated by solid (electrons) and dashed (muons) black contours.

\boldmath
\subsubsection*{$R_{K^\ast}$}
\unboldmath
Analogous considerations to those for $R_K$  will apply then to $R_{K^\ast}$, given the similarity between Eqs.~\eqref{BKllHeavyALP} and those above for the semileptonic $R_{K^\ast}$ decays. In consequence, the experimental tension in $R_{K^\ast}$ is expected to allow for an explanation in terms of ALP tree-level exchange only if the electron sector would receive NP contributions. We expatiate next on this point. From Eqs.~(\ref{BKstarllHeavyALPHigh}) and (\ref{BKstarllHeavyALPLow}) it follows that 
\be
R_{K^\ast}=
\begin{cases}
1+ \dfrac{0.02\,C_{P_-}^e-3.89\,C_{P_-}^\mu+3.95(\,C_{P_-}^{\mu2}-C_{P_-}^{e2})}{100-0.02\,C_{P_-}^e+3.95\,C_{P_-}^{e2}}\qquad\qquad & \text{central bin}
\\[2mm]
0.923+ \dfrac{0.03\,C_{P_-}^e-7.15\,C_{P_-}^\mu+1.15\,C_{P_-}^{\mu2}-1.13\,C_{P_-}^{e2}}{1000-0.04\,C_{P_-}^e+1.23\,C_{P_-}^{e2}}& \text{low bin,}
\end{cases}
\label{RKastHeavyALP}
\ee
with theoretical errors that are expected to be negligible with respect to the corresponding experimental ones. 

Let us first extract the values for $C^{e,\mu}_{P_-}$ that could solve the tension in $R_{K^\ast}$ in case  NP  enters only in either the electron sector or the muon sector. For $C^\mu_{P_-}=0$, the $2\sigma$ error bands for $R_{K^\ast}$, $R_{K^\ast}\in[0.519,\,0.911]$ (central bin) and $R_{K^\ast}\in[0.504,\,0.875]$~\cite{LHCb:2017avl} (low bin) lead to
\be
\text{For $C^\mu_{P_-}=0$:}\quad
\begin{cases}
C^e_{P_-}\in[1.6,\,4.8]\vee[-4.8,\,-1.6]\qquad\qquad & \text{central bin}
\\[2mm]
C^e_{P_-}\in[6.7,\,26.1]\vee[-26.1,\,-6.7]&\text{low bin.}
\end{cases}
\label{NaiveCepmRKstar}
\ee
These two sets of solutions do not overlap even partly and therefore there is no possible explanation in terms of a heavy ALP for the deviations in both energy bins. Alternatively, for $C^e_{P_-}=0$, there is no  $C^\mu_{P_-}$ value  that can explain $R_{K^\ast}$ with the sensitivity considered.

Let us finally consider the ALP explanations to $R_{K^\ast}$ within the two-dimensional  parameter space of couplings 
$\{C^e_{P_-},\, C^\mu_{P_-}\}$. The solutions are depicted in Fig.~\ref{fig:RKstarHALP} (in green, yellow and orange). This figure also shows, though, that when the data on $B\to K^\ast e^+ e^-$  $B\to K^\ast \mu^+ \mu^-$ are taken into account, the regions where the low bin anomaly can be explained are ruled out and those for the central bin one are reduced to 
 \be
1.6<|C^e_{P_-}| <4\,,\qquad\qquad
-3<|C^\mu_{P_-}| <4 \,.
\label{ClpmRKstar}
\ee
\boldmath
\subsubsection*{$B_s\to \ell^+ \ell^-$}
\unboldmath 
A second observable that directly depends on the operator $\cO^\ell_{P_-}$ for a given charged lepton $\ell$ is the branching ratio ${\BR(B_s\to \ell^+ \ell^-)}$. 
The corresponding experimental measurements are in good agreement with the SM and therefore any NP effect should be at most marginal. By implementing the Flavio software, and after performing an interpolation procedure, we obtain the contribution of the SM plus those mediated by  tree-level exchange of a heavy ALP: 
\be
\begin{aligned}
\ov{\BR}(B_s\to \mu^+\mu^-)
& = 10^{-9}\times\left(3.67-1.15\times 10^2\, C^\mu_{P_-} +9.04\times 10^{2}\,C^{\mu2}_{P_-}\right)\,,
\\
\ov{\BR}(B_s\to  e^+e^-)
&=10^{-14}\times\Big(8.58-5.57\times 10^{4} C^e_{P_-} +9.05\times 10^{7}\,C^{e2}_{P_-}\Big)
\,,
\end{aligned}
\label{BsllHeavyALP}
\ee
where the bar over the symbol for the branching ratio denotes untagged decays, that is, the time-integrated quantities which include the probability for the meson to oscillate before decaying (the tagged quantity is $\cO(15\%)$ smaller than the results shown). If the EOS software is used instead of Flavio, the numerical output is $9\%$ smaller than that in Eq.~\eqref{BsllHeavyALP}; this difference is most probably due to some loop contributions considered in Flavio, as discussed in Ref.~\cite{Beneke:2019slt}. The theoretical error on the SM prediction for these quantities is much smaller than that for semileptonic $B$ decays and it is of $\mathcal{O}(4\%)$  at the $1\sigma$ level.

The size of the numerical factors appearing in front of the Wilson coefficients $C^{e,\mu}_{P_-}$ indicates that the latter should not exceed values of about $|C^{e,\mu}_{P_-} |\sim0.1$. More precisely, the  regions allowed  in order to remain within the $2\sigma$ confidence level of the $B_s\to\ell^+\ell^-$ measurements,  $\cB\left(B_s\to\mu^+\mu^-\right)\in[2.2,\,4.1]\times10^{-9}$~\cite{LHCb:2021vsc} and $\cB\left(B_s\to e^+e^-\right)<11.2\times10^{-9}$~\cite{LHCb:2020pcv}, are
\be
\begin{aligned}
C^e_{P_-}&\in[-0.11,\,0.11]\,,\\
C^\mu_{P_-}&\in[-0.0033,\,0.014]\vee[0.11,\,0.13]\,.
\end{aligned}
\label{LimitsBs}
\ee
These solutions are  incompatible with the naive values in Eq.~\eqref{ClpmRKstar}. In summary, the data from purely leptonic $B_s$ decays precludes an explanation of $R_{K^\ast}$ in terms of a heavy ALP, even when the complete parameter space for ALP-electron and ALP-muon couplings is considered.\footnote{A combined analysis of the ATLAS, CMS and LHCb results on $\cB\left(B_s\to\mu^+\mu^-\right)$ using data between 2011 and 2016, showed a small tension with the SM predictions at the $2\sigma$ level~\cite{ATLAS:2020acx}. Was this combined result included,
 the conclusions above would not change. In any case, the more recent analysis  by the LHCb collaboration which includes data till 2018~\cite{LHCb:2021vsc}  
 shows a smaller deviation from the SM prediction. } 
This is illustrated for $R_{K^\ast}$ in Fig.~\ref{fig:RKstarHALP}:  the bounds from purely leptonic $B_s$ meson decays only allow 
 very narrow (white) strips in the parameter space; the impact of $B_s\to e^+ e^-$ in particular leaves no region to explain $R_{K^\ast}$, not even for the central energy bin window.

The comparison of our two-parameter space survey above can be contrasted with those in the one-parameter analysis in Ref.~\cite{Bauer:2021mvw}. While we find that an explanation for $R_{K^*}$ in terms of a heavy ALP exchange is excluded, $R_K$ could be accounted for technically, albeit at  a heavy theoretical cost: to go out of the range of validity of the EFT. If the latter condition was nevertheless disregarded, it would be possible to accommodate at the same time the bound on $\Delta a_e$, but not the $\Delta a_\mu$ anomaly.

%
%
\section{Light ALP}
\label{sec:LIGHTALP}

This section explores the option of an ALP lighter than the $B$ mesons and whose mass is in the ballpark of the energy bin windows considered for the neutral $B$-anomalies. Therefore, the ALP field cannot be integrated out and resonant effects may become relevant. The analysis strongly depends on the precise value of $m_a$. We explore below two distinct scenarios:
\begin{description}
\item[-] ALP mass well within the energy range of the bin under consideration.
\item[-] ALP mass outside the bin window but close to it.
\end{description}

\boldmath
\subsection{ALP mass within the bin window}
\unboldmath
For the $B\to K^{(\ast)} \ell^+ \ell^-$ processes,  we rely on analytic computations of three body $B$ decays which use the relativistic Breit-Wigner expression for the ALP propagator under the condition that the ALP decay width $\Gamma_a$ is smaller than its mass, $\Gamma_a<m_a$. The matrix elements as computed in Refs.~\cite{Bobeth:2007dw,Bharucha:2015bzk} will be used, together with  the inputs in Tab.~\ref{tab:parameters} of App.~\ref{sec:Input} for the SM Wilson coefficients. The form factors -- which are the main source of theoretical uncertainties -- are taken from Refs.~\cite{Bailey:2015dka} and \cite{Bharucha:2015bzk}. A detailed account of our computations can be found in App.~\ref{sec:formulas}, which includes a comparison between our results for the relevant decay widths with those obtained numerically via the Flavio software: we find a very accurate agreement in the energy bin regions relevant for the $R_K$ and $R_{K^\ast}$ anomalies. 

Before presenting the numerical results, it is pertinent though to discuss analytically the validity of the narrow width approximation (NWA), which justifies that the ALP can be safely taken on-shell. In this approximation, the total branching ratio can be decomposed as
\be
\BR(B\to K^{(\ast)}\ell^+\ell^-)=\BR(B\to K^{(\ast)}\ell^+\ell^-)^\text{SM}+\BR(B\to K^{(\ast)}a)\times \BR(a\to \ell^+\ell^-)\,,
\label{NWABKll}
\ee
where the SM contributions can be found in Tab.~\ref{tab:smBR} while the expressions for $\BR(B \to K a)$  and $\BR(B \to K^\ast a)$ are respectively given as a function of $m_a$ by
\be
\begin{aligned}
\BR(B \to K a)&= \tau_B \dfrac{M_B\,\left[ (\bc_d + \bc_Q)_{sb}\right]^2}{64\,  \pi\, f_a^2} f_0^2[m_a^2]\, \lambda^{1/2}_{BKa}\left(1-\dfrac{M_K^2}{M_B^2}\right)^2\,,\\
\BR(B \to K^\ast a) &= \tau_B \dfrac{\left[ (\bc_d - \bc_Q)_{sb}\right]^2}{64\,  \pi\, f_a^2\, M_B^3} A_0^2[m_a^2]\, \lambda^{3/2}_{B K^* a}\,,
\end{aligned}
\ee
where $\tau_B$ and $M_B$ denote respectively the  lifetime and mass of the $B$ mesons (i.e. $B^{0,\pm}$) and $M_{K^{(*)}}$ is the neutral or charged kaon mass (see Tab.~\ref{tab:parameters}). In turn, $f_0[m_a^2]$ and $A_0[m_a^2]$ are two form factors whose dependence on the ALP  mass can be extracted from Refs.~\cite{Bailey:2015dka} and \cite{Bharucha:2015bzk}, respectively,
\be
\begin{aligned}
f_0[m_a^2] &\approx 3.45\times10^{-1}+2.84\times10^{-3}\, \dfrac{m_a^2}{\GeV^2}+6.97\times10^{-4} \,\dfrac{m_a^4}{\GeV^4}\,,\\
A_0[m_a^2] &\approx \, 0.37 + 2.18 \times 10^{-2} \, \dfrac{m_a^2}{\GeV^2} + 8.83 \times 10^{-4} \, \dfrac{m_a^4}{\GeV^4}\,,
\end{aligned}
\ee
where $\lambda_{BK^{(\ast)}a}$ are the K\"{a}ll\'en triangle functions $\lambda_{BK^{(\ast)}a} \equiv \lambda(M_B^2 , \, M_{K^{(\ast)}}^2 , \, m_a^2)$ such that
\begin{equation}
\lambda (a, \, b, \, c) \equiv a^2 + b^2 + c^2 - 2 ab -2 bc - 2 ca \,.
\end{equation}

In turn, the purely leptonic decay width of an ALP  reads, 
\be
\Gamma(a\to\ell^+\ell^-)=\dfrac{m_a\,m_\ell^2}{8\,\pi\,f_a^2}\left[(\bc_e-\bc_L)_{\ell\ell}\right]^2\left(1-\dfrac{4m_\ell^2}{m_a^2}\right)^{1/2}\,.
\label{Gamma-all}
\ee
Given the energy windows of the bins relevant for the $R_K$ and $R_{K^\ast}$ anomalies, an explanation in terms of the exchange of an on-shell ALP requires
\be
m_a \ge 2 m_\mu\,,
\ee
and in consequence, both leptonic decay channels are kinematically open. Nevertheless,  in order to explain the neutral anomalies via an on-shell ALP, the electron-ALP coupling should dominate. Indeed,  it follows from Eq.~\eqref{NWABKll} that 
\be
R_{K^{(\ast)}}\simeq1+
\dfrac{\BR(B\to K^{(\ast)}a)}{\BR(B\to K^{(\ast)}e^+e^-)^\text{SM}}\dfrac{
\Big(m_\mu^2\left[(\bc_e-\bc_L)_{\mu\mu}\right]^2-m_e^2\left[(\bc_e-\bc_L)_{ee}\right]^2\Big)}{
\Big(m_\mu^2\left[(\bc_e-\bc_L)_{\mu\mu}\right]^2+m_e^2\left[(\bc_e-\bc_L)_{ee}\right]^2\Big)}\,,
\ee
which requires for $R_{K^{(\ast)}}<1$ that
\be
\frac{|(\bc_e-\bc_L)_{ee}|}{|(\bc_e-\bc_L)_{\mu\mu}|}\ge \frac{m_\mu}{m_e} \sim 200\,. 
\label{AlPinBinConditioneeLarge}
\ee
It is therefore a good approximation to neglect the ALP-muon couplings in the solutions to the neutral $B$-anomalies.\footnote{The hierarchy suggests a UV structure with all lepton couplings vanishing, but the electron one. We have verified that this condition is RGE stable, with the induced ALP-muon coupling being two-loop suppressed with respect to the ALP-electron coupling.} This has a most important consequence: {\it the solutions to $R_K$ and $R_{K^\ast}$ in terms of resonant ALP exchange are basically independent of the precise values of the ALP coupling to leptons, because $\BR(a\to e^+ e^-)\sim 1$}, see Eq.~(\ref{NWABKll}). This is in stark contrast to the $B$ anomaly solutions via a heavy ALP discussed earlier on, or a very light ALP (to be discussed in the next section), for which lepton couplings  scale inversely proportional to quark couplings in the solutions to $R_K$ and $R_{K^\ast}$, sourcing strong violations of the EFT validity conditions once other independent observables are considered.

\subsubsection*{On the validity of the NWA}

As the use of the Breit-Wigner expression for the ALP propagator is meaningful only as far as the ALP decay rate is smaller than its mass, let us assume a conservative $\Gamma_a/m_a<1/5$ condition. Given the constraint in Eq.~(\ref{AlPinBinConditioneeLarge}), it is reasonable as a working hypothesis to neglect the muon sector ALP couplings, $(\bc_e-\bc_L)_{\mu\mu}=0$. It then follows from Eq.~(\ref{Gamma-all}) the constraint
\be
\dfrac{\left|(\bc_e-\bc_L)_{ee}\right|}{f_a}\lesssim\sqrt{\dfrac{8\,\pi}{5\,m_e^2}}\simeq 4.4\times 10^3\GeV^{-1}\,.
\label{OnShellALPeeCondition}
\ee
This result is fairly independent of the ALP mass and is only slightly modified when considering non-vanishing ALP couplings to both electrons and muons. The corresponding numerical analysis is shown in Fig.~\ref{fig:OnShell_RKvsLeptonCouplings}, in which the region excluded by the NWA validity is depicted in red. Its vertical border corresponds to Eq.~(\ref{OnShellALPeeCondition}). The horizontal border stems instead from the analogous upper limit that can be set for the ALP-muon couplings by formally setting to zero those for electrons, $(\bc_e-\bc_L)_{ee}=0$,  
\be
\dfrac{\left|(\bc_e-\bc_L)_{\mu\mu}\right|}{f_a}
\lesssim\sqrt{\dfrac{8\,\pi}{5\,m_\mu^2}}\simeq 21\GeV^{-1}\,.
\label{OnShellALPmumuCondition}
\ee

\subsubsection*{Prompt ALP decay}
The final leptons in the semileptonic $B$-decays are observed to come from the same point in which the $K^{(\ast)}$ meson is produced, and therefore the ALP needs to have a prompt decay. Considering the typical boost factors at LHCb, this leads to the requirement~\cite{Altmannshofer:2017bsz} 
\be
\Gamma_a\geq0.02\eV\,.
\ee
Accordingly to the previous discussion, assuming that the ALP decays only into electrons, we find a lower bound on the ALP-electron couplings given by:
\be
\dfrac{\left|(\bc_e-\bc_L)_{ee}\right|}{f_a}\gtrsim\sqrt{\dfrac{0.16\,\pi\eV}{m_a\,m_e^2}}\simeq 4.4\times 10^{-2}\left(\dfrac{1\GeV}{m_a}\right)\GeV^{-1}\,.
\label{OnShellALPeeConditionPrompt}
\ee
This determines the vertical frontier of the region excluded by the condition of prompt decay, depicted in grey in Fig.~\ref{fig:OnShell_RKvsLeptonCouplings} for the solutions to $R_K$, see further below. The horizontal frontier in that figure results similarly from  the lower bound on  muon couplings that follows  by formally disregarding the electron contribution in the ALP total decay rate,
\be
\dfrac{\left|(\bc_e-\bc_L)_{\mu\mu}\right|}{f_a}
\gtrsim\sqrt{\dfrac{0.16\,\pi\eV}{m_a\,m_\mu^2}}\simeq 2.1\times 10^{-4}\left(\dfrac{1\GeV}{m_a}\right)\GeV^{-1}\,.
\label{OnShellALPmumuConditionPrompt}
\ee

\boldmath 
\subsubsection*{$\Delta {M_{s}}$}
\unboldmath 
We will refrain below from determining the impact of the meson oscillation data on $R_K$ and $R_{K^\ast}$ in the present case of an on-shell ALP, because the bounds to be obtained from semileptonic $B$ decays are much stronger.

\boldmath
\subsubsection{$B\to K\ell^+\ell^-$,  $R_{K}$ and magnetic moments}
\unboldmath

\boldmath
\subsubsection*{$B\to K \ell^+\ell^-$}
\unboldmath

\begin{figure}[tbh] 
\centering%
\includegraphics[width=0.46\textwidth]{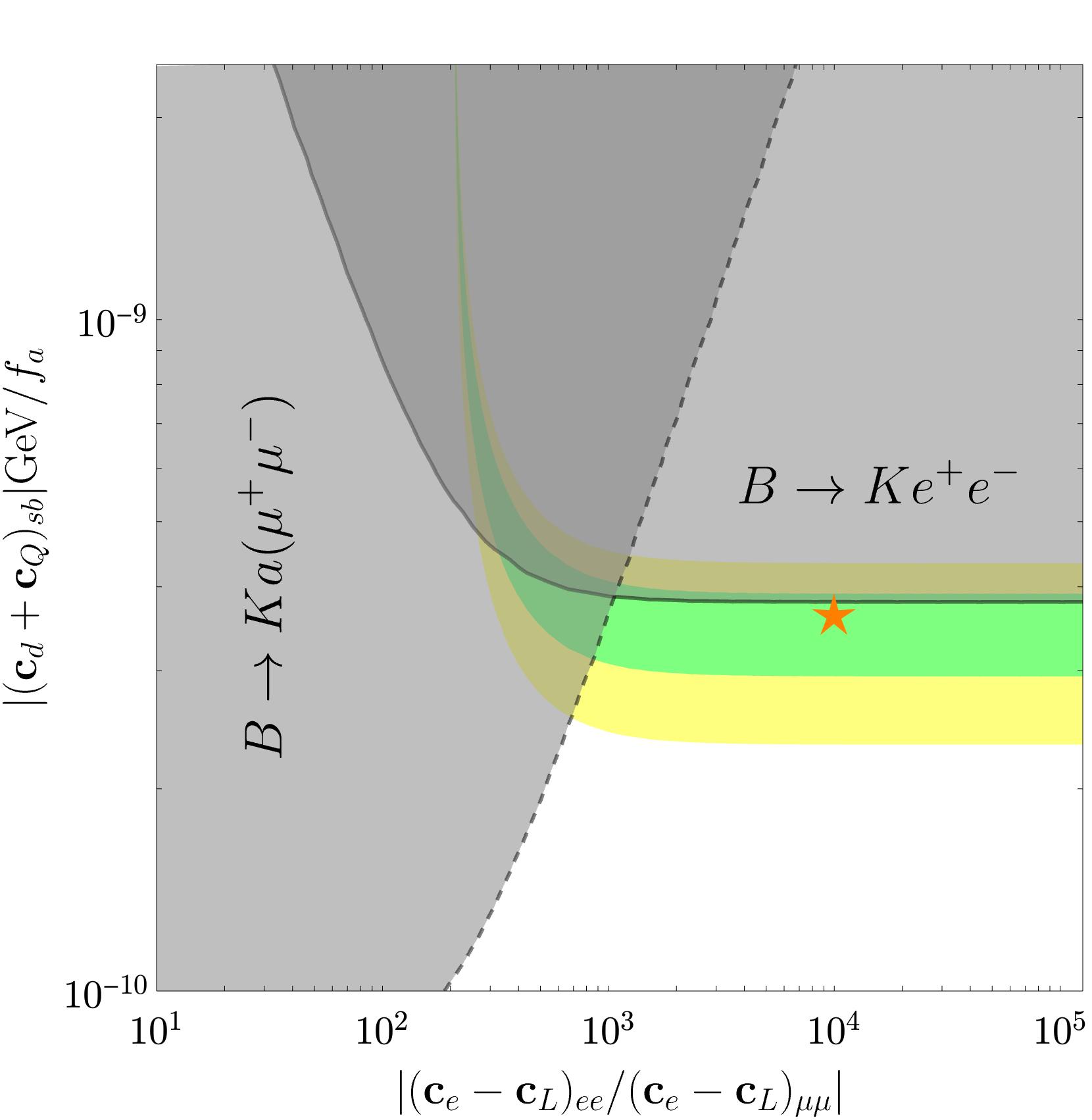}
\caption{\em  ALP mass within the central bin range. Constraints from semileptonic $B$-decays on the parameter space of ALP couplings to quarks and leptons. In grey the excluded regions, while in green (yellow) the solutions to $R_K$ at $1\sigma$ ($2\sigma$). The orange star corresponds to the  illustrative benchmark point $m_a=1.2\GeV$ with  $(\left|(\bc_e-\bc_L)_{ee}\right|/f_a,\,\left|(\bc_e-\bc_L)_{\mu\mu}\right|/f_a)=(10^{-1},\,10^{-5})\GeV^{-1}$. }
\label{fig:limitsQuarkCouplingsRK}
\end{figure}

For the range of ALP masses within the central bin range, the data on $B\to K e^+e^-$ determined in the kinematic region of that bin, ${1.1<q^2 < 6.0}\,\GeV^2$ see Tab.~\ref{tab:Proposal}, result in the $2\sigma$ bound
\be 
\dfrac{\left|(\bc_d + \bc_Q)_{sb}\right|}{f_a} \sqrt{\mathcal{B} (a \to e^+ e^-)} \lesssim3.8\times 10^{-10}\,
\GeV^{-1}\,,
\label{ALPquarkBoundFromBKee}
\ee
This result is fairly independent of the precise value of $m_a$ as it enters only through a very mild dependence in the  $f_0$ form factor.
In the approximation $\mathcal{B} (a \to e^+ e^-)\sim1$, Eq.~(\ref{ALPquarkBoundFromBKee}) would directly imply $|(\bc_d + \bc_Q)_{sb}|/f_a \lesssim3.8\times 10^{-10} $, a bound that gets slightly relaxed though as a consequence of the branching ratio of $a\to e^+e^-$ being different from $1$. This can be appreciated in Fig.~\ref{fig:limitsQuarkCouplingsRK}: the excluded region for $\left|(\bc_d + \bc_Q)_{sb}\right|$ as a function of the ratio of lepton couplings ${|(\bc_e-\bc_L)_{ee}}/{(\bc_e-\bc_L)_{\mu\mu}|}$ is shown in grey.

The same combination of ALP-quark couplings can be independently bounded from analogous data on $B\to K \mu^+\mu^-$, which stem from dedicated searches at LHCb for exotic resonances as reported in Tab.~\ref{tab:Proposal}, 
\be
\dfrac{\left|(\bc_d + \bc_Q)_{sb}\right|}{f_a} \sqrt{\mathcal{B} (a \to \mu^+ \mu^-)} \lesssim 7.4\times 10^{-11}
\GeV^{-1}\,.
\label{ALPquarkBoundFromBKmumu}
\ee
This does not translate into stronger bounds on $|(\bc_d + \bc_Q)_{sb}|$ than those stemming from Eq.~(\ref{ALPquarkBoundFromBKee}), once the  values of $\cB (a\to \mu^+\mu^-)$ are taken into account in the $m_a$ range under discussion, except in the region where the ratio of leptonic couplings acquires the smallest values, see Fig.~\ref{fig:limitsQuarkCouplingsRK}.
 
\boldmath
\subsubsection*{$R_K$}
\begin{figure}[h!]
\centering
\subfigure[Quark couplings \vs $m_a$ for $R_K$.
\label{fig:OnShell_RKvsQuarkCouplings}]%
{\includegraphics[width=0.47\textwidth]{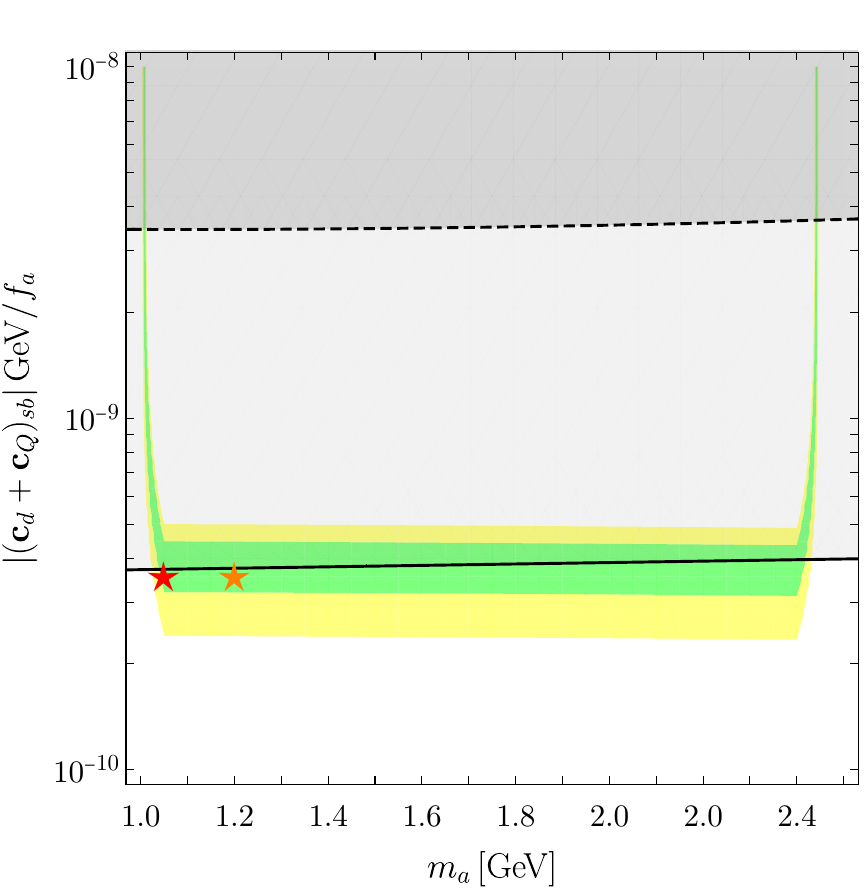}}
\qquad
\subfigure[Lepton couplings for $R_K$. 
\label{fig:OnShell_RKvsLeptonCouplings}]%
{\includegraphics[width=0.47\textwidth]{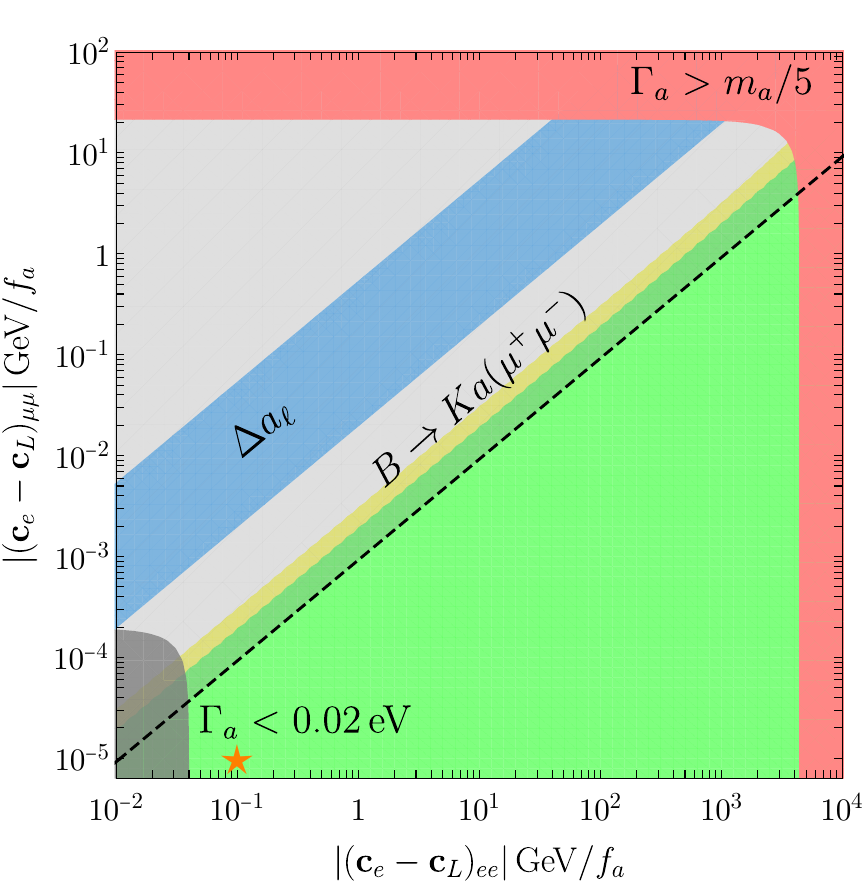}}
\caption{\em  ALP mass on-shell within the central bin range.  In green (yellow) the $1\sigma$ $(2\sigma)$ solutions to $R_K$. The stars correspond to the benchmark ALP-lepton couplings  $(\left|(\bc_e-\bc_L)_{ee}\right|/f_a,\,\left|(\bc_e-\bc_L)_{\mu\mu}\right|/f_a)=(10^{-1},\,10^{-5})\GeV^{-1}$, for two different values of the ALP mass as discussed in the text. On the left: parameter space for ALP-quark couplings \vs $m_a$. In grey the experimental bounds from $B\to K\mu^+\mu^-$ (enclosed by the dashed line) and $B\to Ke^+e^-$ (enclosed by the solid line). On the right: parameter space $\left|(\bc_e-\bc_L)_{ee}\right|/f_a$ \vs $\left|(\bc_e-\bc_L)_{\mu\mu}\right|/f_a$, for $m_a=1.2\GeV$ and 
$\left|(\bc_d + \bc_Q)_{sb}\right|/f_a=3.8\times 10^{-10}$. The shaded red region corresponds to the exclusion condition $\Gamma_a<m_a/5$ in Eqs.~\eqref{OnShellALPeeCondition} and ~\eqref{OnShellALPmumuCondition}, while the dark grey one to the prompt decay condition in Eqs.~\eqref{OnShellALPeeConditionPrompt} and \eqref{OnShellALPmumuConditionPrompt}. The light grey region is excluded by the LHCb search for an exotic resonance decaying to muons. The blue band shows the parameter space compatible with $\Delta a_\mu$ once the photon coupling is fixed to comply with the $\Delta a_e$ bound, both quantities taken at the $2\sigma$ level.}
\label{Fig:OnShell_RK}
\end{figure}

\unboldmath
The parameter space in which the $R_K$ anomaly can be explained through the on-shell exchange of an ALP within one (two) sigma  is depicted in green (yellow) in the plots that follow. In all of them, the orange star corresponds to the  illustrative benchmark point $m_a=1.2\GeV$ and  $(\left|(\bc_e-\bc_L)_{ee}\right|/f_a,\,\left|(\bc_e-\bc_L)_{\mu\mu}\right|/f_a)=(10^{-1},\,10^{-5})\GeV^{-1}$. The parameter space is depicted as a function of:
\begin{itemize}
\item[-] Quark couplings \vs lepton couplings in Fig.~\ref{fig:limitsQuarkCouplingsRK}, for an ALP mass $m_a=1.2\GeV$.
\item[-] Quark couplings \vs ALP mass in Fig.~\ref{fig:OnShell_RKvsQuarkCouplings}. The limit on quark couplings obtained above, Eq.~(\ref{ALPquarkBoundFromBKee}), is depicted as a continuous line.  
\item[-]  Muon couplings \vs electron couplings in Fig.~\ref{fig:OnShell_RKvsLeptonCouplings}, also for $m_a=1.2 \GeV$ and for quark coupling values which saturate Eq.~(\ref{ALPquarkBoundFromBKee}). The upper-left half of the parameter space in this plot (in light grey) is excluded by  the constraint in Eq.~\eqref{ALPquarkBoundFromBKmumu}; this constraint turns out to be stronger than that in Eq.~(\ref{AlPinBinConditioneeLarge}).
\end{itemize}

These figures indicate that indeed $R_K$ could be explained by the on-shell exchange of an ALP and furthermore that the validity of the ALP EFT is maintained for those solutions which are located towards the lower left corner of Fig.~\ref{fig:OnShell_RKvsLeptonCouplings}, an example being the benchmark point indicated by the orange star.

\subsubsection*{Anomalous magnetic moment of the electron and the muon}
The analysis of anomalous magnetic moments for a heavy ALP applies as well to the resonant ALP considered in this section, because $m_a$ is still larger than the electron and muon masses, and in consequence the expression in Eq.~\eqref{eq:heavyALP-g-2} holds. It results that, taking now the value $m_a = 1.2$ GeV, the bounds on the right hand side of Eqs.~(\ref{ElectronG2ee-bounds})  and Eq.~(\ref{ElectronG2mumu-bounds}) are now multiplied by a factor $\sim 0.8$. Once again, it would be possible to remain within the EFT validity range and account simultaneously for the data in the set $\{R_K, \Delta a_e\}$ or  for those of the two anomalies, $\{R_K, \Delta a_{\mu}\}$, while it would not be possible to account simultaneously for the data on the three observables $\{R_K, \Delta a_{e}, \Delta a_{\mu}\}$. Indeed, the blue region in Fig.~\ref{fig:OnShell_RKvsLeptonCouplings}, for which the ALP-couplings are required within $2\sigma$ to both respect the $\Delta a_e$ bound and to account for the  $a_\mu$ anomaly falls outside the parameter space that would explain $R_K$.

\boldmath
\subsubsection{$B\to K^\ast \ell^+\ell^-$,   $R_{K^\ast}$, $B_s\to \ell^+\ell^-$ and magnetic moments}
\unboldmath
\begin{table}[t]
\centering{}
\renewcommand{\arraystretch}{1.4}
\resizebox{\textwidth}{!}{
\begin{tabular}{c|c|c|c}
Observable&
$m_a^2$ [GeV$^2$] & 
Values& 
$|(\bc_d - \bc_Q)_{sb}|\sqrt{\mathcal{B}_{a\to \ell^+\ell^-}}/f_a$[GeV$^{-1}$]  \\[1mm]
\hline
&&&\\[-4mm]
$\mathcal{B}(B^0 \to K^{0\ast} a(e^+ e^-))$&$(0.0004,0.05)$
&$<1.344\times 10^{-7}$& $<7.96\times10^{-10}$\\
&$(0.05,0.15)$&$<1.22\times 10^{-8}$&$<2.40\times10^{-10}$\\
&$(0.25,0.4)$&$<1.97\times 10^{-8}$&$<3.05\times10^{-10}$\\
&$(0.4,0.7)$&$<1.74\times 10^{-8}$&$<2.87\times10^{-10}$\\
&$(0.7,1)$&$<6.5\times 10^{-9}$&$<1.75\times10^{-10}$\\[1mm]
\hline
&&&\\[-4mm]
$\mathcal{B}(B^0 \to K^{0\ast} a (\mu^+ \mu^-))$ & $(0.05,18.9)$ & $<3\times 10^{-9}$\cite{LHCb:2015nkv} & $< 1.19\times10^{-10}$ \\[1mm]
\hline
&&&\\[-4mm]
$\mathcal{B}(B^0\to K^{0\ast} e^+ e^-)$ &$(1.1,6)$&$(1.8\pm 0.6)\times 10^{-7}$\cite{Belle:2019oag}& $<6.46\times10^{-10}$\\
&$(0.1,8)$&$(3.7\pm1.0)\times 10^{-7}$\cite{Belle:2019oag}& $<8.71\times10^{-10}$\\[1mm]
\hline
\end{tabular} 
}
\caption{\em Constraints on the ALP-quark coupling from $B \to K^\ast \ell^+ \ell^-$ and $B \to K^\ast a(a\to\ell^+ \ell^-)$ decay processes used in the following figures of this section. The bounds in the last column are expressed at the $2\sigma$ level. For each bound, the value of $m_a$ considered lies in the middle of the corresponding energy bin window. The values presented in the third column are those in Tab.~\ref{tab:Proposal} and are copied here for convenience.}
\label{tab:dedicatedsearchesBKstarll}
\end{table}


While $B\to K \ell^+\ell^-$ offered light on $|(\bc_d + \bc_Q)_{sb}|$, $B\to K^\ast \ell^+\ell^-$ tests the orthogonal combination $|(\bc_d - \bc_Q)_{sb}|$.

The experimental information on the decay $B\to K^\ast e^+e^-$  is more detailed than for its $B\to K$ counterpart, see Tab.~\ref{tab:dedicatedsearchesBKstarll}. The bounds on NP presented in the first row of this table and divided in several small-energy bins were not provided by the experimental collaborations, but are instead a recast from bounds on the differential distribution of the total number of events $N$,  $dN/dq^2 (B\to K^* e^+ e^-)$, provided by the LHCb collaboration in their search of resonant new particles~\cite{LHCb:2015ycz}; see App.~\ref{sec:BtoKstaree-limits} for details. 

\begin{figure}[t] 
\centering%
\subfigure[\em Quark couplings \vs $m_a$ for $R_{K^\ast}$.
\label{fig:OnShell_BenchMarkeemumu}]%
{\includegraphics[width=0.46\textwidth]{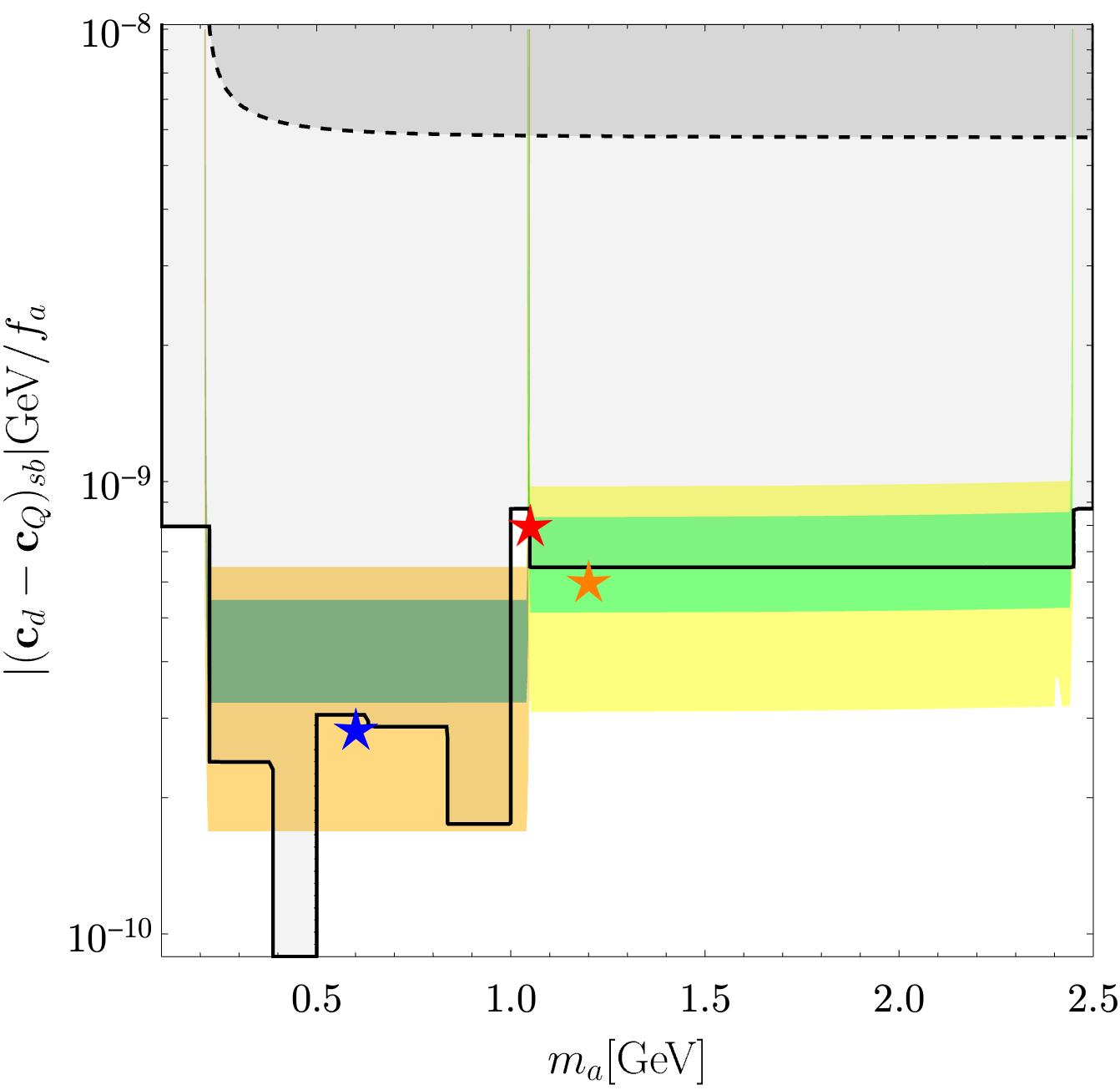}}\qquad
\subfigure[\em Quark couplings \vs lepton coupling ratio for $R_{K^\ast}$.
\label{fig:limitsQuarkCouplingsRKstar}]%
{\includegraphics[width=0.46\textwidth]{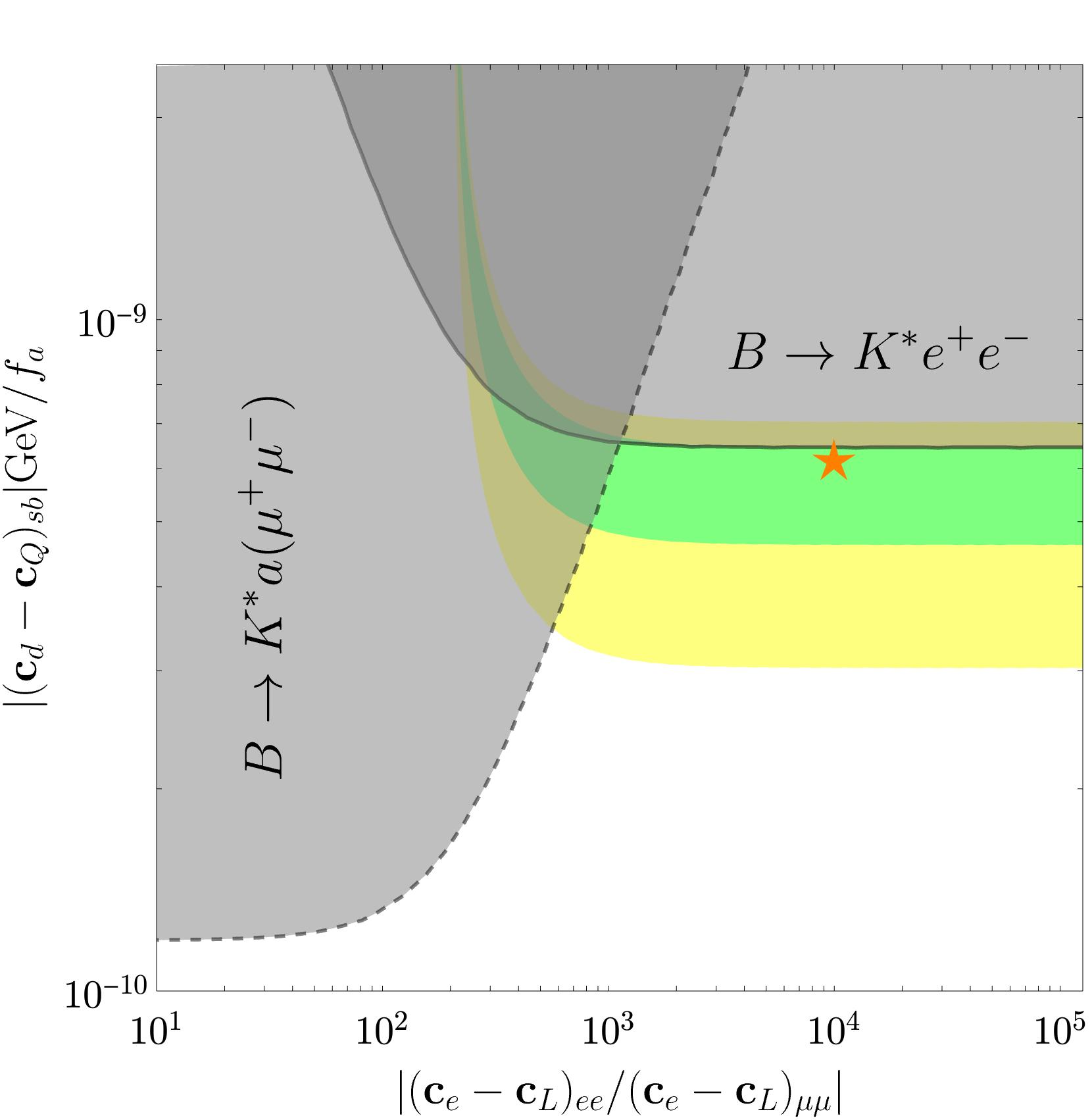}}
\caption{\em Solutions to  $R_{K^\ast}$ with ALP masses within the bin window ranges.  In green (yellow) the $1\sigma$ $(2\sigma)$ solutions to the central bin. The coloured stars correspond to the benchmark ALP-lepton couplings  $(\left|(\bc_e-\bc_L)_{ee}\right|/f_a,\,\left|(\bc_e-\bc_L)_{\mu\mu}\right|/f_a)=(10^{-1},\,10^{-5})\GeV^{-1}$, for different values of $m_a$. On the left: parameter space of ALP-quark couplings \vs $m_a$  excluded by $B\to K^\ast \mu^+\mu^-$ data (enclosed by the dashed line) and by $B\to K^\ast e^+e^-$ data (enclosed by the solid line). For $m_a\in[0.39,\,0.5]\GeV$ data show a tension of more than $2\sigma$ with respect to the SM prediction and the additional contribution of the ALP could only worsen it (see Fig.~\ref{Fig:LHCb_BtoKstaree}). The dark green (dark yellow) shaded areas  indicate ALP solutions to $R_{K^\ast}$ low bin at  $1\sigma$  $(2\sigma)$. On the right:  Constraints from semileptonic $B$-decays on the parameter space of ALP couplings to quarks and leptons, for the reference value $m_a=1.2\GeV$. In grey the  regions excluded by $B\to K^{(\ast)}\mu^+\mu^-$ data (enclosed by the dashed line) and $B\to K^{(\ast)}e^+e^-$ data (enclosed by the solid line).}
\label{fig:limitsQuarkCouplingsRKast}
\end{figure}

Still regarding the electron channel,  the constraints in the third block of Tab.~\ref{tab:dedicatedsearchesBKstarll}
result form the integration over two large windows in energy-bins~\cite{Belle:2019oag}, and they apply to the total branching ratio which includes both the SM and the NP contributions. The limits involving the  combination of ALP-quark couplings $|(\bc_d - \bc_Q)_{sb}|$  derived from the data and shown in the table have been extracted using the complete dependence on them. Finally, once again, the apparently stronger limits on those  couplings in the last column which stem from muon channels turn out to be in fact weaker  in a large fraction of the parameter space, once the true values of $\cB_{a\to \ell^+\ell^-}$ are taken into account.  This is illustrated in Fig.~\ref{fig:OnShell_BenchMarkeemumu} for the specific value of the ALP-lepton couplings $(\left|(\bc_e-\bc_L)_{ee}\right|/f_a,\,\left|(\bc_e-\bc_L)_{\mu\mu}\right|/f_a)=(10^{-1},\,10^{-5})\GeV^{-1}$. Fig.~\ref{fig:limitsQuarkCouplingsRKstar} focuses on the  central energy bin and for the illustrative case $m_a=1.2\GeV$: it shows that the constraint due to $B\to K^* e^+ e^-$ depends only mildly  on the ratio of  ALP-lepton couplings. Once this ratio gets smaller enough, the dominant bound is provided by the $B\to K^* a(\mu^+ \mu^-)$ decay instead. The plots in Fig.~\ref{fig:limitsQuarkCouplingsRKast} are the siblings of those in Figs.~\ref{fig:OnShell_RKvsQuarkCouplings} and \ref{fig:limitsQuarkCouplingsRK}, respectively, and the same colour code has been used.

\begin{figure}[h!]
\centering
\subfigure[\em Low bin $R_{K^\ast}$. 
\label{fig:OnShell_RKstarLEBin}]%
{\includegraphics[width=0.42\textwidth]{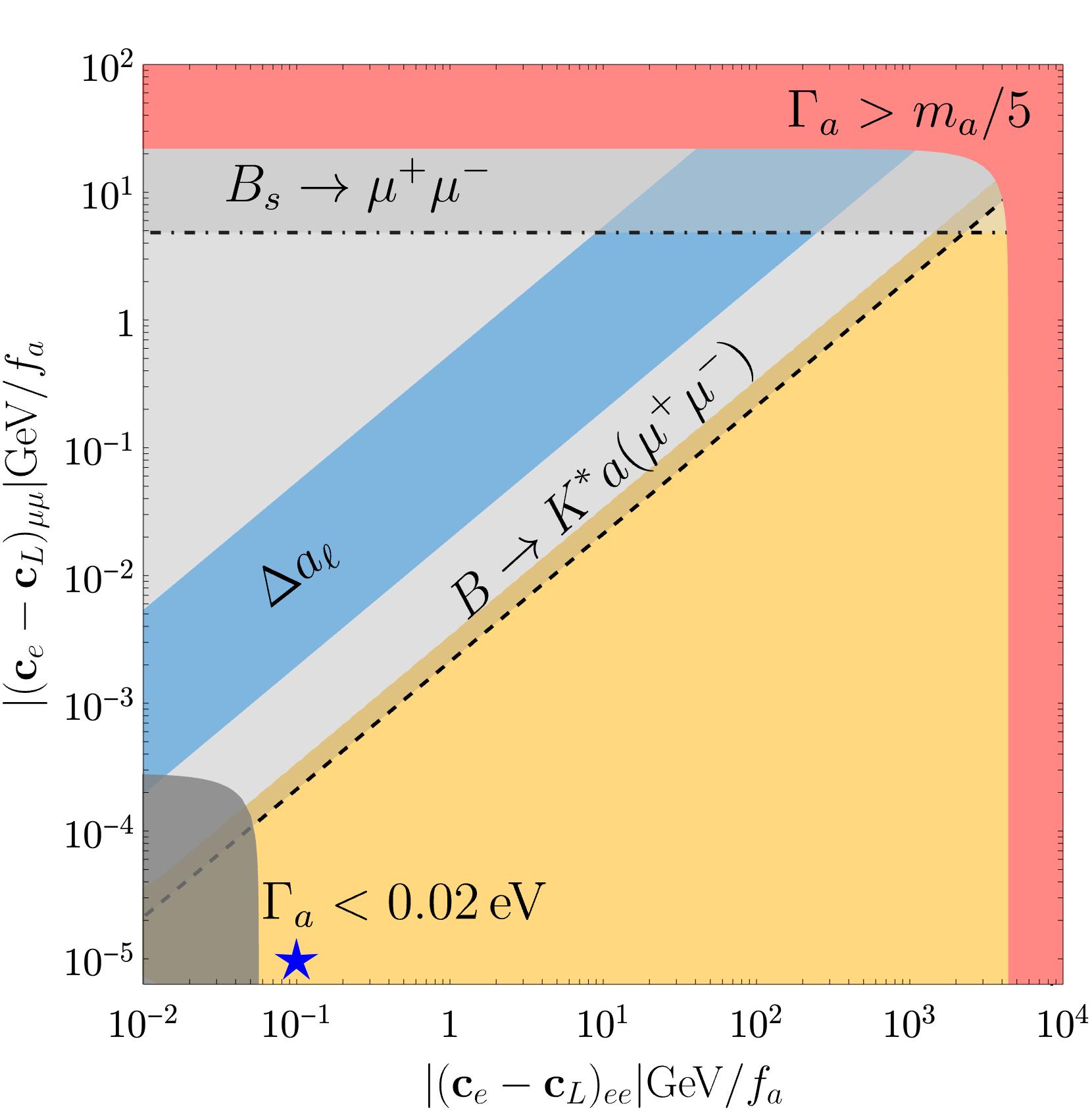}}
\qquad
\subfigure[\em Central bin $R_{K^\ast}$. 
\label{fig:OnShell_RKstarCEBin}]
{\includegraphics[width=0.42\textwidth]{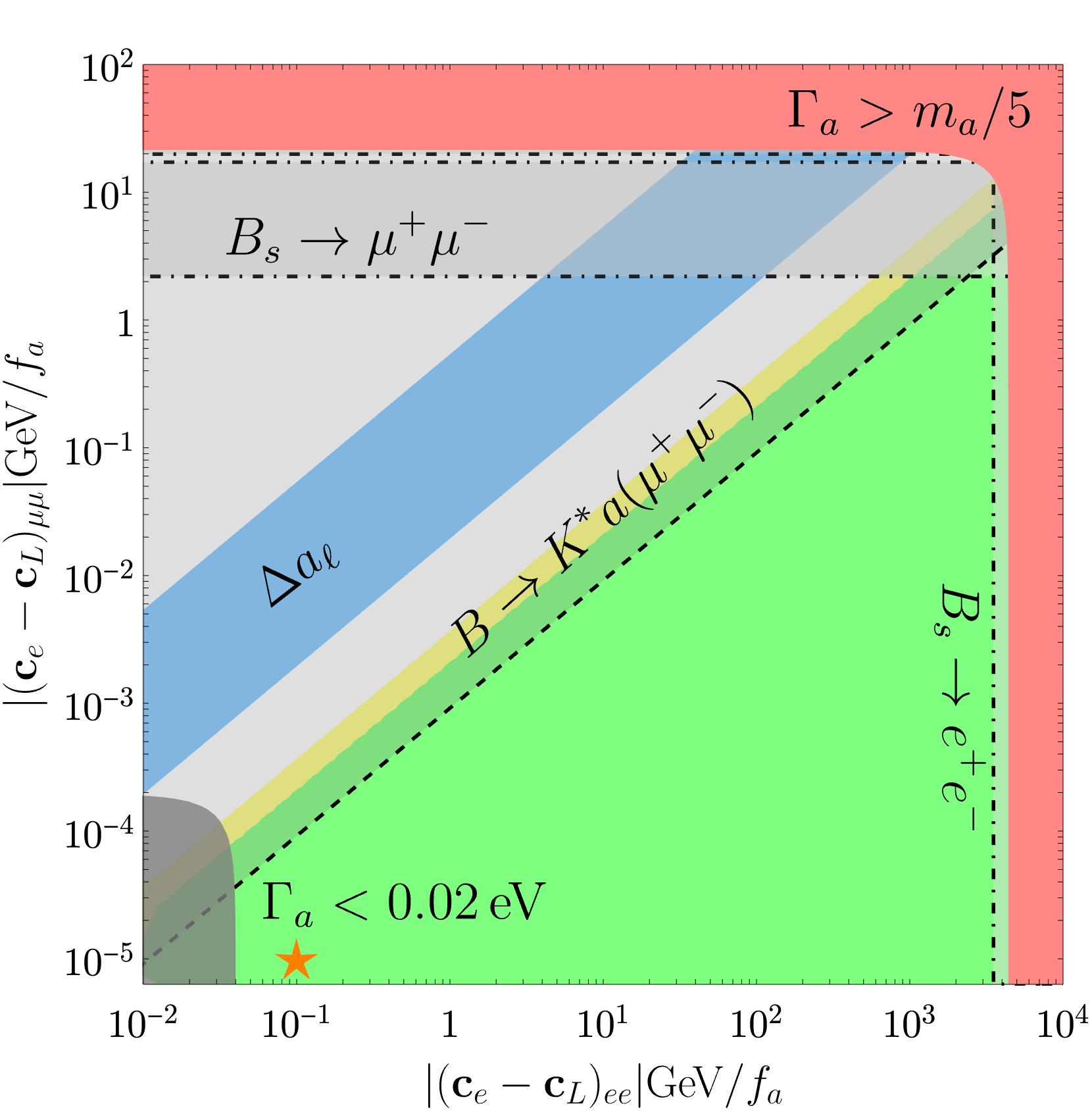}}
\caption{\em  ALP mass within the $R_{K^\ast}$ bin windows. Parameter space $\left|(\bc_e-\bc_L)_{ee}\right|/f_a$ \vs $\left|(\bc_e-\bc_L)_{\mu\mu}\right|/f_a$ that solves the $R_{K^\ast}$ anomaly, in the low energy-bin on the left and in the central energy-bin on the right. In green (yellow) the $1\sigma$ ($2\sigma$) sensitivity. The ALP mass  is chosen to be $m_a=0.6\GeV$ ($m_a=1.2\GeV$) on the left (right) plot  together with, respectively, the values $\left|(\bc_d - \bc_Q)_{sb}\right|/f_a=3.05 \cross 10^{-10}\text{GeV}^{-1}$ and $\left|(\bc_d - \bc_Q)_{sb}\right|/f_a=6.46\cross 10^{-10}\text{GeV}^{-1}$, chosen to comply with the $B\to K^\ast e^+ e^-$ bounds. The shaded red region corresponds to the exclusion condition $\Gamma_a<m_a/5$, while the dark grey one to the prompt decay condition. The light grey region delimited by an oblique dashed line is excluded by the LHCb search for an exotic resonance decaying to muons. The light grey regions delimited by horizontal and vertical dot-dashed lines are excluded by $B_s\to \mu^+\mu^-$ and $B_s\to e^+e^-$ data, respectively. The blue band shows the parameter space compatible with $\Delta a_\mu$ once the photon coupling is fixed to comply with bounds on $\Delta a_e$.} 
\label{Fig:OnShell_RKstar}
\end{figure}

The bounds obtained from $B \to K^{\ast} e^+ e^-$  --see  Tab.~\ref{tab:dedicatedsearchesBKstarll}-- can be used as conservative benchmarks for the ALP-quark couplings in our numerical analysis. 
Specifically, Fig.~\ref{fig:OnShell_RKstarLEBin} and Fig.~\ref{fig:OnShell_RKstarCEBin} illustrate the parameter space of ALP-couplings to leptons which can explain $R_{K^\ast}$ via resonant ALP exchange, for the benchmark values  
\be
\dfrac{\left|(\bc_d - \bc_Q)_{sb}\right|}{f_a}=3.05 \cross 10^{-10}\text{GeV}^{-1}\quad ,\quad \dfrac{\left|(\bc_d - \bc_Q)_{sb}\right|}{f_a}=6.46\cross 10^{-10}\text{GeV}^{-1}\,,
\ee
respectively for the low bin ($m_a=0.6\GeV$, blue star)   and the central bin ($m_a=1.2\GeV$, orange star).   In both figures, the stars correspond to the  (previously used) values of leptonic couplings $(\left|(\bc_e-\bc_L)_{ee}\right|/f_a,\,\left|(\bc_e-\bc_L)_{\mu\mu}\right|/f_a)=(10^{-1},\,10^{-5})\GeV^{-1}$.

These plots in Fig.~\ref{Fig:OnShell_RKstar} for  $R_{K^\ast}$  are  very similar to that for $R_K$ in Fig.~\ref{fig:OnShell_RKvsLeptonCouplings}, and use the same colour code.  Once again, the limits on ALP leptonic couplings from $B\to K^\ast a (\ell^+ \ell^-)$ severely limit the allowed parameter space, in addition to those resulting from the validity conditions for the NWA and for prompt ALP decays in Eqs.~(\ref{OnShellALPeeCondition})-(\ref{OnShellALPmumuConditionPrompt}). On the other hand, the bounds from $B_s\to \ell^+\ell^-$ are at best of the same order of magnitude than the ones just mentioned. All in all, the lower-left area of the parameter space is the region where possible explanations to the  $R_{K^\ast}$ anomaly can be found within the validity range of the ALP EFT. \\

Finally, the compatibility of the data on leptonic anomalous magnetic moments and the solutions to the $R_{K^\ast}$ anomaly through on-shell ALP exchange parallels the analysis for $R_K$  in the previous subsection: when all data available are taken into account, it is possible to account simultaneously for the data in the set $\{R_{K^\ast}, \Delta a_e\}$ within the theoretical region of validity of the ALP EFT. In contrast, the ensemble of the three observables $\{R_{K^\ast}, \Delta a_e, \Delta a_{\mu}\}$ cannot be simultaneously explained through such an ALP, see Fig.~\ref{Fig:OnShell_RKstar}, and thus the $\Delta a_{\mu}$ anomaly would remain unexplained.

\subsubsection{The golden mass}

\begin{figure}[h!]
\centering
\subfigure[\em Low bin $R_{K^\ast}$. Golden $m_a$.
\label{fig:OnShell_RKstarLEBinGolden}]
{\includegraphics[width=0.42\textwidth]{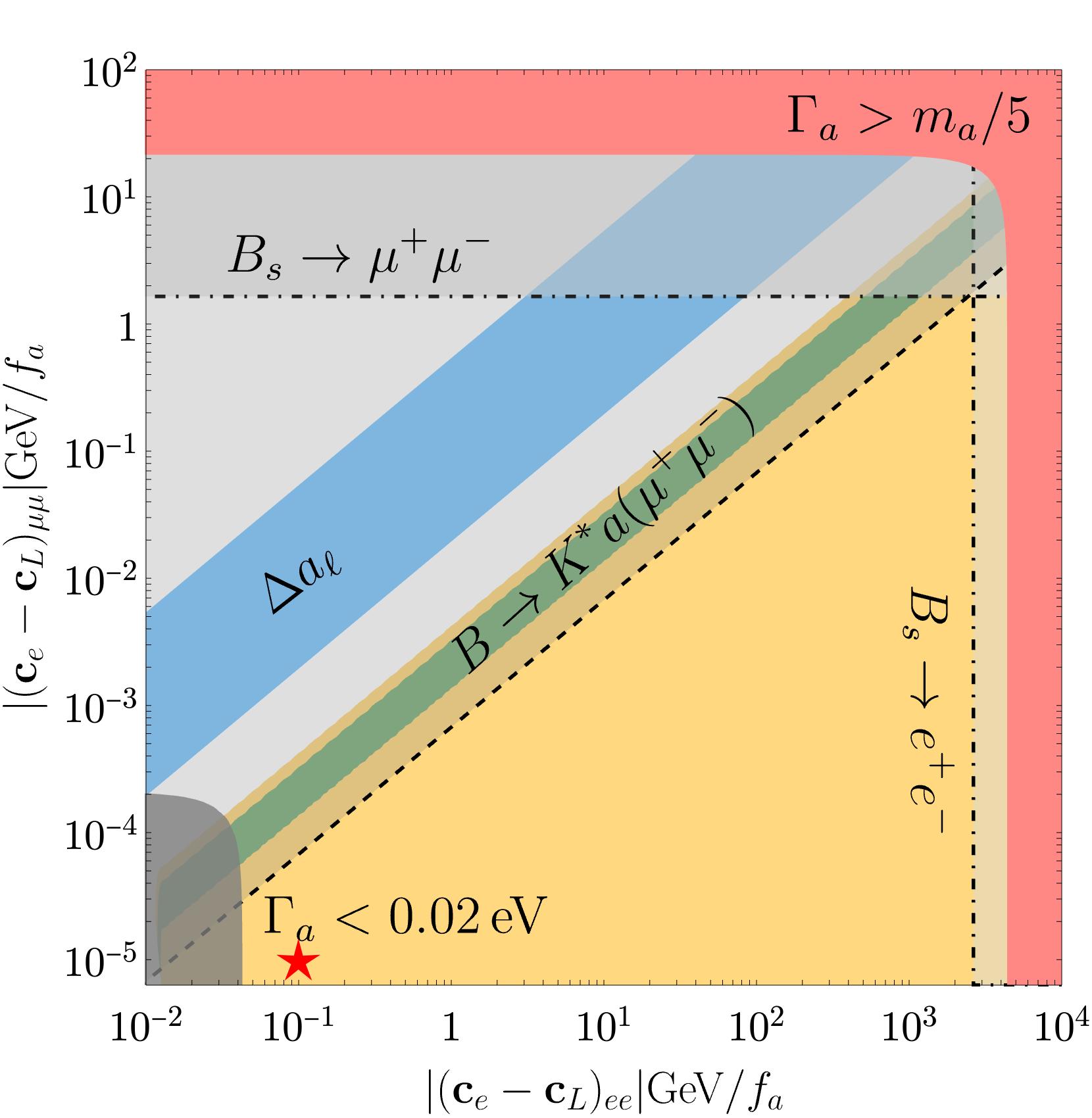}}
\qquad
\subfigure[\em Central bin $R_{K^\ast}$. Golden $m_a$.
\label{fig:OnShell_RKstarCEBinGolden}]
{\includegraphics[width=0.42\textwidth]{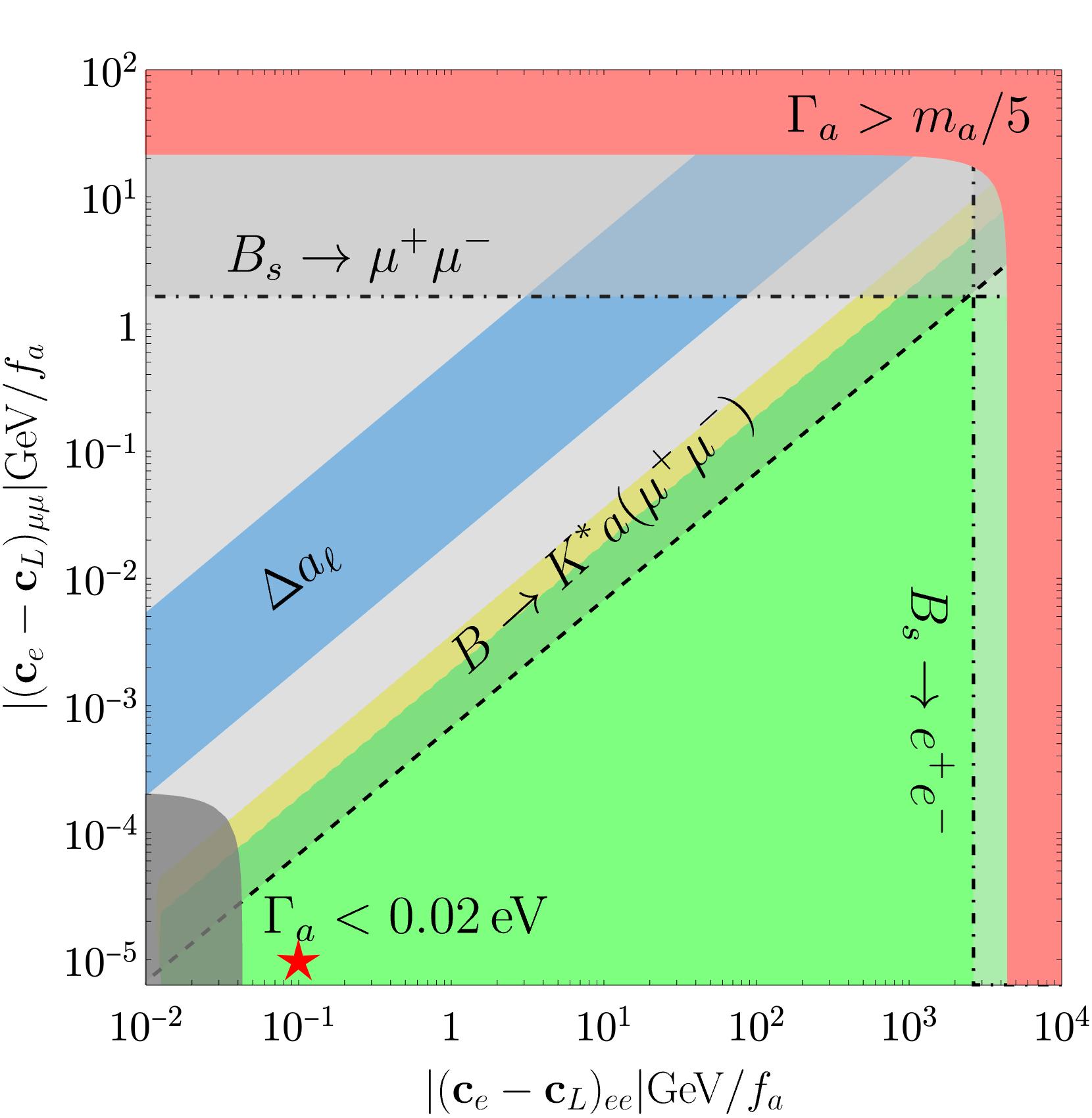}}
\caption{\em Parameter space $\left|(\bc_e-\bc_L)_{ee}\right|/f_a$ \vs $\left|(\bc_e-\bc_L)_{\mu\mu}\right|/f_a$ interesting to explain the $R_{K^\ast}$ anomaly, in the low energy-bin on the left and in the central energy-bin on the right, for the golden mass $m_a=\sqrt{1.1}\GeV$. In green (yellow) the $1\sigma$ ($2\sigma$) sensitivity. The ALP-quark coupling is fixed to $\left|(\bc_d - \bc_Q)_{sb}\right|/f_a=8.71\cross 10^{-10} \text{GeV}^{-1}$. The colour code and lines follow the same description as in Fig.~\ref{Fig:OnShell_RKstar}.} 
\label{Fig:OnShell_RKstarGolden}
\end{figure}

The analysis above  explored whether  $R_{K}$ and the central-energy bin of $R_{K^\ast}$ could be explained via ALP exchange, while the low-energy bin of the $R_{K^\ast}$ anomaly was analysed separately. The respective benchmarks points were $m_a=1.2\GeV$ (orange star in Figs.~\ref{fig:limitsQuarkCouplingsRK}, \ref{Fig:OnShell_RK}, \ref{fig:limitsQuarkCouplingsRKast} and \ref{fig:OnShell_RKstarCEBin}) and $m_a=0.6\GeV$ (blue star in Figs.~\ref{fig:OnShell_BenchMarkeemumu} and \ref{fig:OnShell_RKstarLEBin}). It is a pertinent question, though, whether there exists some value of the ALP mass which could explain the data on all three neutral $B$ anomalies, i.e.   $R_{K}$ and the two energy bins for $R_{K^\ast}$. We have identified a ``golden mass'' solution which could satisfy these three requirements within $2\sigma$ sensitivity (in yellow), located right at the edge of the two energy-bin windows, and which corresponds to 
\be
m_a=\sqrt{1.1}\GeV\,.
\label{goldenmass}
\ee
This point is indicated by a red star in Figs.~\ref{fig:OnShell_RKvsQuarkCouplings}, \ref{fig:OnShell_BenchMarkeemumu} and \ref{Fig:OnShell_RKstarGolden}.

In particular, Fig.~\ref{Fig:OnShell_RKstarGolden} depicts the same plots as those in Fig.~\ref{Fig:OnShell_RKstar} except that $m_a$ is taken to be the golden mass value, indicated by the red star.\footnote{For this $m_a$ value, the benchmark quark couplings that saturate the constraints  from $B\to K^\ast e^+ e^-$ data are slightly different than those used previously: $\left|(\bc_d - \bc_Q)_{sb}\right|/f_a=8.71\cross 10^{-10} \text{GeV}^{-1}$. 
} This figure shows that the exchange of an ALP with the mass given by Eq.~(\ref{goldenmass}) could {\it a priori} account for the anomalies in  both energy bins of  $R_{K^\ast}$  within $2\sigma$ (in yellow).  This result is strongly dependent on the value of the ALP-quark couplings, which ultimately regulates the impact of the on-shell contribution. Indeed, for smaller ALP-quark couplings, the resonant contributions disappear and no-overlap region is left between the low and central energy-bin anomalies. 

The dependence on the ALP mass of the  solutions to $R_{K^\ast}$ is further scrutinised going back to Fig.~\ref{fig:OnShell_BenchMarkeemumu}.  It shows that:
\begin{itemize} 
\item Within $1\sigma$ sensitivity (in green), all ALP solutions  with masses within the low-energy bin of the $R_{K^*}$ anomaly are excluded by other data.  This conclusion agrees with that in Ref.~\cite{Altmannshofer:2017bsz}, where the parameter space of a generic resonance compatible only with this low bin anomaly was studied. 
\item The comparison with Fig.~\ref{fig:OnShell_RKvsQuarkCouplings} shows  that any ALP mass within the central bin range of  $R_{K^{*}}$ can accommodate a combined explanation of the two anomalies in the set $\{R_K, R_{K^{*}}\}$ within less than $2\sigma$, for $\mathbf{c}_d \approx 3\times 10^{-10}$ (which corresponds to ${\mathcal{B}(B\to K a) \sim 10^{-8}}$). This possible explanation for the two neutral $B$ anomalies via a resonance on the bin is a novel aspect of our work.
\item Finally, the  location of the red star in Fig.~\ref{fig:OnShell_RKvsQuarkCouplings} and Fig.~\ref{fig:OnShell_BenchMarkeemumu} 
illustrates that the on-shell exchange of a golden mass ALP could simultaneously account for the $R_K$ anomaly  {\it and}  for  the two anomalies in the two different $R_{K^\ast}$ energy bins. The details of the mass dependence can be appreciated in the zoom-in view around the golden mass value depicted in Fig.~\ref{fig:npspecialnqqq}.
\end{itemize}

Nevertheless, in spite of this last encouraging result, explanations of physics anomalies located at the frontier of energy bins are suspicious. The take-away message is that a  different binning of the data is well-motivated and can quickly clarify the issue.

\subsection{ALP mass close to the bin window:  the smearing function}

The aim of this section is twofold: the first is to consider the case in which the ALP mass is outside, but close to, a given energy-bin window; the second is to include in the previous analysis the finite experimental sensitivity. Indeed, if the value of the ALP mass lies outside the energy-bin window, the ALP is technically off-shell and  its contribution to observables gets thus suppressed. On the other side, the experimental resolution in terms of bin distribution is not infinite and therefore it is possible that certain events with a  $q^2$ close to the borders of a chosen window are simply not correctly taken into account.

To take into consideration these two sources of systematic errors, a Gaussian smearing function is traditionally adopted to modify the  NWA expression. For the case of the semileptonic $B$-meson decays in Eq.~(\ref{NWABKll}), it reads~\cite{Altmannshofer:2017bsz}
\begin{equation}
\BR(B\rightarrow K^{(\ast)} a(\ell^+\ell^-))= \BR(B\rightarrow K^{(
\ast)} a)\times\BR(a\rightarrow \ell^+ \ell^-)\times  \cG^{(r_\ell)}(q_\text{min.},q_\text{max.})\,,
    \label{eq:BtoKstaree}
\end{equation}
where $\cG^{(r_\ell)}$ is a Gaussian smearing function defined as
\begin{equation}
\cG^{(r_\ell)}(q_\text{min.},q_\text{max.}) \equiv \dfrac{1}{\sqrt{2\pi}r_\ell}\int\limits_{q_\text{min.}}^{q_\text{max.}} \diff{|q|}\, e^{-\frac{(|q|-m_a)^2}{2r_\ell^2}}\,,
\end{equation}
where $r_e = 10\MeV$~\cite{Ilten:2015hya} and $r_\mu = 2\MeV$~\cite{LHCb:2014set} refer to the di-lepton mass resolution of the LHCb detector, and  the boundaries of the integration range correspond to the extremes of the considered energy-bin window. The net effect of this function is to broaden the distributions found in the previous section near the borders of the energy-bin windows. We will explicitly show the impact of this smearing on the analysis of $R_{K^\ast}$ in two mass ranges corresponding respectively to:  the golden mass solution in between the two energy-bin windows, and the lowest energies within the low-energy bin region.  
 
\subsubsection*{The golden mass solution}

\begin{figure}[h!] 
\centering
\subfigure[{Without smearing function.}
\label{fig:npsp1nqnn}]
{\includegraphics[width=0.47\textwidth]{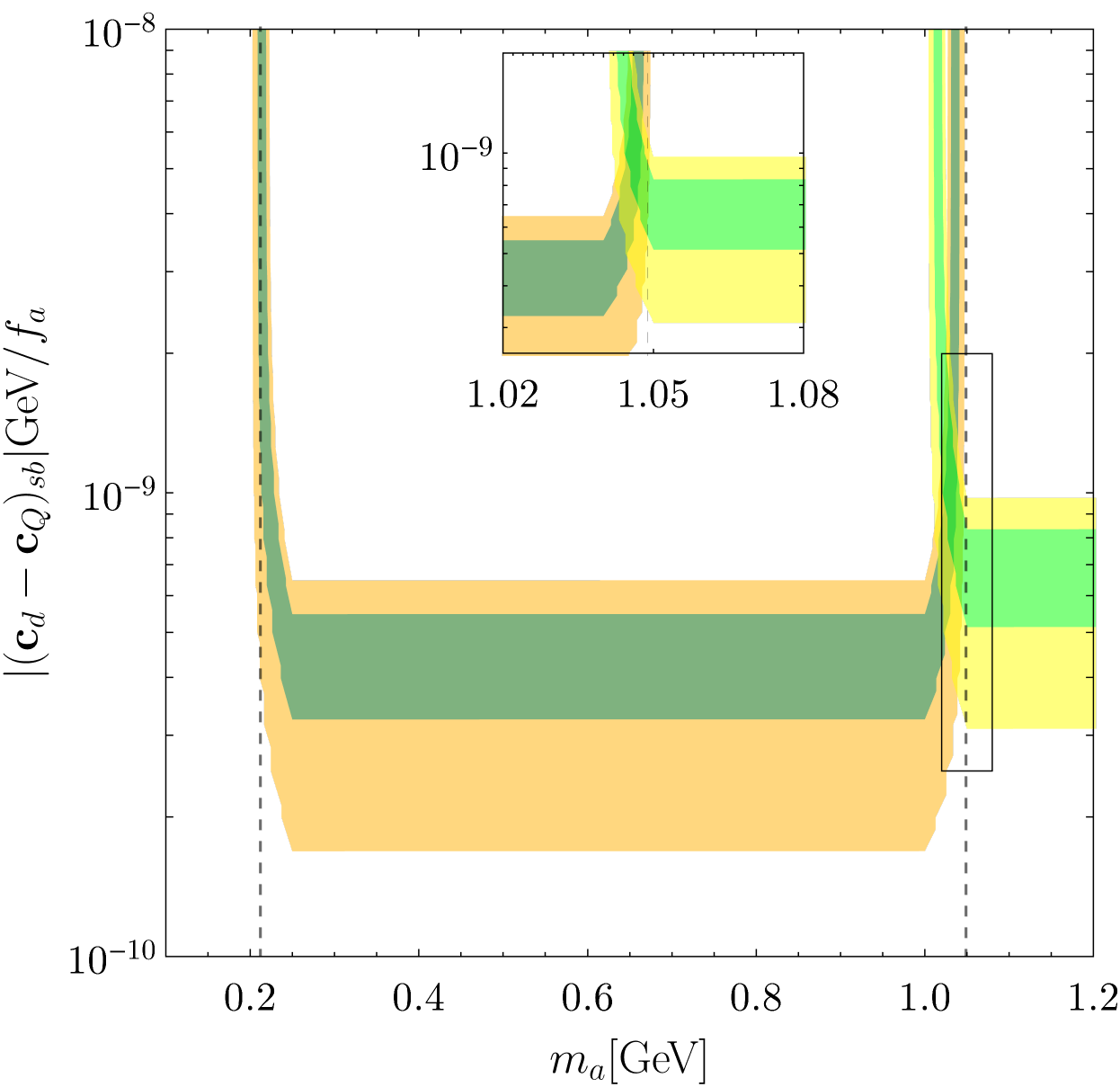}}
\subfigure[{With smearing function.}
\label{fig:npsp2nqnn}]
{\includegraphics[width=0.47\textwidth]{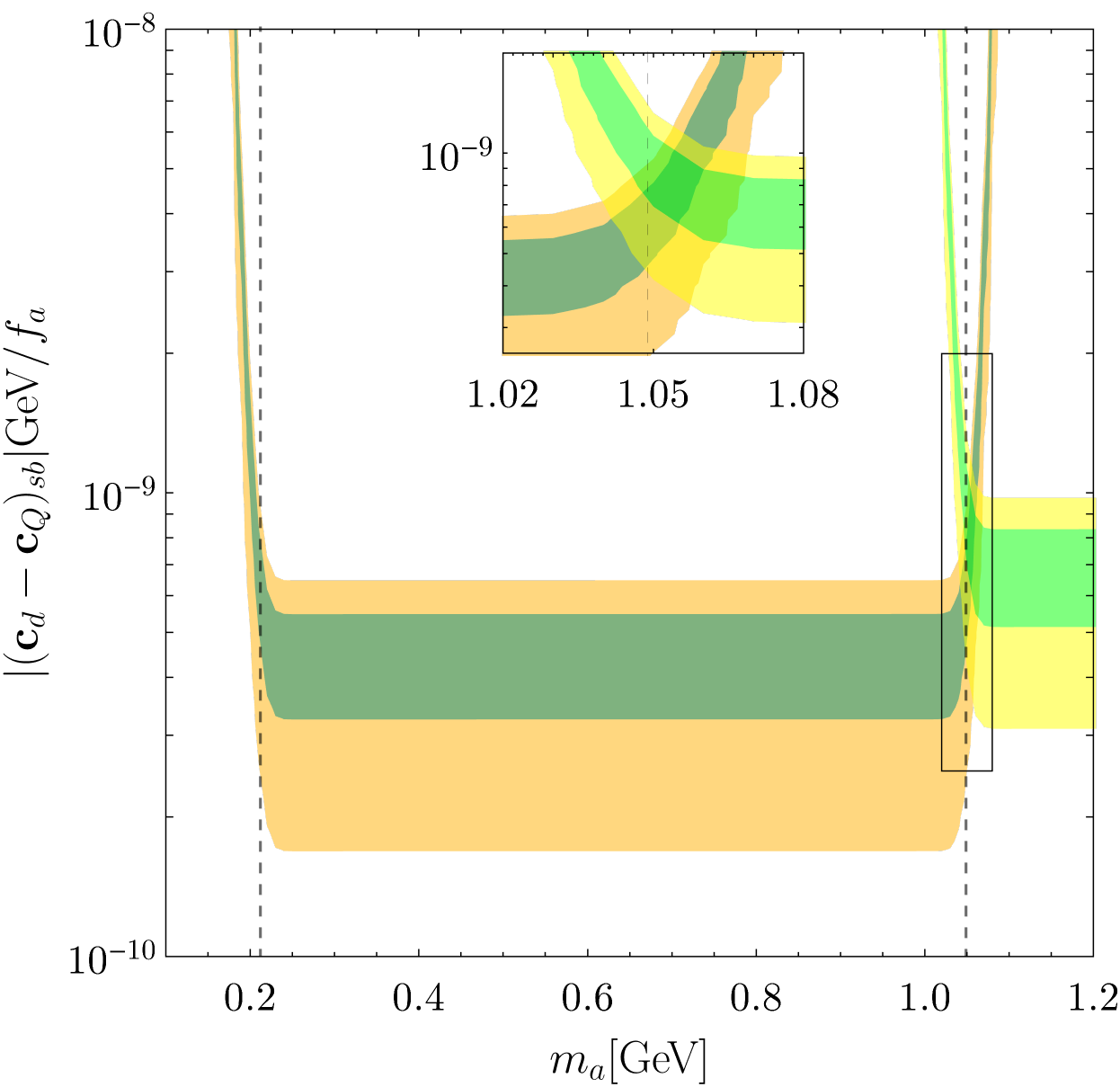}}
\caption{\em  Golden ALP mass. Impact of the smearing function in (a selected region of) the parameter space $m_a$ \vs $|(\bc_d-\bc_Q)_{sb}|/f_a$ for the $R_{K^\ast}$ anomaly. On the left (right) the case without (with) the effect of the smearing function. The benchmark point for the ALP-lepton couplings is $(\left|(\bc_e-\bc_L)_{ee}\right|/f_a,\,\left|(\bc_e-\bc_L)_{\mu\mu}\right|/f_a)=(10^{-1},\,10^{-5})\GeV^{-1}$. In green (yellow) the $1\sigma(2\sigma)$ allowed region for the low and central energy-bin window, with the darker colours corresponding to the low bin.
\label{fig:npspecialnqqq}
}
\end{figure}

In  Fig.~\ref{fig:npspecialnqqq} we zoom in the relevant part of the parameter space for $R_{K^*}$ showed in Fig.~\ref{fig:OnShell_BenchMarkeemumu}, for the same benchmark point of the ALP-lepton couplings. The impact of the smearing function around the golden mass region can be appreciated in Fig.~\ref{fig:npsp2nqnn}, as compared to Fig.~\ref{fig:npsp1nqnn} which does not include smearing effects. The overlap at the $2\sigma$ level of the ALP solutions common to the low and central energy-bin anomalies broadens now to an interval  around the precise value $m_a=\sqrt{1.1}\GeV$, given by
\be
m_a\in[1.04\,,1.07]\GeV\,.
\ee

\subsubsection*{The kinematic solution to the low-bin anomaly}
Let us focus now instead on the lower boundary of the low-energy bin of $R_{K^*}$.  This is of particular interest because this boundary is higher than the di-muon threshold: an ALP with mass $m_a<2m_\mu$ cannot decay into muons but only into electrons, which {\it a priori} opens the possibility of a kinematic explanation of the low energy-bin $R_{K^\ast}$ anomaly~\cite{Altmannshofer:2017bsz}.  While this is not possible without the effect of the smearing function, as shown in the previous section, now it appears to be a viable possibility, see Fig.~\ref{fig:npsp2nqnn}.

\begin{figure}[h!] 
\centering
\subfigure[{Without smearing function.}
\label{fig:compres1Bis}]
{\includegraphics[width=0.4\textwidth]{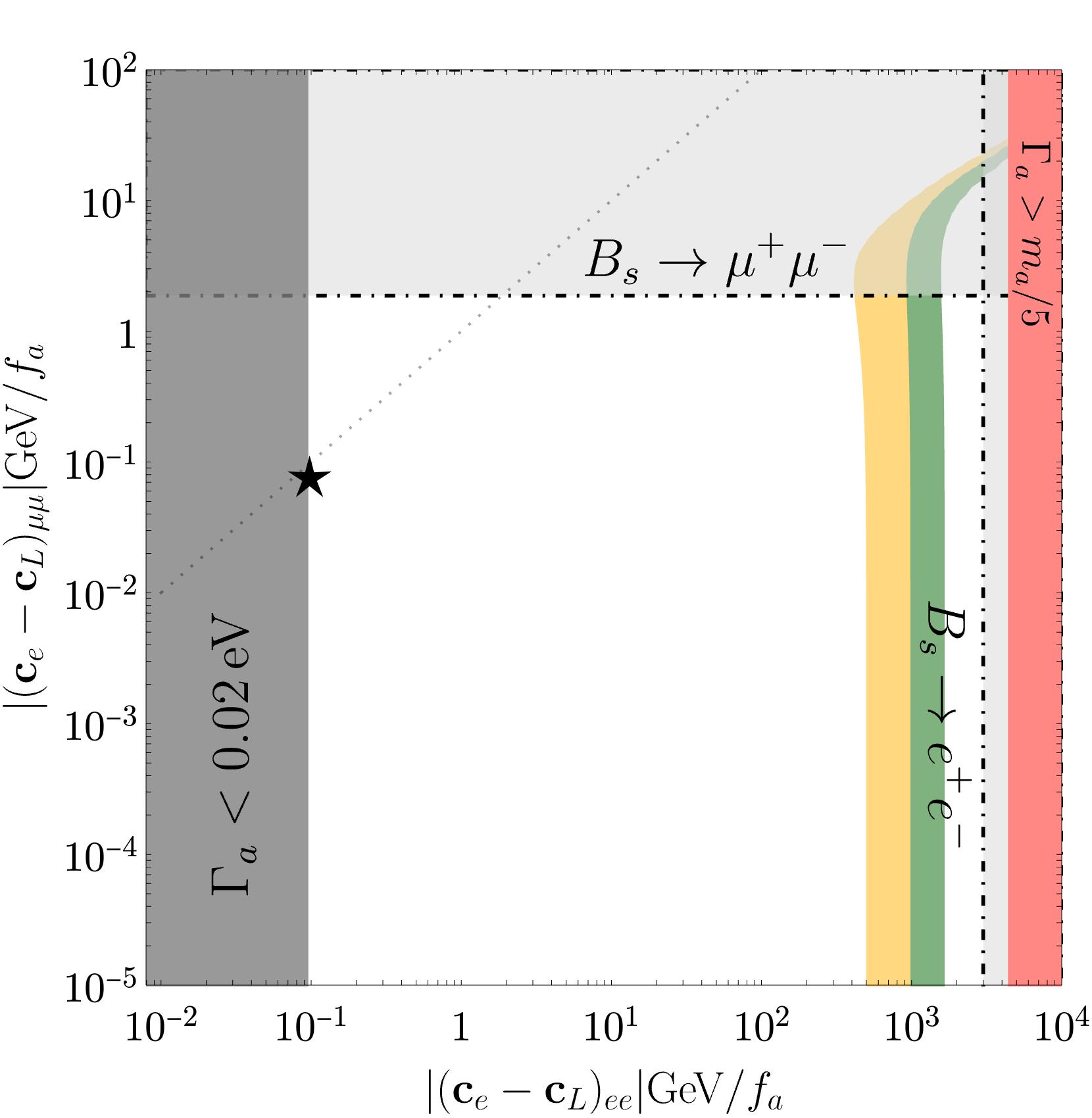}}\qquad
\subfigure[{With smearing function.}
\label{fig:compres2Bis}]
{\includegraphics[width=0.4\textwidth]{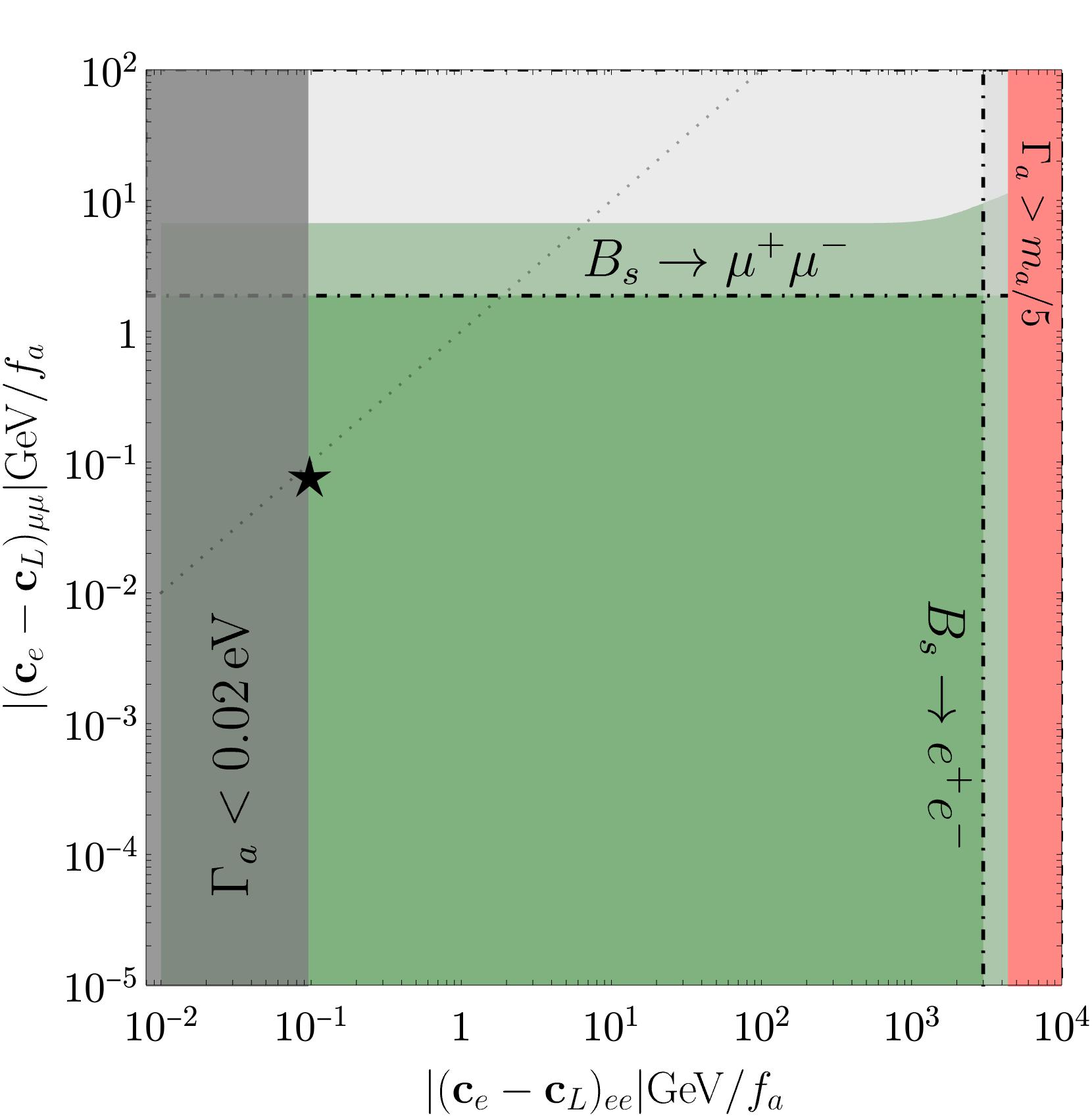}}
\caption{\em ALP mass just under the low bin window. 
Impact of the smearing function in the parameter space $|(\bc_e-\bc_L)_{ee}|/f_a$ \vs $|(\bc_e-\bc_L)_{\mu\mu}|/f_a$ of solutions to the low bin $R_{K^\ast}$ anomaly;   $1\sigma(2\sigma)$ allowed regions depicted in green (yellow), for $m_a=210\MeV$ and $|(\bc_d- \bc_Q)_{sb}|/f_a=7.96 \cross 10^{-10}\GeV^{-1}$. The figure on the left (right), does not (does) include the smearing function.  The dotted line represents the flavour universal setup for the lepton couplings and the black star the smallest allowed value. The dark grey regions are excluded by the prompt decay condition, while the light grey ones are excluded by $B_s\to \ell^+\ell^-$ data.}
\label{fig:comparison_resonance}
\end{figure}

The kinematic solution is further illustrated in Fig.~\ref{fig:comparison_resonance} for the particular case of an ALP mass slightly below the di-muon threshold, $m_a=210\MeV$. The allowed parameter space for lepton couplings is depicted (the ALP-quark coupling has been fixed to a reference value  
 that allows us to avoid conflict with  the semileptonic $B$-decay constraints). The comparison of the left and right plots of this figure shows that the effect of the smearing is to substantially enlarge the relevant $2\sigma$ region that explains the anomaly. Furthermore, an oblique dotted line in these plots indicates the location of the flavour universal solutions: the particular case analysed in Ref.~\cite{Bauer:2021mvw} is represented by a black star and shown to be excluded without smearing effects and viable once smearing effects are included. The expectation is that the future experimental improvements in these observables will increase the sensitivity and thus the effect of the smearing will get progressively reduced; in the absence of a discovery, the realistic analysis should ultimately converge towards  Fig.~\ref{fig:compres1Bis} as  final result.

\subsection{Impact of sizeable couplings}

We have previously discussed that a large ALP-electron coupling induces a non-negligible ALP-photon coupling in the context of $(g-2)$ data. A similar effect occurs for flavour-violating couplings of the ALP to quarks, which are generated by the ALP-electron coupling  at the two-loop level.\footnote{We thank the referee for recalling the relevance of the two-loop effects in these couplings.} Indeed, given the large $|(\bc_e-\bc_L)_{ee}|$ values required to explain the neutral $B$-anomalies and the very high precision on some data, these effects might become, in some cases, relevant for the present study. It is beyond the scope of the present work to include such effects in the analysis. However, following the discussion in Ref.~\cite{Bauer:2021mvw}, we remark that the loop induced ALP-$bs$ coupling could become larger than the value in Eq.~\eqref{ALPquarkBoundFromBKee}. 
The experimental bound on $\mathcal{B}(B\to K e^+ e^-)$ in fact applies to the effective coupling
\begin{equation}
\frac{(\bc_d + \bc_Q)_{sb}}{f_a} \sim \frac{(\bc_d + \bc_Q)_{sb}^{\rm tree}}{f_a} + \frac{\alpha_{em}^2}{s_w^4(4\pi)^2} \frac{(\bc_L)_{ee}}{f_a} \log{\frac{\Lambda}{m_B}}\,,
\end{equation}
as it can be estimated from the renormalization group equations of the ALP EFT. Hence, as we discussed in the context of magnetic moments, a cancellation between tree and loop level contributions can become necessary in order to satisfy this experimental constraint. Note that this type of fine-tuning can be avoided in certain UV models, such as those producing only right-handed lepton couplings to the ALP, where this loop contribution can be naturally suppressed. 

\section{Very light ALP}
\label{sec:VeryLightALP}

We address next whether the neutral $B$ anomalies could be mediated by ALPs even lighter than those discussed in the previous section, that is lighter than twice the muon mass.  In consequence, the ALP cannot decay into two muons but it can decay into two electrons.

Astrophysical constraints on non-negligible ALP-electron couplings~\cite{Bauer:2021mvw}  exclude ALPs lighter than $1\MeV$, though, and in consequence, the range of masses to be explored in this section is
\be
1\MeV < m_a \lesssim2\,m_\mu\,.
\ee
For this range of masses, additional bounds from Beam Dump experiments and from supernova data analysis constrain the possible values for the ALP-electron coupling to be outside of a small interval~\cite{Bauer:2021mvw}, approximately $|(c_e-c_L)_{ee}|/f_a\not\in[10^{-4},\,10^{-1}]\GeV^{-1}$ and $[10^{-6},\,10^{-4}]\GeV^{-1}$, respectively.

Furthermore, for $m_a <2m_\mu$, astrophysical upper limits on the ratio between the effective ALP-photon couplings and the scale $f_a$ would be very strong, of the order of $10^{-11} \GeV^{-1}$~\footnote{Even stronger bounds could follow from cosmological constraints, which however depend on the specific assumption of the cosmological model considered.}, but they can be evaded using the freedom on the value of the tree-level $c_{a\gamma\gamma}$ coefficient in Eq.~(\ref{ferm-Lag}).

A peculiar feature of this mass regime is the similarity of the final expressions for $R_K$ and $R_{K^\ast}$ with those in the heavy ALP scenario discussed in Sect.~\ref{sec:HEAVYALP}. Indeed, for $m_a^2\ll q^2$, the $m_a$ dependence in the ALP propagator can be neglected, which leaves only its $q^2$ dependence. Once the integration over $q^2$ is performed with Flavio and EOS, the final expressions for the SM plus ALP contributions to $B\to K^{(\ast)} \ell^+ \ell^-$ read,  for the central bin of $B\rightarrow K$ semileptonic decays,
\be
\begin{aligned}
\BR(B\to K \mu^+ \mu^-)_{1.1\GeV^2}^{6.0\GeV^2} & =10^{-7}\times\left(1.5-2.4\times 10^{-2}\,\widetilde{C}^\mu_{P_+} +6.6\times 10^{-3}\,\widetilde{C}^{\mu2}_{P_+}\right)\,,\\
\BR(B\to K e^+ e^-)_{1.1\GeV^2}^{6.0\GeV^2} & =10^{-7}\times\left(1.5-1.2\times 10^{-4}\,\widetilde{C}^e_{P_+} +6.7\times 10^{-3}\,\widetilde{C}^{e2}_{P_+}\right)\,,
\end{aligned}
\label{BKllVeryLightALP}
\ee
while for the central bin of $B\rightarrow K^\ast$ semileptonic transitions, it results
\be
\begin{aligned}
\BR(B\to K^\ast \mu^+ \mu^-)_{1.1\GeV^2}^{6.0\GeV^2} & =10^{-7}\times\left(1.9-2.3\times 10^{-2}\,\widetilde{C}^\mu_{P_-} +6.2\times 10^{-2}\,\widetilde{C}^{\mu2}_{P_-}\right)\,,\\
\BR(B\to K^\ast e^+ e^-)_{1.1\GeV^2}^{6.0\GeV^2} & =10^{-7}\times\left(1.9-1.1\times 10^{-4}\, \widetilde{C}^e_{P_-} +6.3\times 10^{-3}\,\widetilde{C}^{e2}_{P_-}\right)\,,
\end{aligned}
\label{BKstarllVeryLightALPHigh}
\ee
and for its low-energy bin they read
\be
\begin{aligned}
\BR(B\to K^\ast \mu^+ \mu^-)_{0.045\GeV^2}^{1.1\GeV^2} & =10^{-7}\times\left(1.2-2.4\times 10^{-2}\,\widetilde{C}^\mu_{P_-} +6.3\times 10^{-3}\,\widetilde{C}^{\mu2}_{P_-}\right)\,,\\
\BR(B\to K^\ast e^+ e^-)_{0.045\GeV^2}^{1.1\GeV^2} & =10^{-7}\times\left(1.3-1.4\times 10^{-4}\, \widetilde{C}^e_{P_-} +7.7\times 10^{-3}\,\widetilde{C}^{e2}_{P_-}\right)\,.
\end{aligned}
\label{BKstareeVeryLightALPLow}
\ee
For all these quantities, the theoretical error is estimated to be $\mathcal{O}(15)\%$.  The comparison of these expressions with their corresponding siblings for the heavy ALP case in Eqs.~(\ref{BKllHeavyALP}), (\ref{BKstarllHeavyALPHigh}) and  (\ref{BKstarllHeavyALPLow}) shows that the numerical coefficients in front of the NP Wilson coefficients have a similar order of magnitude, and similar considerations can be applied to the analysis of both cases. 
 Using the available data on non-resonant searches presented in Tab.~\ref{tab:Proposal}, the $2\sigma$ bounds on the corresponding Wilson coefficients read:
\begin{equation}
\begin{aligned}
\widetilde{C}_{P_+}^e &\in [-8.3,8.3]\,, \qquad \qquad &&\widetilde{C}_{P_+}^\mu \in [-4.2,7.8]\,,\\
\widetilde{C}_{P_-}^e &\in [-14.0,14.0]\,, &&\widetilde{C}_{P_-}^\mu \in [-4.8,5.1]\,.
\end{aligned}
\end{equation}
Before proceeding to compare with other observables, it is useful to rewrite these $\widetilde{C}_{P_\pm}^\ell$  coefficients in terms of ALP-fermion couplings. They can be expressed in terms of  the $C_{P_\pm}^\ell$ coefficients defined in Eq.~\eqref{DefinitionCellPpm} as
\be
\widetilde{C}_{P_\pm}^\ell\equiv- 
m_a^2 \dfrac{C_{P_\pm}^\ell}{\GeV^2}=-
\dfrac{2\sqrt{2}\pi}{\alpha_\text{em} G_F  V_{tb}V_{ts}^* }\dfrac{m_\ell}{(f_a \GeV)^2}(m_s\mp m_b) \left({\bf K}^{S,P}_d\right)_{sb}\left({\bf K}^P_e\right)_{\ell\ell}\,.
\label{DefinitionCtildeellPpm}
\ee
which can be simplified to  
\be
\begin{aligned}
\widetilde{C}^e_{P_\pm} \approx&\, \pm1.3\times 10^6\GeV^2 \,\dfrac{\left(\bc_d\pm\bc_Q\right)_{sb}}{f_a}\dfrac{\left(\bc_e-\bc_L\right)_{ee}}{f_a}\,,\\
\widetilde{C}^\mu_{P_\pm} \approx&\, \pm2.7\times 10^8\GeV^2 \,\dfrac{\left(\bc_d\pm\bc_Q\right)_{sb}}{f_a}\dfrac{\left(\bc_e-\bc_L\right)_{\mu\mu}}{f_a} \,.
\end{aligned}
\label{Ctildes-simplified}
\ee
Note that these relations are independent of the specific value of the ALP mass, in contrast with the case of a heavy ALP,  
see Eq.~(\ref{Cs-simplified}).

\boldmath
\subsection{$R_{K}$, $\Delta M_s$ and magnetic moments}
\unboldmath

\boldmath
\subsubsection*{$R_K$}
\begin{figure}[h!] 
\centering%
\subfigure[\em $R_K$\label{fig:RKHALPUL}]%
{\includegraphics[width=0.48\textwidth]{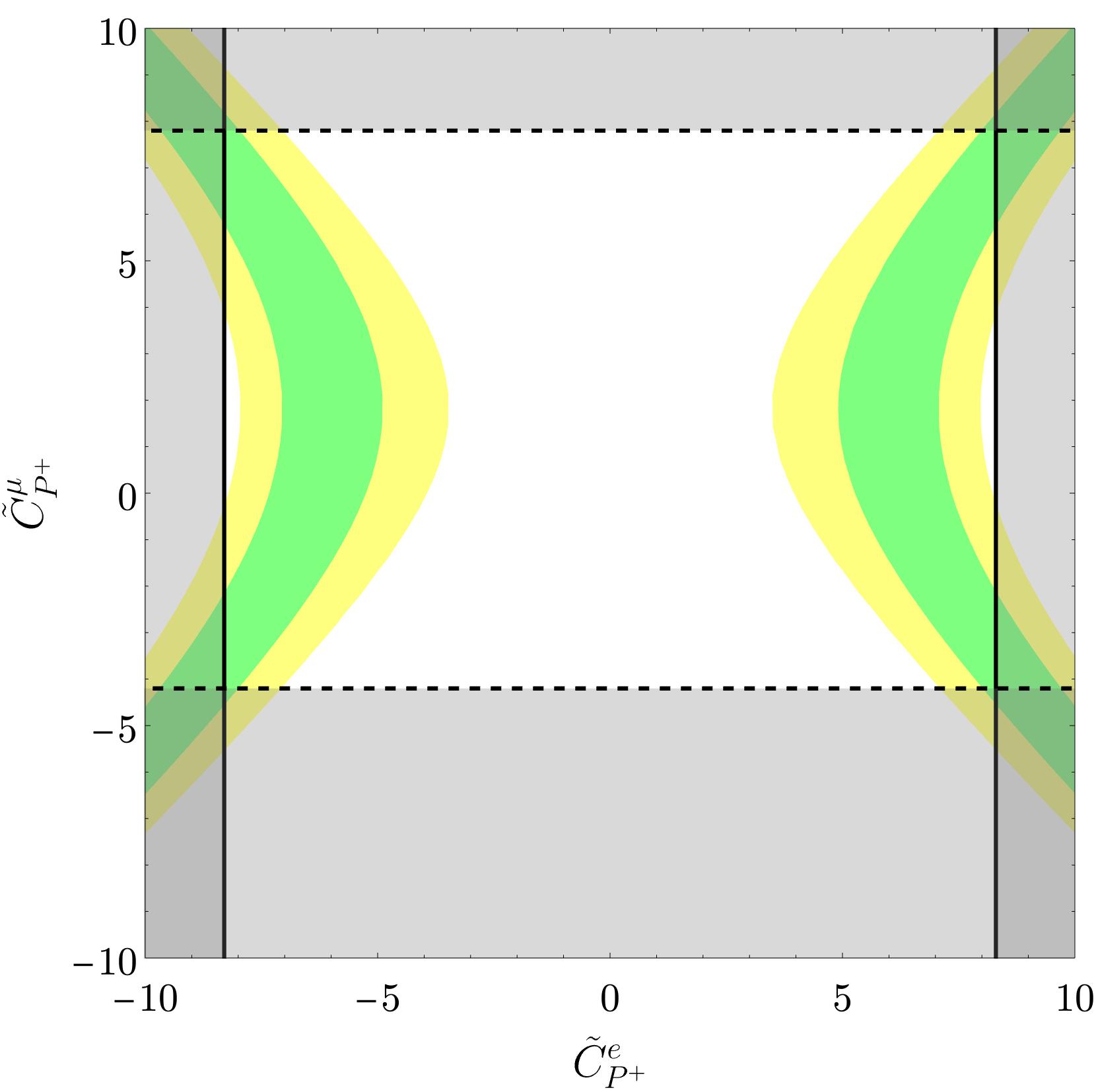}}\quad
\subfigure[\em $R_{K^\ast}$\label{fig:RKstarHALPUL}]%
{\includegraphics[width=0.48\textwidth]{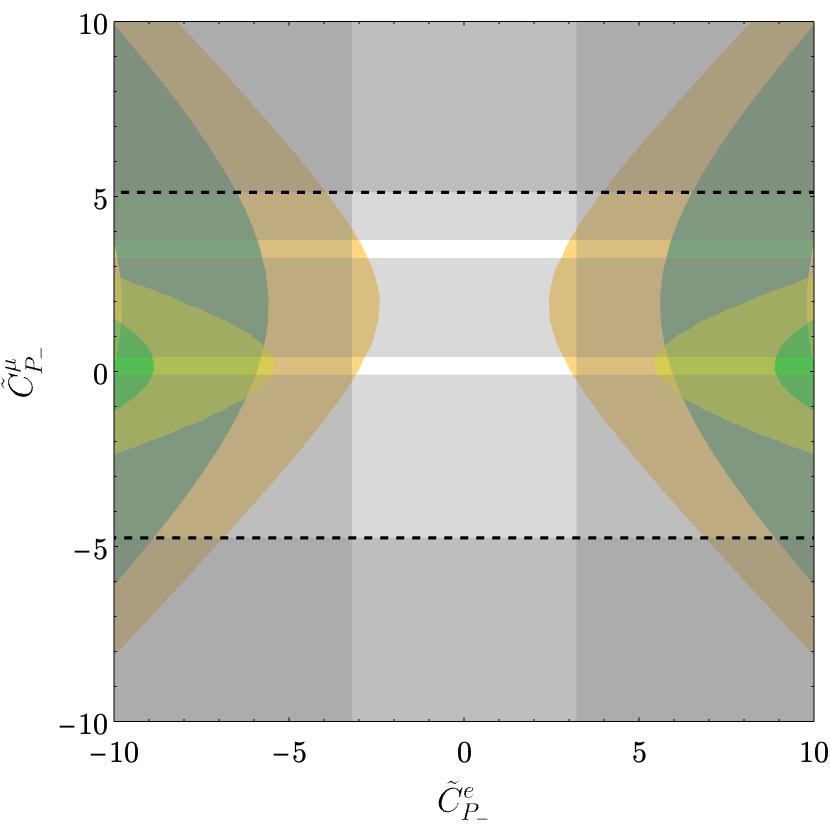}}
\caption{\em Parameter space for $R_K$ (left) and $R_{K^{\ast}}$(right) for an ALP lighter than $2m_\mu$.  In light yellow and  light green  are respectively depicted the $1\sigma$ and $2\sigma$ solutions to the central bin of $R_{K^\ast}$, while darker shades of those colours on the right plot denote the corresponding solutions for the low bin of $R_{K^{\ast}}$. The grey regions around the frame of the figures are excluded  at $2\sigma$  by data on semileptonic $B\to K^{(\ast)}\mu^+\mu^-$ (dashed black contours) decays, and on the left plot also by $B\to K e^+e^-$ data (solid black contours). On the right plot, regions excluded by purely leptonic $B_s$ decays reach the central area and are also depicted in grey.}
\label{fig:limitsCP2sTilde}
\end{figure}

\unboldmath
In the illustrative case 
$\widetilde{C}^\mu_{P_+}=0$, the regions in parameter space which now allow to explain $R_K$ within the $2\sigma$ region are $\widetilde{C}^e_{P_+}\in[-8.2,\,-3.9]\vee[4.0,\,8.2]$. When the complete bi-dimensional parameter space $\{C^e_{P_+}, C^\mu_{P_+}\}$ is considered, the allowed regions in parameter space can be seen in Fig.~\ref{fig:RKHALPUL}, leading to the $2\sigma$ allowed range:
\be
\begin{split}
& \widetilde{C}^e_{P_+}\in[-8.3,\, -3.5]\vee[3.5,\,8.3]\,,\\
& \widetilde{C}^\mu_{P_+}\in[-4.2,\,7.8]\,,
\end{split}
\label{LimitsCl-RK_LALP}
\ee
where the bounds which stem from the data on  semileptonic $B\to K$ decays have already been taken into account. Comparing these results with the expressions in Eq.~\eqref{Ctildes-simplified}, a naive estimation for the ratio between electron-ALP and muon-ALP  couplings is obtained: 
\be
\left|\dfrac{\left(\bc_e-\bc_L\right)_{\mu\mu}}{\left(\bc_e-\bc_L\right)_{ee}}\right|\approx4.8\times 10^{-3}\left|\dfrac{\widetilde{C}^\mu_{P_+}}{\widetilde{C}^e_{P_+}}\right|\lesssim10^{-2}\,.
\label{HierarchymueLALP}
\ee

\boldmath 
\subsubsection*{$\Delta {M_{s}}$}
\unboldmath
The data on meson oscillations provide bounds on quark-ALP  couplings similar to those previously obtained for the heavy ALP,\footnote{Stronger bounds will be obtained further below from semileptonic resonant searches for $\left(\bc_d-\bc_Q\right)_{sb}$ in a particular range of $m_a$, see Eq.~\eqref{VeryLightALPSemileptonicQuarkBound}.}
\be
\dfrac{\left(\bc_d\pm\bc_Q\right)_{sb}}{f_a} \lesssim 10^{-5} \GeV^{-1} \,,
\label{Ms-verylight}
\ee
which, taking into account Eq.~(\ref{Ctildes-simplified}) and the range of solutions for $R_K$ in Eq.~\eqref{LimitsCl-RK_LALP}, implies 
\be
\begin{aligned}
\dfrac{|\left(\bc_e-\bc_L\right)_{ee}|}{f_a} \gtrsim 0.3\GeV^{-1}\,, \qquad 
\dfrac{|\left(\bc_e-\bc_L\right)_{\mu\mu}|}{f_a} \gtrsim 1.6
\times 10^{-3}\GeV^{-1}\,,
\end{aligned}
\label{LeptonCouplingSolutionsRKLightALP}
\ee
 in order to solve $R_K$. At least for the electron case, these large  values for ALP-lepton coupling are borderline with respect to the validity of the EFT.  

\subsubsection*{Anomalous magnetic moment of the electron and the muon}

\begin{figure}[h!] 
\centering%
\includegraphics[width=0.6\textwidth]{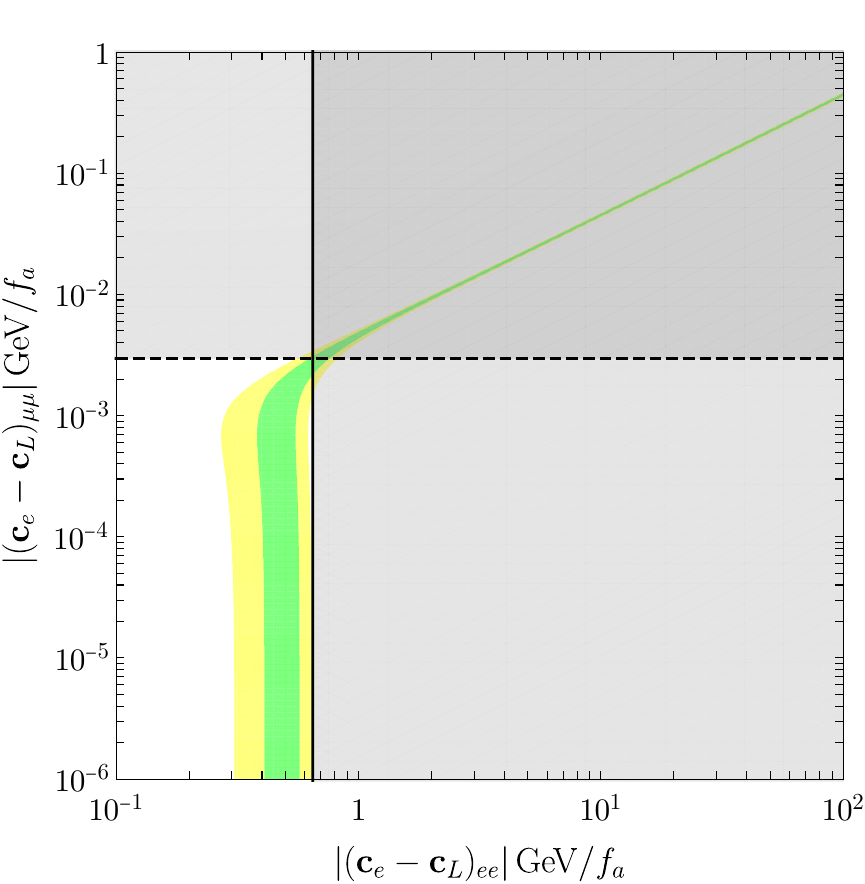}
\caption{\em  Very light ALP ($m_a\leq 2 m_\mu$). 
Parameter space $(\bc_e-\bc_L)_{ee}/f_a$ \vs $(\bc_e-\bc_L)_{\mu\mu}/f_a$ that solves the $R_K$ anomaly assuming ${(\bc_Q+ \bc_d)_{sb}/f_a=10^{-5}\GeV^{-1}}$.  The green (yellow) regions correspond to the allowed parameter space at $1\sigma$ ($2\sigma$). In grey are represented the experimental bounds from $B\to K\mu^+ \mu^-$ (enclosed by the dashed line) and $B\to K e^+ e^-$ (enclosed by the solid line).}
\label{fig:RKlig}
\end{figure}

Similarly to the case of larger $m_a$ values explored in previous sections, for the very light ALP the contribution from the diagram in Fig.~\ref{fig:DiagramsMagneticMomentLH} would be largely insufficient by itself to either saturate the $\Delta a_e$ bound or to account for the $\Delta a_\mu$ anomaly. The dominant contribution would then stem from the insertion of gauge anomalous ALP-couplings in Fig.~\ref{fig:DiagramsMagneticMomentRH}, and in particular from the photon-photon one. 

Given the ALP mass range under consideration, $1\MeV<m_a< 2 m_\mu$, the expression in Eq.~\eqref{eq:heavyALP-g-2} and the ensuing  bound from $\Delta a_e$ still applies to this very light ALP case because $m_a>m_e$. In contrast, a new analysis is in order for the ALP contribution to $\Delta a_\mu$, which results in 
\be
\Delta a_\mu^{\text{ALP}} \simeq \, \widetilde{A}_\mu \, 
\dfrac{c_{a\gamma\gamma}+\widetilde{\Delta}c_{a\gamma\gamma}}{f_a}\,
\dfrac{\left(\bc_e-\bc_L\right)_{\ell\ell}}{f_a}\,
 \,,
\label{ElectronG2eeLIGHT}
\ee
where  
\be
\widetilde{A}_{\mu} \equiv  \dfrac{m_\mu^2}{2 \pi^2}\left( \log \dfrac{\Lambda^2}{m_\mu^2} - 1 \right) = 
1.0\times 10^{-2} \GeV^2\,,
\ee
for $\Lambda=4\pi f_a\approx 1\TeV$.  In the $m_a \ll m_\mu$ limit the $\Delta c_{a\gamma\gamma}$ coefficient --induced by the anomalous chiral rotations needed to reach the mass basis-- is modified, as the anomalous contribution proportional to $\left(\bc_e-\bc_L\right)_{\mu\mu}$ cancels exactly with the one-loop corrections from the pseudoscalar ALP-muon coupling to the $a F \tilde{F}$ coupling~\cite{Bonilla:2021ufe}. Thus, a different coefficient, $\widetilde{\Delta} c_{a\gamma\gamma}$, is defined, which only contains the anomalous contribution proportional to the ALP-electron coupling constant $\left(\bc_e - \bc_L\right)_{ee}$, i.e.
\be
\widetilde{\Delta} c_{a\gamma\gamma} \equiv \,  - \dfrac{\alpha_{em}}{4 \pi}  \left(\bc_e-\bc_L\right)_{ee}  \,,
\label{ElectronG2ee}
\ee
and replaces $\Delta c_{a\gamma\gamma}$ in the $m_a \ll m_\mu$ limit. This means that the strong astrophysical bounds apply to the combination
\be
c_{a\gamma\gamma}+\widetilde{\Delta} c_{a\gamma\gamma} \simeq \, c_{a\gamma\gamma}  - \dfrac{\alpha_{em}}{4 \pi}  \left(\bc_e-\bc_L\right)_{ee}< 10^{-11}\GeV^{-1}\,.
\label{astroph}
\ee
It follows that an explanation at  the $2\sigma$ level of the data on $g-2$ of the electron and the muon in terms of  the exchange of  a very light ALP  leads respectively to the requirements
\be
\dfrac{1}{f_a} \left[     
\left(\bc_e-\bc_L\right)_{ee} \left( \left(\bc_e-\bc_L\right)_{ee} - \dfrac{4\pi}{\alpha_{em}} \, c_{a\gamma\gamma} \right)\right]^{1/2}\in [0.03,\,0.10]\GeV^{-1}\,,
\label{ElectronG2ee-bounds-very-light}
\ee
and 
\be
\dfrac{1}{f_a} \left[ \left(\bc_e-\bc_L\right)_{\mu\mu} \left( \dfrac{4\pi}{\alpha_{em}} c_{a\gamma\gamma} - \left(\bc_e-\bc_L\right)_{ee} \right)\right]^{1/2} \in [0.02,\,0.03] \GeV^{-1}\,,
\label{ElectronG2mumu-bounds-very-light}
\ee
where $m_a = 10\MeV$ has been used as illustration.  

Given the constraint in Eq.~(\ref{astroph}), the very large values  of $\left(\bc_e-\bc_L\right)_{ee}$  required to satisfy  $\Delta a_e$  in Eq.~(\ref{ElectronG2ee-bounds-very-light}) are incompatible with the possible explanations of $R_K$ in terms of the exchange of a very light ALP; see Fig.~\ref{fig:RKlig}. Thus, experimental data  exclude by themselves such a solution to $R_K$. Note that even if this had not been the case, the very large values of $\left(\bc_e-\bc_L\right)_{ee}$ required would have lied outside the regime of validity of the EFT by several orders of magnitude.
 
\boldmath
\subsection{$R_{K^\ast}$, $B\to K^\ast a(e^+ e^-)$, $B_s\to\ell^+\ell^-$ and magnetic moments}
\unboldmath

\boldmath 
\subsubsection*{$R_{K^\ast}$}
\unboldmath
An explanation of the $R_{K^*}$ anomalies in terms of the exchange of a very light ALP leads to 
\be
\text{for $\widetilde{C}^\mu_{P_-}=0$:}\qquad\begin{cases}
\widetilde{C}^e_{P_-}\in[-14.0,\,-5.4]\vee[5.4,\,15.3]\qquad&\text{central bin}\\ 
\widetilde{C}^e_{P_-}\in[-14.0,\,-3.0]\vee[3.0,\,15.3]&\text{low bin}
\end{cases}
\ee
while Fig.~\ref{fig:RKstarHALPUL} illustrates the enlarged range for $\widetilde{C}^\mu_{P_-}\ne0$, which translates into a parameter space of solutions $\{\widetilde{C}^e_{P_-}, \widetilde{C}^\mu_{P_-}\}$ with 
\be
\widetilde{C}^\mu_{P_-} \in[-4.8,\,5.1]\qquad
\text{central  and low bin}
\label{CtildelpmRKstarmu}
\ee
and 
\be
\begin{cases}
\widetilde{C}^e_{P_-} \in[-14.0,\,-5.4]\vee[5.4,\,14.0]\qquad&\text{central bin}\\[2mm]
\widetilde{C}^e_{P_-} \in[-14.0,\,-2.4]\vee[2.4,\,14.0]\qquad&\text{low bin}
\end{cases}
\label{CtildelpmRKstare}
\ee
As before, these bounds already take into account  the non-resonant semileptonic $B \to K^*$ constraints.  It follows from Eq.~(\ref{Ctildes-simplified}) that
\be
\left|\dfrac{\left(\bc_e-\bc_L\right)_{\mu\mu}}{\left(\bc_e-\bc_L\right)_{ee}}\right|\approx4.8\times 10^{-3}\left|\dfrac{\widetilde{C}^\mu_{P_-}}{\widetilde{C}^e_{P_-}}\right|\lesssim\begin{cases}
5\times10^{-3}\qquad&\text{central bin}\\[2mm]
10^{-2}\qquad&\text{low bin}
\end{cases}
\label{HierarchymueLALPRKstar}
\ee
which shows the necessity of a moderate hierarchy between the electronic and muonic couplings of the ALP.

\boldmath 
\subsubsection*{$B\to K^\ast a(e^+e^-)$}
\unboldmath
For  a light ALP with mass value within the $q^2$ bin $(0.0004,0.05)\GeV^2$, that is with $m_a>10\MeV$, data from  resonant $B\to K^\ast a(e^+e^-)$ searches  are available. As the ALP is on-shell, it is possible to use the NWA as in the previous section. 
Because $\cB (a \to e^+ e^-)=1$, it is then possible to infer directly from those data a very strong bound on  ALP-quark couplings, given by
\begin{equation}
\dfrac{\left(\bc_d - \bc_Q\right)_{sb}}{f_a} \lesssim 8 \times 10^{-10}\,\text{GeV}^{-1}\,,
\label{VeryLightALPSemileptonicQuarkBound}
\end{equation}
which in order to account now for the $R_{K^\ast}$ anomaly leads to the following constraints on  ALP-lepton couplings, in that range of $m_a$, 
\be
\dfrac{|\left(\bc_e-\bc_L\right)_{\mu\mu}|}{f_a} \in[-23.8,\,22.4] \qquad
\text{central and low bin}\,,
\ee
together with
\be
\begin{cases}
\dfrac{|\left(\bc_e-\bc_L\right)_{ee}|}{f_a} \in[-13.5,\,-5.2]\times 10^3 \vee[5.2,\,13.5]\times 10^3\qquad&\text{central bin}\\[2mm]
\dfrac{|\left(\bc_e-\bc_L\right)_{ee}|}{f_a} \in[-13.5,\,-2.3]\times 10^3\vee[2.3,\,13.5]\times 10^3\qquad&\text{low bin}
\end{cases}
\ee
again strongly at odds with the range of validity of the EFT.

\boldmath 
\subsubsection*{$B_s\to  \ell^+\ell^-$}
\unboldmath

The contributions of the SM plus ALP exchange to the branching ratios for the purely leptonic decays of the $B_s$ meson are given by 
\be
\begin{aligned}
\ov{\BR}(B_s\to \mu^+\mu^-)
& = 10^{-9}\times\left(3.67-3.99\, \tilde{C}^\mu_{P_-} +1.09\,\tilde{C}^{\mu2}_{P_-}\right)\,,
\\
\ov{\BR}(B_s\to  e^+e^-)
&=10^{-14}\times\Big(8.58-1.93\times 10^{3} \tilde{C}^e_{P_-} +1.09\times 10^{5}\,\tilde{C}^{e2}_{P_-}\Big)
\,,
\end{aligned}
\label{BKstarmumuVeryLightALPLow}
\ee
with a theoretical error of $4\%$ at the $1\sigma$ level. It follows that the  $2\sigma$ allowed regions in parameter space are
\be
\begin{cases}
\widetilde{C}^e_{P_-}&\in[-3.2,\,3.2]\,,\\[2mm]
\widetilde{C}^\mu_{P_-}&\in[-0.095,\,0.41]\vee[3.2,\,3.8]\,.
\end{cases}
\ee
This is an interesting point for this very light ALP case, as illustrated in Fig.~\ref{fig:RKstarHALPUL}. It shows that -- in contrast with the scenario for a heavy ALP -- there is $2\sigma$ {\it compatibility between the solutions to the $R_{K^\ast}$ anomaly --low bin-- and the data on $B_s\to  \mu^+\mu^-$ and $B_s\to  e^+e^-$}: the four small yellow square regions with white background in Fig.~\ref{fig:RKstarHALPUL} survive, corresponding to 
\be
\left(\widetilde{C}^e_{P_-},\,\widetilde{C}^\mu_{P_-}\right)\simeq\Big\{(-3,\,0),\,(-3,\,3.4),\,(3,\,0),\,(3,\,3.4)\Big\}\,.
\label{fourpoints}
\ee
The prize in this case is again theoretical, as these solutions imply ALP-lepton couplings outside the range of validity of the EFT. Indeed, applying the bounds from dedicated resonant searches in  Eq.~\eqref{VeryLightALPSemileptonicQuarkBound}, the four points listed above translate into the following unacceptably large values for ALP-lepton couplings: 
\be
\left(\dfrac{\left(\bc_e-\bc_L\right)_{ee}}{f_a},\,\dfrac{\left(\bc_e-\bc_L\right)_{\mu\mu}}{f_a}\right)\simeq\Big\{
(\pm 3 \times 10^3,\,0),\,(\pm 3\times 10^3,\,17)\Big\}\GeV^{-1}\,.
\label{LeptonCouplingSolutionsRKstarLightALP}
\ee

\subsubsection*{Anomalous magnetic moment of the electron and the muon}
The analysis   of ALP exchange on leptonic magnetic moments, compared with the solutions to $R_{K^\ast}$,  parallels that for the solutions to $R_K$ above. In particular,  the data on the set of observables  $\{R_{K^\ast}, \Delta a_{e}\}$ require tree-level photon couplings that are too large to comply with the existent astrophysical constraints~\cite{Bauer:2017ris} (and are also incompatible with the EFT validity conditions).

%
%
\section{Conclusions}
\label{sec:Concls}

 We have analysed  the technical {\it and} the theoretical cost required to explain the neutral anomalies in $B$-meson decays via the tree-level exchange of an ALP.  Within the ALP effective field theory and assuming ALP-$bs$ couplings, the complete two-dimensional  parameter space for flavour-diagonal ALP  couplings to electrons and muons is explored (considering, in addition, ALP-photon couplings when certain loop-level effects require it).  The range of ALP masses contemplated sweeps from heavy ALPs, i.e. heavier than the $B$ mesons, to very light ALPs down to 1 MeV --which is the lower value allowed by astrophysical constraints on the ALP-electron coupling.

The predictions for $R_K$ and the two bins of $R_{K^\ast}$ are confronted with the impact of ALP exchange on other observables, namely meson oscillations ($\Delta M_s$),  $B_s\to \ell^+ \ell^-$ decays, ${B\to K^{(\ast)} \ell^+ \ell^-}$ decays --including searches for new resonances-- and astrophysical constraints. The data on these observables severely limit the available parameter space.  Furthermore, we have analysed the impact of the solutions found on the $g-2$ of the electron and of the muon. 

The solutions allowed are then compared with the theoretical conditions for the validity of the ALP EFT, requiring to remain within the perturbative domain of the effective theory on the assumption that the ALP scale is at least of the order of the electroweak scale, and the ALP mass under it. 

For a heavy ALP, no viable explanation of the neutral anomalies in terms of tree-level ALP exchange survives. Solutions to $R_K$ compatible with other observables --except the  $\Delta a_{\mu}$ anomaly-- are found. Nevertheless, they are in strong conflict with the EFT validity conditions: in order to account for $R_K$, the very small ALP-quark couplings required by $\Delta M_s$ data require in turn  ALP-lepton effective couplings  unacceptably large from the theoretical point of view.   In the case of $R_{K^\ast}$, all ALP mediated solutions  are directly excluded by the data on $B_s \to \ell^+ \ell^-$ irrespective of EFT consistency considerations.  
 
A similar fate applies to the other extreme of the ALP mass range:  ALPs with mass smaller than the energies of the low-energy bin of  $R_{K^\ast}$ are also excluded. In fact, we do find solutions to the $R_{K}$  or  the $R_{K^\ast}$ anomalies  allowed within $2\sigma$ by the other observables mentioned; for instance the explanation of $R_{K^\ast}$ and the $B_s \to \ell^+ \ell^-$ data are in this case compatible within $2\sigma$.  From the theoretical point of view, the validity constraints of the EFT are (in)compatible by themselves with the values required by the $R_{K}$ ($R_{K^\ast}$). Nevertheless, all solutions via these very light ALPs are excluded by the experimental bound on $\Delta a_e$, since in this case, astrophysical bounds set strong constraints on the effective photon coupling.

In contrast to the above, an ALP lighter than the $B$ mesons but with a mass value within any of the bin windows provides an altogether different perspective. The ALP exchanged can then be on-shell and enter a resonant regime: $B\to K^{(\ast)}\ell^+\ell^-$ processes factorise into ALP on-shell production followed by decay. In this situation, the ALP coupling to muons must be much smaller than that  to electrons to explain the neutral $B$-anomalies and thus $\BR(a\to e^+ e^-)\sim \mathcal{O}(1)$. The latter implies in turn that $R_K$ and/or $R_{K^\ast}$ become rather independent of the precise values of ALP leptonic couplings, and the solutions, therefore, escape from the theoretical problems with the EFT validity encountered for either heavy or extremely light ALPs. In this mass regime, we have also taken into account the validity requirements for the narrow-width approximation  and  for prompt ALP decays. The latter defines a minimum electron coupling for the solutions to $R_{K^{(\ast)}}$  and hence  the parameter space compatible with the EFT validity constraints is reduced even in this on-shell regime.  

Within the allowed parameter space for on-shell ALP exchange, we have furthermore identified a golden ALP mass value which lies at the frontier between the two energy bins for $R_{K^\ast}$, $m_a= \sqrt{1.1}\GeV$, and which becomes a broader mass range when smearing effects --associated to the finite experimental precision-- are estimated.  These golden mass values  provide solutions which could {\it a priori } explain the three anomalies, i.e.  $R_K$ together with the two bins of  $R_{K^\ast}$, always remaining compatible with the observables mentioned above and with the EFT validity constraints. While solutions in-between bins are always suspect, they are technically allowed  and prompt the convenience to perform a slightly different experimental binning, which could easily clean up this avenue. 

When the loop-level impact of the Lagrangian couplings are considered for an on-shell ALP, it is also possible to  account simultaneously for the data in the sets $\{R_{K^{(\ast)}}, \Delta a_e\}$, while once again the $ \Delta a_\mu$ anomaly cannot be then accounted for.  Nonetheless, given the large electron couplings required by the analysis, their loop-level impact becomes relevant for some set of data. Correspondingly, some level of fine-tuning is called for to comply with the experimental bounds on $\Delta a_e$, as well as those obtained via $B\to K^{(\ast)} a(e^+e^-)$ searches. This adds to the already established theoretical cost of  the ALP solution to the neutral $B$-anomalies. A complete loop-analysis is beyond the scope of this paper. Along the same line, we have not addressed the so-called charged flavour $B$-anomalies, as they cannot be explained by tree-level ALP exchanges.

We have exposed the high cost and conditions required to explain the neutral $B$-anomalies via tree-level ALP exchange. This is furthermore within the assumption --customary in the literature-- that the only new physics couplings  present in Nature are non-diagonal $bs$-ALP couplings and diagonal electron and muon ALP couplings as defined in the mass basis, instead of the most natural flavour basis. Nevertheless, the potential groundbreaking implications of the flavour $B$-anomalies, would they turn to be definitely confirmed by experiment, prompt to let no stone unturned. The broad ALP arena is a generic and compelling option to explore.

\section*{Acknowledgements}

The authors acknowledge Cristopher Bobeth, Gudrun Hiller and Jos\'e Miguel No for useful discussions. The authors acknowledge as well partial financial support by the Spanish Research Agency (Agencia Estatal de Investigaci\'on) through the grant IFT Centro de Excelencia Severo Ochoa No CEX2020-001007-S and by the grant PID2019-108892RB-I00 funded by MCIN/AEI/ 10.13039/501100011033, by the European Union's Horizon 2020 research and innovation programme under the Marie Sk\l odowska-Curie grant agreement No 860881-HIDDeN. 
 The work of J.B. was supported by the Spanish MICIU through the National Program FPU (grant number FPU18/03047). The work of A.d.G. and M.R. was supported by the European Union's Horizon 2020  Marie Sk\l odowska-Curie grant agreement No 860881-HIDDeN.

\appendix
\section{The input data and SM predictions}
\label{sec:Input}

The parameters used for the computations, as well as the SM predictions used to derive the constraints along this work,  are shown in Tab.~\ref{tab:parameters}~and Tab.~\ref{tab:smBR}, respectively. 

\begin{table}[h!]
\centering
\begin{tabular}{c|c|c}
Parameter & Value & Unit of Measure \\
\toprule
$\alpha_\text{em} (m_b)$ & $0.007518797$ & - \\
$G_F$ &  $1.1663787(6)\times10^{-5}$ & GeV$^{-2}$ \\
\midrule
$m_e$ & 0.000510999 & GeV\\
$m_\mu $ & 0.105658 & GeV\\
$\bar{m}_s(2\ \text{GeV})$ & $0.093^{+ 0.011}_{-0.005}$ & GeV\\
$\bar{m}_b({m}_b)$ & $4.18^{+ 0.03}_{-0.02}$ & GeV\\
$M_{B_s}$ & $5.36688 \pm 0.00014$ & GeV\\
$M_{B^0}$ & $5.27965 \pm 0.00012$ & GeV\\
$M_{B^\pm}$ & $5.27934 \pm 0.00012$  & GeV\\
$M_{K^\pm}$ & $0.493677\pm 0.000016$ & GeV\\
$M_{K^{0*}}$ & $0.89555\pm 0.0008$ & GeV\\
\midrule
$\tau_{B_s}$ & $(1.516 \pm 0.004) \cross 10^{-12}$ & $s$\\
$\tau_{B^0}$ & $(1.519 \pm 0.004) \cross 10^{-12}$ & $s$\\
$\tau_{B^\pm}$ & $(1.638 \pm 0.004) \times 10^{-12}$ & $s$\\
\midrule
$|V_{ts}|$& $0.04065^{+0.00040}_{-0.00055}$ & -\\
$|V_{tb}|$& $0.999142^{+0.000018}_{-0.000023}$ & -\\
$C_{7}$ & $-0.33726473$ & -\\
$C_9$ &$4.27342842$ & -\\
$C_{10}$ & $-4.16611761$& - \\
\bottomrule
\end{tabular}
\caption{\em Parameters used for the computations. The quark masses are estimates of the $\overline{\text{MS}}$ scheme at the given renormalisation scale~\cite{Workman:2022ynf}. The values of the WET Wilson coefficients are those used in EOS~\cite{vanDyk:2021sup}.
}
\label{tab:parameters}
\end{table}

\begin{table}[h!]
\centering
\begin{tabular}{c|c|c}
SM Prediction & $q^2~\text{[GeV]}^2$ & Value \\
\toprule
$\mathcal{B} (B\to K \ell^+ \ell^-)$ & [1.1, 6.0] & $(1.71\pm0.29) \times 10^{-7}$  \\
\midrule
$\mathcal{B} (B\to K^* \ell^+ \ell^-)$ & [0.0004, 0.05] & \begin{tabular}{c} $e^+ e^-$: $(1.65 \pm 0.31) \times 10^{-7}$  \\ $\mu^+ \mu^-$: $(1.28 \pm 0.24) \times 10^{-9}$ \end{tabular} \\
\midrule
$\mathcal{B} (B\to K^* \ell^+ \ell^-)$ & [0.05, 0.15] & \begin{tabular}{c} $e^+ e^-$: $(3.94 \pm 0.69) \times 10^{-8}$  \\ $\mu^+ \mu^-$: $(3.28 \pm 0.60) \times 10^{-8}$ \end{tabular} \\
\midrule
$\mathcal{B} (B\to K^* \ell^+ \ell^-)$ & [0.15, 0.25] & \begin{tabular}{c} $e^+ e^-$: $(1.96 \pm 0.34) \times 10^{-8}$  \\ $\mu^+ \mu^-$: $(1.92 \pm 0.33) \times 10^{-8}$ \end{tabular} \\
\midrule
$\mathcal{B} (B\to K^* \ell^+ \ell^-)$ & [0.25, 0.4] & \begin{tabular}{c} $e^+ e^-$: $(1.94 \pm 0.31) \times 10^{-8}$ \\ $\mu^+ \mu^-$: $(1.92 \pm 0.30) \times 10^{-8}$ \end{tabular} \\
\midrule
$\mathcal{B} (B\to K^* \ell^+ \ell^-)$ & [0.4, 0.7] & \begin{tabular}{c} $e^+ e^-$: $(2.62 \pm 0.38) \times 10^{-8}$  \\ $\mu^+ \mu^-$: $(2.61 \pm 0.37) \times 10^{-8}$ \end{tabular} \\
\midrule
$\mathcal{B} (B\to K^* \ell^+ \ell^-)$ & [0.7, 1] & \begin{tabular}{c} $e^+ e^-$: $(1.98 \pm 0.26) \times 10^{-8}$  \\ $\mu^+ \mu^-$: $(1.97 \pm 0.29) \times 10^{-8}$ \end{tabular} \\
\midrule
$\mathcal{B} (B\to K^* \ell^+ \ell^-)$ & [1.1, 6.0] & $(2.53 \pm 0.36) \times 10^{-7}$ \\
\midrule
$\mathcal{B} (B\to K^* \ell^+ \ell^-)$ & [0.1, 8.0] & \begin{tabular}{c} $e^+ e^-$: $(4.87 \pm 0.65) \times 10^{-7}$  \\ $\mu^+ \mu^-$: $(4.82 \pm 0.68) \times 10^{-7}$ \end{tabular} \\
\midrule
$\ov{\BR} (B_s \to \mu^+ \mu^-) $ & - & $(3.67 \pm 0.15) \times 10^{-9}$ \\
\midrule
$\ov{\BR} (B_s \to e^+ e^-) $ &  - & $(8.58\pm 0.35) \times 10^{-14}$ \\
\bottomrule
\end{tabular}
\caption{\em SM predictions relevant for the analyses discussed in this work, computed directly from flavio~\cite{Straub:2018kue}. }
\label{tab:smBR}
\end{table}

\boldmath
\section{Details of $B\to K^{(*)} \ell \ell$ computations}
\unboldmath
\label{sec:formulas}

The differential semileptonic decay rates considered in this work depend strongly on the values of the form factors, which have been calculated using different models and methods in the literature. The central values of the $B\to K$ form factors presented in Ref.~\cite{Bailey:2015dka} have been used in this work, under the standard BCL parameterisation~\cite{Bourrely:2008za}.
In the $B\to K^*$ case the central values reported in Ref.~\cite{Bharucha:2015bzk} have been used instead.

\begin{figure}[h!]
\centering
{\includegraphics[scale=0.54]{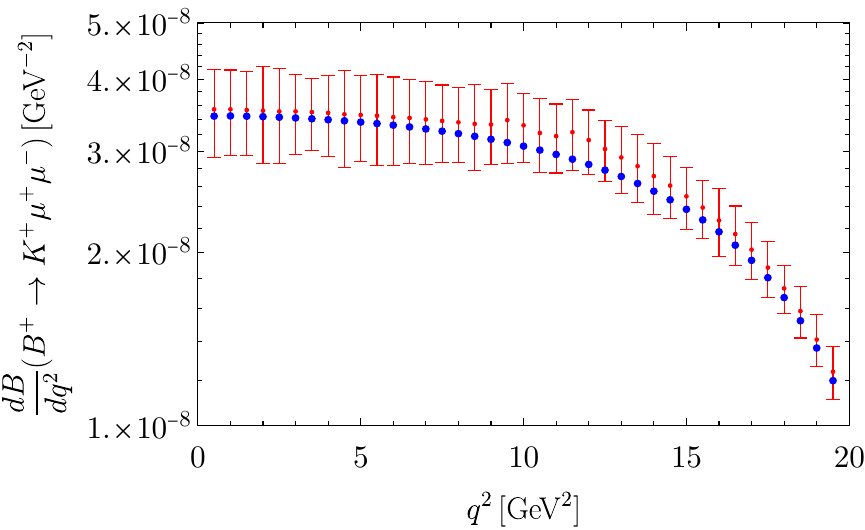}}
{\includegraphics[scale=0.54]{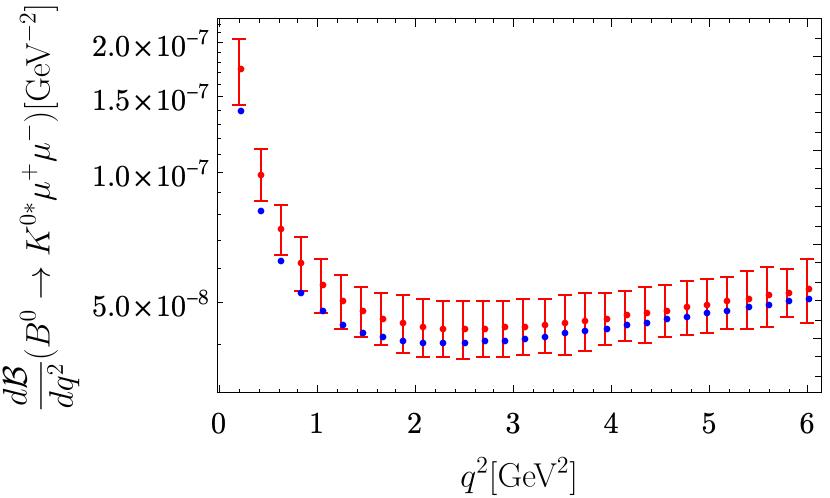}}
\caption{\em The SM prediction for the differential distributions obtained using Flavio (red dots) \vs our analytical formulas (blue dots). Error bars include the theoretical uncertainties in the SM prediction.}
\label{fig:comparison}
\end{figure}

We have cross-checked our analytical expressions, used in all figures presented in section~\ref{sec:LIGHTALP} and to cross-check the results in other sections, by comparing the differential distributions $\dd \mathcal{B}(B\to K^{(*)} \ell^+ \ell^-)/\dd q^2$ assuming only the SM with the output from Flavio; see Fig.~\ref{fig:comparison}. The accuracy between the two results is evident. The corresponding error bands, obtained with the same software, are also shown. Such theoretical uncertainties, together with the experimental ones, have been included to estimate the bounds on the new physics couplings. On the other hand, we neglect the theory errors associated to the new physics branching ratios, as they are expected to have a negligible impact compared to the SM ones. In fact, the former are typically $\mathcal{O}(15\%)$ of the latter.

Using the new form factors presented by the FLAG collaboration~\cite{Aoki:2021kgd}, the results from our analytical expressions remain compatible with the Flavio output, although the central values in Fig.~\ref{fig:comparison} show variations of $\mathcal{O}(7\%)$.

\boldmath
\section{Bounds from binned $B\to K^* e^+ e^ -$ data}
\unboldmath
\label{sec:BtoKstaree-limits}

\begin{figure}[t]
\centering
\includegraphics[width=0.6\textwidth]{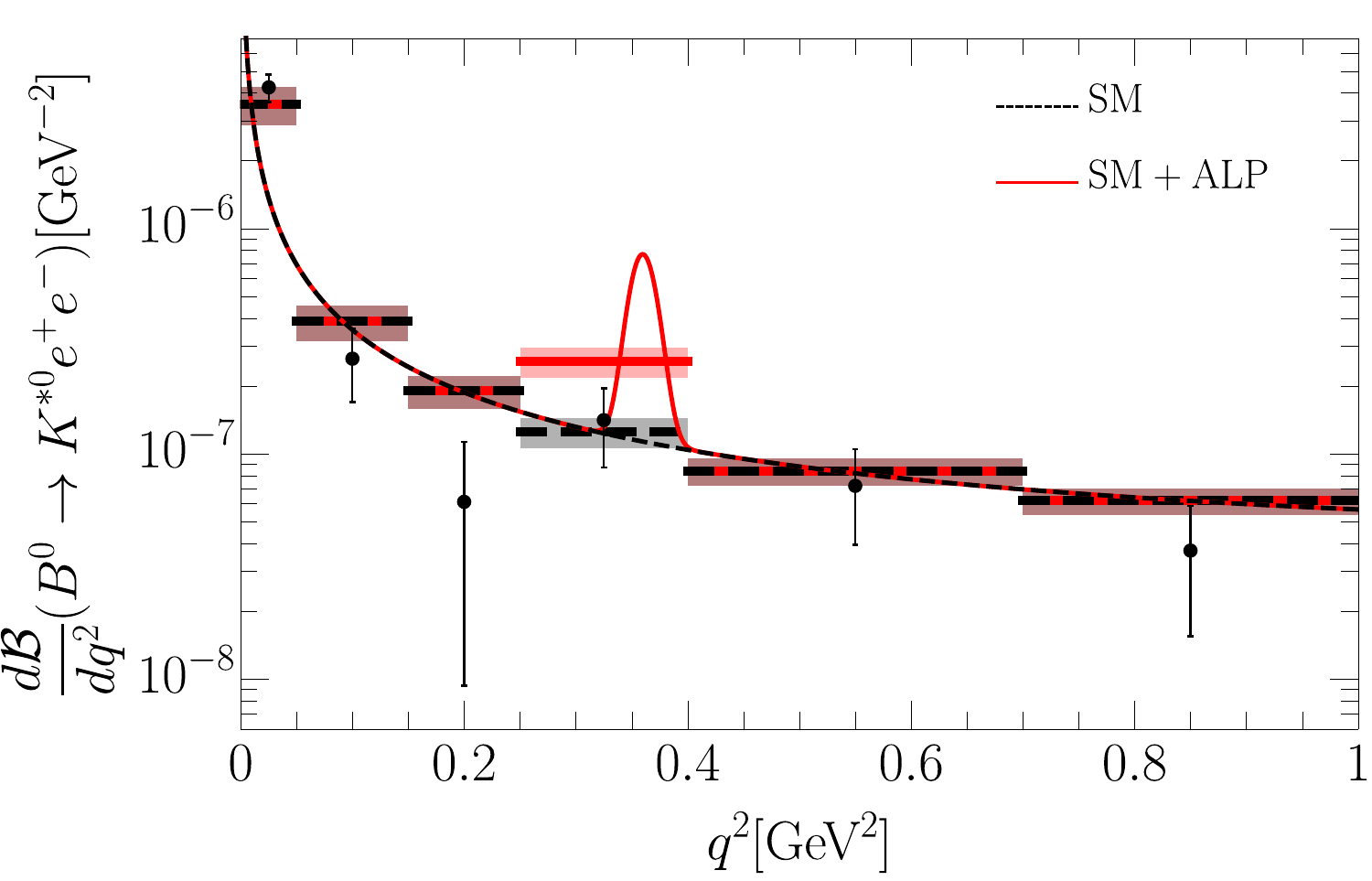}
\caption{\em The experimental limits on the differential branching ratio ${\dd \mathcal{B}/\dd q^2 (B\to K^* e^+ e^-)}$ obtained from Ref.~\cite{LHCb:2015ycz}, together with the SM (in black) and the NP (in red) predictions. Both the continuous and binned distributions, divided in the 6 measured bins of $q^2$, are presented. For the NP, we have assumed an ALP with $m_a= 0.6$ GeV and decaying $\approx 100\%$ into electrons. The quark coupling is set to $|(\mathbf{c}_d - \mathbf{c}_Q)_{sb}|/f_a =3.05 \times10^{-10}\,\text{GeV}^{-1}$.}
\label{Fig:LHCb_BtoKstaree}
\end{figure}

Bounds from the differential distribution of the observed number of events in the ${B\to K^* e^+ e^-}$ decay, ${\dd N/\dd q^2 (B\to K^* e^+ e^-)}$, measured at LHCb~\cite{LHCb:2015ycz} constrain the product $\mathcal{B}(B \to K^* a) \times \mathcal{B} (B \to e^+e^-)$ for ALP masses within the 6 measured bins of $q^2$. In order to obtain such constraints, we have estimated the efficiency effects by comparing the number of events resulting from the Monte Carlo simulation of the SM, reported in the experimental paper, with the predictions from Flavio~\cite{Straub:2018kue}; see Tab.~\ref{tab:smBR}.
In this way, we can relate the two relevant quantities $N$ and $\mathcal{B}$ by
\begin{equation}
N(B\to K^* e^+ e^-) = \mathcal{B}(B\to K^* e^+ e^-) \, \sigma^B \, \mathcal{L} \, \epsilon \,,
\end{equation}
where $\sigma^B$ is the production cross-section of a $B$ meson at LHCb, $\mathcal{L}$ is the integrated luminosity and $\epsilon$ is the detector efficiency for a given energy bin. Hence, knowing the SM predictions for the branching fraction and the expected number of events, the quotient $N^\text{data} / \mathcal{B}^\text{data}$ can be obtained from $N^\text{SM} / \mathcal{B}^\text{SM}$.

The resulting bounds, taking into account both experimental and the SM theoretical errors\footnote{Correlations across bins are ignored in this procedure. The former are taken into account in Ref.~\cite{Altmannshofer:2017bsz} where a similar procedure was adopted using only the first two bins of $q^2$ determined in the experimental analysis; the resulting bounds are very similar to those we have obtained.}, are reported in Tab.~\ref{tab:dedicatedsearchesBKstarll} with exception of one interval, $q^2 \in [0.15,0.25]\,\text{GeV}^2$, where a tension of more than $2\sigma$ with respect to the SM prediction is observed in data, that would be worsened by the presence of NP. No beyond SM contributions are therefore considered in this bin; see Fig.~\ref{fig:OnShell_BenchMarkeemumu}.

These bounds are represented in Fig.~\ref{Fig:LHCb_BtoKstaree} along with the SM predictions. In addition, the contribution from a resonant ALP with a mass of $0.6$ GeV is also shown. For the fermionic couplings, we have adopted the benchmark defined by $(|(\mathbf{c}_e-\mathbf{c_L})_{ee}|/f_a,\,|(\mathbf{c}_e-\mathbf{c_L})_{\mu\mu}|/f_a) = (10^{-1}, 10^{-5})\,\text{GeV}^{-1}$ and  $|(\mathbf{c}_d-\mathbf{c_Q})_{bs}|/f_a| = 3.05 \times 10^{-10}\,\text{GeV}^{-1}$ corresponding to the blue star in Figs.~\ref{fig:OnShell_BenchMarkeemumu} and \ref{fig:OnShell_RKstarLEBin}.
 It becomes clear the potential of these measurements to probe the ALP parameter space relevant to the $B$ anomalies and, in particular, the advantages of using a smaller binning in order to resolve the ALP resonant peak.

\footnotesize


\providecommand{\href}[2]{#2}\begingroup\raggedright\endgroup

\end{document}